\documentclass[11pt]{article}
\usepackage{eurosym}
\usepackage{amsfonts}
\usepackage{amssymb}
\usepackage{graphicx}
\usepackage{amsmath}
\usepackage{makeidx}
\usepackage{indentfirst}
\usepackage[T1]{fontenc}
\usepackage[utf8]{inputenc}

\newcounter{resultnum}[section]
\newcounter{conclusionnum}[section]
\newcounter{conditionnum}[section]
\newcounter{conjecturenum}[section]
\newcounter{examplenum}[section]
\newcounter{exercisenum}[section]
\newcounter{lemmanum}[section]
\newcounter{notationnum}[section]
\newcounter{theoremnum}[section]
\newcounter{definitionnum}[section]
\newcounter{corollarynum}[section]
\newcounter{remarknum}[section]
\newcounter{propositionnum}[section]
\newcounter{acknowledgementnum}[section]
\newcounter{algorithmnum}[section]
\newcounter{axiomnum}[section]
\newcounter{casenum}[section]
\newcounter{claimnum}[section]
\newcounter{summarynum}[section]
\newcounter{problemnum}[section]
\begin{document}

\title{Geometric information flows and G. Perelman entropy\\
for relativistic classical and quantum mechanical systems}
\date{July 5, 2020}
\author{ \textbf{\Large Sergiu I. Vacaru}
\thanks{emails: sergiu.vacaru@gmail.com and sergiuvacaru@mail.fresnostate.edu ;\newline
\textit{Address for post correspondence in 2019-2020 as a visitor senior researcher at YF CNU Ukraine is:\ }
37 Yu. Gagarin street, ap. 3, Chernivtsi, Ukraine, 58008; the UAIC affiliation is preserved for a former hosted project IDEI during 2012-2015}  \\
{\small \textit{Physics Department, California State University at Fresno,
Fresno, CA 93740, USA;  }}\\
 {\small \textit{ Project IDEI, University "Al. I. Cuza" Ia\c si, Romania; and }} \\
 {\small \textit{Dep. Theoretical Physics and Computer Modelling, 101  Storozhynetska street;}}\\
 {\small \textit{ Yuriy Fedkovych Chernivtsi National University,  Chernivtsi, 58029, Ukraine}} 
}
\maketitle

\begin{abstract}
This work consists an introduction to the classical and quantum information theory of geometric flows of (relativistic) Lagrange--Hamilton mechanical systems. Basic geometric and physical properties of the canonical nonholonomic deformations of G. Perelman entropy functionals and geometric flows evolution equations of classical mechanical systems are described. There are studied projections of such F- and W-functionals on Lorentz spacetime manifolds and three-dimensional spacelike hypersurfaces. These functionals are used for elaborating relativistic thermodynamic models for Lagrange--Hamilton geometric evolution and respective generalized R. Hamilton geometric flow and nonholonomic Ricci flow equations. The concept of nonholonomic W-entropy is developed as a complementary one for the classical Shannon entropy and the quantum von Neumann entropy. There are considered geometric flow generalizations of the approaches based on classical and quantum relative entropy, conditional entropy, mutual information, and related thermodynamic models. Such basic ingredients and topics of quantum geometric flow information theory are elaborated using the formalism of density matrices and measurements with quantum channels for the evolution of quantum mechanical systems.

\vskip3pt


\vskip3pt

PACS2010:\ 02.40.-k, 02.40.Yy, 02.90.+p, 03.67.-a, 05.90.+m, 45.10.Na

\end{abstract}

\tableofcontents



\section{Introduction}

One of the most remarkable success in modern mathematics is the proof of the Poincar\'{e}--Thurston conjecture due to G. Perelman \cite{perelman1,perelman2,perelman3}. We cite here most important related works on W. Thurston's classification of three dimensional, 3-d, manifolds,
\cite{thurston1,thurston2,thurston3}; then D. Friedman's  geometric flow evolution equations derived for renorm-group considerations in quantum field theory and condensed matter physics, see \cite{friedan1,friedan2,friedan3}; and R. Hamilton \cite{hamilt1,hamilt2,hamilt3} fundamental contributions to Ricci flow theory. The monographs \cite{monogrrf1,monogrrf2,monogrrf3} can be considered for rigorous proofs and reviews of results in geometric analysis and topology.\footnote{We emphasize that the terms Hamilton mechanics and Hamilton equations for Ricci flows are related to the names of two different famous scientists. In the first case, it refers to William R. Hamilton who formulated in 1834 his Hamiltonian mechanics starting from Lagrangian mechanics (a previous reformulation for classical mechanics introduced by Joseph Louis Lagrange in 1788). On mathematical and physical approaches and historical remarks on
Lagrange and Hamilton mechanics, see \cite{abraham,arnold,deleon85}. In the second case, Richard Hamilton is known because of his achievements on the Ricci flows theory and applications in topology and geometric analysis \cite{hamilt1,hamilt2,hamilt3}.} A series of our works were elaborated in a
'geometry and physics' style involving generalizations for relativistic systems and applications in modern physics and cosmology. We cite \cite{vjmp08,vrmp09,vacaru11,vacaru13}, for geometric flows of Lagrange-Finsler spaces and nonholonomic manifolds and algebroids; \cite{vacaru09}, on noncommutative geometric flow evolution theories; \cite{rajpoot17,ruchin13}, for respective super-Ricci flows and thermodynamics of relativistic Ricci flows; and a series of works \cite{gheorghiu16,bubuianu19,vacaru19,vacaru19a} related to modified gravity theories, MGTs, and cosmology, see reviews \cite{capozziello10,basilakos13,elghozi15,nojiri17,vacaru18,bubuianu18}.

Above mentioned directions for advanced studies in geometry and mathematical physics were developed using G. Perelman's concepts of F- and W-entropy Perelman. Such values were constructed as A. M. Lyapunov type functionals \cite{lyapunov1892} which for geometric flows of Riemannian metrics are
determined by Ricci tensors and scalars.  We defined their nonholonomic deformations (equivalently, anholonomic, i.e. subjected to non-integrable constraints) for various generalized  geometric and physical models. The W-entropy is like a "minus entropy" and it describes some nonholonomic entropic flows of various classical and quantum physical systems. The concept of W-entropy is different from the Shannon, von Neumann, or other type, entropy used in modern thermodynamics and classical/ quantum information theory, see \cite{preskill,witten18} and references therein. With respect to various developments and applications in modern gravity and black hole, BH, and cosmology information theory (based on area--entropy, holography and similar concepts), the constructions with the G. Perelman entropy and modifications seem to be more general than those based on the Bekenstein--Hawking thermodynamics  \cite{bekenstein72,bekenstein73,bardeen73,hawking75}. On recent research with "non-area and non-holographic" entropies for geometric flows and gravity, see details and discussions in \cite%
{ruchin13,gheorghiu16,bubuianu19,vacaru19,vacaru19a,rajpoot17}.

This paper is the 4th partner one in a series of previous works \cite{bubuianu19,vacaru19,vacaru19a}. The goal is to elaborate on certain most important principles and methods for formulating classical and quantum
information theories encoding geometric flows of relativistic Lagrange-Hamilton mechanical systems. We shall also consider spacetime configurations emerging as nonholonomic Ricci solitons, and their analogous
geometric thermodynamic models. This new approach to formulating geometric information flow, GIF, theories is based on the concept of G. Perelman entropy and a geometrization of physical theories due to J. Kern \cite{kern74} and M. Matsumoto \cite{matsumoto66,matsumoto86}. The Kern-Matsumoto
ideas were that classical mechanics can be formulated as Finsler like geometries without homogeneity conditions on respective Lagrange and/or Hamilton generating functions on (co) vector and tangent bundles, see a modern axiomatic approach and historical remarks in \cite{vacaru18,bubuianu18}. For such a geometric formulation, the classical and quantum field and flow evolution theories can be characterized by certain
generalized Perelman's entropy like functionals. These functionals allow new developments and applications to classical and quantum information theories. In this work, there are not studied emergent (modified) gravity theories even we provide certain generalized classical and quantum mechanical entropic functionals from which generalized Einstein equations can be derived. We cite \cite{gheorghiu16,bubuianu19,vacaru19a} for recent results on exact solutions and modified Ricci flow theories and gravity.

It is assumed that the reader has a background knowledge about mathematical physics and geometric methods in QFT and (modified) gravity theories, and certain familiarity with fiber bundles and (non) linear connections, nonholonomic mechanics and geometric thermodynamics, see \cite{misner,vacaru09,ruchin13} and references therein. Certain other sources of literature on classical and quantum information theory and modern physics \cite{preskill,witten18,nielsen,cover,wilde} are listed for more comprehensive treatments of the subjects that we touch in our developments.

This work is organized as follows:\ In section \ref{s2}, we summarize necessary results on J. Kern's approach to geometrization of classical Lagrange and Hamilton mechanics. There are also defined the geometric objects which are important for relativistic generalizations of such geometric models on (co) tangent Lorentz bundles. Section \ref{s3} is devoted to the theory of nonholonomic geometric flow evolution of classical relativistic mechanical systems. There are introduced the Perelman--Lagrange and Perelman--Hamilton functionals for geometric mechanics flows on curved phase spacetimes and their reductions on 4-d Lorentz manifolds (as certain emergent flow evolution gravity theories) and 3-d space like hypersurfaces. Corresponding relativistic thermodynamics values are defined. Generalized R. Hamilton geometric flow evolution of flow equations are derived for relativistic Lagrange--Hamilton systems. Self--similar configurations are defined as nonholonomic Ricci--Lagrange and Ricci--Hamilton solitons and studied the conditions certain analogous mechanical models define emergent vacuum gravitational configurations. A brief introduction to theories of classical and quantum mechanical geometric information flow, GIF, is provided in section \ref{s4}. We define and study basic properties of GIF entropies and basic ingredients of the quantum geometric flow information, QGIF, theory
and respective thermodynamics for quantum channels. Finally, we draw conclusions and speculate on further perspectives in section \ref{s5}.

\section{A Hessian type geometrization of Lagrange-Hamilton mechanics}

\label{s2}

We develop an approach to geometrization of relativistic Lagrange and Hamilton mechanics on tangent and cotangent Lorentz manifolds (respectively, $TV$ and $T^{\ast }V)$ on a Lorentz manifold $V$ of dimension $\dim V=4$ and with local Euclidean signature $(+++-)$, see \cite{vacaru18,bubuianu18} for details and historical remarks. The concept of Lagrange space was proposed in \cite{kern74} as an alternative geometrization for nonrelativistic mechanics outlined in \cite{abraham,arnold,deleon85}. The main idea in such Hessian geometric models (with a so-called vertical, or covertical, metric determined by a Lagrange, or Hamilton, generating function) is to drop the homogeneity condition for generating functions and apply Finselr and almost K\"{a}hler geometry methods to classical field theories and mechanics
\cite{matsumoto66,matsumoto86}. Here we note that other approaches on geometrization of classical mechanics and fields, for instance, the poly-simplectic formalism (see \cite{deleon85}, references therein and
further developments in modern literature), do not allow an unified formulation of models for geometric flow evolution, thermodynamics and statistics, (modified) gravity theories and classical and quantum information. In our works \cite{vjmp08,vrmp09,vacaru11,vacaru13,vacaru09,gheorghiu16,bubuianu19,vacaru19,
vacaru19a, ruchin13,gheorghiu16,bubuianu19,vacaru19,vacaru19a,rajpoot17}, using constructions with generalized Finsler like Hessian geometrization of Lagrange-Hamilton systems in mathematical relativity, cosmology and particle physics, various directions were developed for classical and quantum (non) commutative / supersymetric field theories, in modified gravity, inhomogeneous cosmology and theory of nonholonomic geometric flows.

\subsection{Canonic nonholonomic models of Lagrange-Hamilton geometry}

Geometrization of classical nonrelativistic and relativistic mechanical systems can be performed on a Riemannian or Lorentz manifold $V$ and it tangent $TV$ and cotangent $T^{\ast }V$ bundles enabled with (pseudo) Riemannian metrics with local (pseudo) Euclidean signature.

\subsubsection{Phase spacetimes with Lagrange -- Hamilton generating functions and Hessian metrics}

We call $TV$ and/or $T^{\ast }V$ as phase spaces or phase spacetimes  depending on signatures of metrics they are enabled. In a typical case, there are considered corresponding quadratic line elements determined by
total phase space metrics with signature $(+++-;+++-)$,
\begin{eqnarray}
ds^{2} &=&g_{\alpha \beta }(x^{k})du^{\alpha }du^{\beta }=g_{ij}(x^{k})dx^{i}dx^{j}+\eta _{ab}dy^{a}dy^{b},\mbox{ for }y^{a}\sim dx^{a}/d\tau ;\mbox{ and/ or }  \label{lqe} \\
d\ ^{\shortmid }s^{2} &=&\ ^{\shortmid }g_{\alpha \beta }(x^{k})d\  ^{\shortmid }u^{\alpha }d\ ^{\shortmid }u^{\beta}=g_{ij}(x^{k})dx^{i}dx^{j}+\eta ^{ab}dp_{a}dp_{b},\mbox{ for }p_{a}\sim
dx_{a}/d\tau .  \label{lqed}
\end{eqnarray}%
In these formulas, the local frame and dual frame (co-frame) coordinates are labeled respectively. We write $u^{\alpha }=(x^{i},y^{a}),$ (or in brief, $u=(x,y)),$ on the tangent bundle $TV;$ and
$\ ^{\shortmid }u^{\alpha }=(x^{i},p_{a}),$ (or in brief, $\ ^{\shortmid }u=(x,p)),$ on the cotangent bundle $T^{\ast }V.$ The total phase space metrics $g_{\alpha \beta }(u)$ and
$\ ^{\shortmid }g_{\alpha \beta }(\ ^{\shortmid }u)$ are determined, for such examples, by a pseudo--Riemannian spacetime metric $g=\{g_{ij}(x)\}$ with the Levi-Civita connection, LC-connection, $\nabla ,$ which is metric compatible and with zero torsion. In diagonal form, the vertical metric
$\eta _{ab}$ and its dual $\eta ^{ab}$ are standard Minkowski metrics, $\eta _{ab}=diag[1,1,1,-1]$ used for computations in typical fibers of respective (co) tangent bundles. The mechanical models can be elaborated for general frame/ coordinate transforms in total spaces when the metric structures can be parameterized equivalently by the same h-components of
$g_{\alpha \beta}(x^{k})$ and $\ ^{\shortmid }g_{\alpha \beta }(x^{k})=g_{\alpha \beta}(x^{k}),$ but different (co) fiber metrics $g_{ab}(x,y)$ and $g^{ab}(x,p)$ than those considered in (\ref{lqe}) and (\ref{lqed}).\footnote{\label{fncoordconv}There are used such conventions for indices:\ the "horizontal"
indices, h--indices, run values $i,j,k,...=1,2,3,4;$ the vertical indices, v-vertical, run values $a,b,c...=5,6,7,8$; respectively, the v-indices can be identified/ contracted with h-indices $1,2,3,4$ for lifts on total (co)
tangent Lorentz bundles, when $\alpha =(i,a),\beta =(j,b),\gamma =(k,c),...=1,2,3,...8.$ We shall consider letters labelled by an abstract left up/low symbol "$\ ^{\shortmid }$" (for instance, $\ ^{\shortmid
}u^{\alpha }$ and $\ ^{\shortmid }g_{\alpha \beta })$ in order to emphasize that certain geometric/ physical objects are defined on $T^{\ast }V.$ In similar forms, we can consider indices for lower and higher dimensions than $4+4,$ or other type signatures.}

A relativistic 4-d model of Lagrange space $L^{3,1}=(TV,L(x,y))$ is determined by a fundamental function (equivalently, generating function) $TV\ni (x,y)\rightarrow L(x,y)\in \mathbb{R},$ i.e. a real valued function
(in brief, called a Lagrangian or a Lagrange density) which is differentiable on $\widetilde{TV}:=TV/\{0\},$ for $\{0\}$ being the null section of $TV,$ and continuous on the null section of $\pi :TV\rightarrow V. $ Such a relativistic model is regular if the Hessian metric (equivalently, v-metric)
\begin{equation}
\widetilde{g}_{ab}(x,y):=\frac{1}{2}\frac{\partial ^{2}L}{\partial
y^{a}\partial y^{b}}  \label{hessls}
\end{equation}%
is non-degenerate, i.e. $\det |\widetilde{g}_{ab}|\neq 0,$ and of constant signature.

In modern literature on geometric mechanics, kinetics and statistical mechanics of locally anisotropic processes (see a review of such results and references in \cite{vacaru18,bubuianu18}), there are used constructions on cotangent bundles with such a concept:\ A 4-d relativistic model of Hamilton space $H^{3,1}=(T^{\ast }V,H(x,p))$ is constructed for a fundamental function (equivalently, generating Hamilton function, in brief, Hamiltonian or Hamilton density) on a Lorentz manifold $V.$ One considers that
$T^{\ast}V\ni (x,p)\rightarrow H(x,p)\in \mathbb{R}$ defines a real valued function being differentiable on $\widetilde{T^{\ast }V}:=T^{\ast }V/\{0^{\ast }\}$, for $\{0^{\ast }\}$ being the null section of  $T^{\ast}V,$ and continuous on the null section of $\pi ^{\ast }:\ T^{\ast }V\rightarrow V.$ Such a
relativistic mechanical model is regular if the Hessian (cv-metric)
\begin{equation}
\ ^{\shortmid }\widetilde{g}^{ab}(x,p):=\frac{1}{2}\frac{\partial ^{2}H}{%
\partial p_{a}\partial p_{b}}  \label{hesshs}
\end{equation}%
is non-degenerate, i.e. $\det |\ ^{\shortmid }\widetilde{g}^{ab}|\neq 0,$ and of constant signature.

For Lagrange and Hamilton spaces, we can consider Legendre transforms
$L\rightarrow H(x,p):=p_{a}y^{a}-L(x,y)$ and $y^{a}$ determining solutions of the equations $p_{a}=\partial L(x,y)/\partial y^{a}.$ In a similar manner, the inverse Legendre transforms can be introduced, $H\rightarrow L,$ when
\begin{equation}
L(x,y):=p_{a}y^{a}-H(x,p)  \label{invlegendre}
\end{equation}%
for $p_{a}$ determining solutions of the equations $y^{a}=\partial H(x,p)/\partial p_{a}.$

The non-Riemannian total phase space geometries are characterized by nonlinear quadratic line elements
\begin{equation}
ds_{L}^{2}=L(x,y),\mbox{ for models on  }TV;\quad d\ ^{\shortmid
}s_{H}^{2}=H(x,p),\mbox{ for models on  }T^{\ast }V.  \label{nqed}
\end{equation}%
We can elaborate on geometric and physical theories with an effective phase spacetime modelled on (co) tangent Lorentz bundles endowed with generalized frame, metric and linear and nonlinear connection structures determined by nonlinear quadratic line elements and (\ref{nqed}). For certain special cases, such values transform correspondingly into quadratic line elements (\ref{lqe}) and (\ref{lqed}).

The Hessians $\widetilde{g}_{ab}$ and $\ ^{\shortmid }\widetilde{g}^{ab}$ are labeled by a tilde "\symbol{126}" in order to emphasize that such conventional v- and cv--metrics are defined canonically by respective Lagrange and Hamilton generating functions. For simplicity, we can work with such regular metrics even, in principle, mechanical models with degenerate Hessians are also studied in modern mechanics and field theories. Considering general frame/ coordinate transforms on phase spaces, we can express any "tilde" Hessian in a general quadratic form, respectively as a vertical metric (v-metric), $g_{ab}(x,y),$ and/or co-vertical metric (cv-metric), $\ ^{\shortmid }g^{ab}(x,p).$ Inversely, if a v-metric (cv-metric) is prescribed, we can introduce respective (co) frame /coordinate systems, when such values can transformed into certain canonical ones, with "tilde" values. In general, a v-metric $g_{ab}$ is different from the inverse of a cv-metric$\ ^{\shortmid }g^{ab},$ i.e. from the $\ ^{\shortmid }g_{ab}.$ Nevertheless, certain relations between such values can be found via Legendre transforms. We shall omit tildes on geometrical/ physical objects on respective phase spaces if certain formulas hold in general (not only canonical) forms and/or that will not result in ambiguities.

For simplicity, the bulk of geometric constructions in this paper will be performed for (effective and/or generalized) Hamilton spaces if that will not result in ambiguities. We shall consider that via corresponding frame and Legendre transforms, or homogeneity conditions, we can generate necessary type Lagrange/ Finsler/ Cartan configurations.\footnote{A relativistic 4-d model of Finsler space is an example of Lagrange space when a regular $L=F^{2}$ is defined by a fundamental (generating) Finsler function subjected to certain additional conditions: 1) $F$ is a real positive valued function which is differential on $\widetilde{TV}$ and continuous on the null section of the projection $\pi :TV\rightarrow V;$ 2)
it is satisfied the homogeneity condition $F(x,\lambda y)=|\lambda |$ $F(x,y),$ for a nonzero real value $\lambda ;$ and 3) the Hessian (\ref{hessls}) is defined by $F^{2}$ in such a form that in any point
$(x_{(0)},y_{(0)})$ the v-metric is of signature $(+++-).$ In a similar form, we can define relativistic Cartan spaces $C^{3,1}=(V,C(x,p)),$ when $H=C^{2}(x,p)$ is 1-homogeneous on co-fiber coordinates $p_{a}.$}

\subsubsection{Nonlinear connections, adapted frames, and distinguished metrics}

A complete geometrization of mechanical models is not possible if we use only Lagrange-Hamilton functions and respective (non) linear quadratic elements. There are necessary additional concepts and definition of new
geometric objects like the nonlinear connection structure, the distinguished linear connection, various distinguished geometric objects etc., see details and motivations in \cite{vacaru18,bubuianu18}.

A nonlinear connection, N--connection, structure for $TV,$ or $T^{\ast }V,$ is defined as a Whitney sum of conventional $h$ and $v$--distributions, or $h $ and $cv$--distributions,
\begin{equation}
\mathbf{N}:TTV=hTV\oplus vTV,\mbox{ or }\ \ ^{\shortmid }\mathbf{N}:TT^{\ast
}V=hT^{\ast }V\oplus vT^{\ast }V.  \label{ncon}
\end{equation}

Parameterizing locally the N-connections with respect to coordinate bases by corresponding coefficients $\mathbf{N}=\{N_{i}^{a}\}$ and $\ ^{\shortmid }\mathbf{N}=\{\ ^{\shortmid }N_{ia}\},$ we obtain by explicit constructions that decompositions/splitting (\ref{ncon}) define respective systems of N--linear (i.e. N-adapted) bases
\begin{eqnarray}
\mathbf{e}_{\alpha } &=&(\mathbf{e}_{i}=\frac{\partial }{\partial x^{i}}%
-N_{i}^{a}(x,y)\frac{\partial }{\partial y^{a}},e_{b}=\frac{\partial }{%
\partial y^{b}}),\mathbf{e}^{\alpha }=(e^{i}=dx^{i},\mathbf{e}%
^{a}=dy^{a}+N_{i}^{a}(x,y)dx^{i}),\mbox{ and/ or }  \label{nadapb} \\
\ ^{\shortmid }\mathbf{e}_{\alpha } &=&(\ ^{\shortmid }\mathbf{e}_{i}=\frac{%
\partial }{\partial x^{i}}-\ ^{\shortmid }N_{ia}(x,p)\frac{\partial }{%
\partial p_{a}},\ ^{\shortmid }e^{b}=\frac{\partial }{\partial p_{b}}),\ \
^{\shortmid }\mathbf{e}^{\alpha }=(\ ^{\shortmid }e^{i}=dx^{i},\ ^{\shortmid
}\mathbf{e}_{a}=dp_{a}+\ ^{\shortmid }N_{ia}(x,p)dx^{i}).  \notag
\end{eqnarray}%
The N--connection coefficients and necessary types of (co) frame/ coordinate transforms can be used for constructing lifts of metric structures $(V,g)$ to respective nonholonomic (co)tangent bundles, $(\mathbf{TV,N,g})$ and $(\mathbf{T}^{\ast }\mathbf{V,\ ^{\shortmid }N,\ ^{\shortmid }g})$.\footnote{%
Boldface symbols are used in order to emphasize that certain geometric/physical objects are considered in N--adapted form for certain phase spaces and/or spacetime enabled with N--connection structure and when the coefficients of tensors, spinors, and fundamental geometric objects can be computed with respect to N-elongated bases of type (\ref{nadapb}).}

We can consider various type of metric structures on a tangent, $\mathbf{TV}, $ and/or cotangent, $\mathbf{T}^{\ast }\mathbf{V},$ Lorentz bundles. This can be used for elaborating mechanical models, thermodynamic and kinetic theories and generalizations of the Einstein gravity. Such metric structures
can be parameterized by frame transforms in N--adapted form, i.e. as distinguished metrics (d-metrics)
\begin{eqnarray}
\mathbf{g} &=&\mathbf{g}_{\alpha \beta }(x,y)\mathbf{\mathbf{e}}^{\alpha }%
\mathbf{\otimes \mathbf{e}}^{\beta }=g_{ij}(x)e^{i}\otimes e^{j}+\mathbf{g}%
_{ab}(x,y)\mathbf{e}^{a}\otimes \mathbf{e}^{a}\mbox{ and/or }  \label{dmt} \\
\ ^{\shortmid }\mathbf{g} &=&\ ^{\shortmid }\mathbf{g}_{\alpha \beta }(x,p)\
^{\shortmid }\mathbf{\mathbf{e}}^{\alpha }\mathbf{\otimes \ ^{\shortmid }%
\mathbf{e}}^{\beta }=g_{ij}(x)e^{i}\otimes e^{j}+\ ^{\shortmid }\mathbf{g}%
^{ab}(x,p)\ ^{\shortmid }\mathbf{e}_{a}\otimes \ ^{\shortmid }\mathbf{e}_{b}.
\label{dmct}
\end{eqnarray}%
In this work, such metrics on conventional 8-d manifolds are of signature $(+,+,+,-,+,+,+,-)$ but for elaborating non-relativistic mechanical/ thermodynamical / statistical models other type signatures can be
considered. For instance, a pseudo--Riemannian metric $g_{ij}(x)$ can be subjected to the condition that it defines a solution of the standard Einstein equations in GR, or a MGT, with a corresponding base Lorentz
manifold $\mathbf{V}.$ For various mechanical and thermodynamical models, there are necessary additional geometrically and physically motivated assumptions on how nonlinear quadratic elements of type or (\ref{nqed}), and/or (\ref{dmt}), or (\ref{dmct}), encode local anisotropies, inhomogeneous structures, modified dispersion relations etc.

\subsubsection{Hamilton-Jacoby, Euler-Lagrange, and semi-spray equations and N--connections}

Let us consider that a spacetime Lorentzian (or a space Riemannian) manifold $\mathbf{V} $ is endowed with a metric $hg=\{g_{ij}(x)\}$ of signature $(3,1) $ (or of Euclidean signature). Using frame/generalized coordinate transforms on base and total spaces, metrics can be deformed to off-diagonal metrics depending on velocity/ momentum coordinates, including horizontal components of Hessian type.

Considering a regular curve $c(\tau )$ defined
$c:\tau \in \lbrack 0,1]\rightarrow x^{i}(\tau )\subset U\subset V,$ for a real parameter $\tau , $ we can construct a lifted to $\pi ^{-1}(U)\subset \widetilde{TV}$ defining a curve in the total space, when $\widetilde{c}(\tau ):\tau \in \lbrack 0,1]\rightarrow \left( x^{i}(\tau ),y^{i}(\tau )=dx^{i}/d\tau \right) $ with a non-vanishing v-vector field $dx^{i}/d\tau .$ Using a canonical symplectic structure
$\theta :=dp_{i}\wedge dx^{i}$ on $T^{\ast }V$ \ and a unique vector filed
$ \widetilde{X}_{H}:=\frac{\partial \widetilde{H}}{\partial p_{i}}\frac{\partial }{\partial x^{i}} - \frac{\partial \widetilde{H}}{\partial x^{i}}\frac{\partial }{\partial p_{i}}$ defined by $\widetilde{H},$ we construct an equation $i_{\widetilde{X}_{H}}\theta =-d\widetilde{H}.$ We write $\wedge $ for the antisymmetric product where $i_{\widetilde{X}_{H}}$ denotes the interior produce defined by $\widetilde{X}_{H}.$ This allows us to formulate and prove using an explicit calculus for any functions
$\ ^{1}f(x,p)$ and $\ ^{2}f(x,p)$ on $T^{\ast }V$ and a canonical Poisson structure
$\{\ ^{1}f, \ ^{2}f\}:=\theta (\widetilde{X}_{^{1}f},\widetilde{X}_{^{2}f}).$

The canonical Hamilton-Jacobi equations are defined using above canonical Poisson structure,
\begin{equation*}
\frac{dx^{i}}{d\tau }=\{\widetilde{H},x^{i}\}\mbox{ and }\frac{dp_{a}}{d\tau
}=\{\widetilde{H},p_{a}\}.
\end{equation*}%
The dynamics of a probing point particle in $L$-dual effective phase spaces $%
\widetilde{H}^{3,1}$ and $\widetilde{L}^{3,1}$ is described equivalently by
the Hamilton equations $\frac{dx^{i}}{d\tau }=\frac{\partial \widetilde{H}}{%
\partial p_{i}}$ and $\frac{dp_{i}}{d\tau }=-\frac{\partial \widetilde{H}}{%
\partial x^{i}},$ or as Euler-Lagrange equations, $\ \frac{d}{d\tau }\frac{%
\partial \widetilde{L}}{\partial y^{i}}-\frac{\partial \widetilde{L}}{%
\partial x^{i}}=0.$ In their turn, these equations are equivalent to the
\textit{nonlinear geodesic (semi-spray) equations}
\begin{equation}
\frac{d^{2}x^{i}}{d\tau ^{2}}+2\widetilde{G}^{i}(x,y)=0,  \label{ngeqf}
\end{equation}%
for $\widetilde{G}^{i}=\frac{1}{2}\widetilde{g}^{ij}(\frac{\partial ^{2}%
\widetilde{L}}{\partial y^{i}}y^{k}-\frac{\partial \widetilde{L}}{\partial
x^{i}}),\,\ $ with $\widetilde{g}^{ij}$ being inverse to $\widetilde{g}_{ij}$
(\ref{hessls}).

The equations (\ref{ngeqf}) show that point like probing particles move not along usual geodesics as on Lorentz manifolds but follow some nonlinear geodesic equations determined by generating Lagrange functions and their Hessians.

Using the constructions from above subsection, we prove there are canonical N--connections determined by generating Lagrange/ Hamilton functions following formulas
\begin{equation}
\ \ \ \ ^{\shortmid }\widetilde{\mathbf{N}}=\left\{ \ ^{\shortmid }%
\widetilde{N}_{ij}:=\frac{1}{2}\left[ \{\ \ ^{\shortmid }\widetilde{g}_{ij},%
\widetilde{H}\}-\frac{\partial ^{2}\widetilde{H}}{\partial p_{k}\partial
x^{i}}\ ^{\shortmid }\widetilde{g}_{jk}-\frac{\partial ^{2}\widetilde{H}}{%
\partial p_{k}\partial x^{j}}\ ^{\shortmid }\widetilde{g}_{ik}\right]
\right\} \mbox{ and }\widetilde{\mathbf{N}}=\left\{ \widetilde{N}_{i}^{a}:=%
\frac{\partial \widetilde{G}}{\partial y^{i}}\right\} ,  \label{cannc}
\end{equation}%
where $\ \ ^{\shortmid }\widetilde{g}_{ij}$ is inverse to $\ \ ^{\shortmid }%
\widetilde{g}^{ab}$ (\ref{hesshs}). Introducing these canonical
N--connection coefficients into formulas (\ref{nadapb}), we prove that there
are canonical N--adapted (co) frames
\begin{eqnarray}
\widetilde{\mathbf{e}}_{\alpha } &=&(\widetilde{\mathbf{e}}_{i}=\frac{%
\partial }{\partial x^{i}}-\widetilde{N}_{i}^{a}(x,y)\frac{\partial }{%
\partial y^{a}},e_{b}=\frac{\partial }{\partial y^{b}});\widetilde{\mathbf{e}%
}^{\alpha }=(\widetilde{e}^{i}=dx^{i},\widetilde{\mathbf{e}}^{a}=dy^{a}+%
\widetilde{N}_{i}^{a}(x,y)dx^{i});\mbox{and \ }  \label{cnddapb} \\
\ ^{\shortmid }\widetilde{\mathbf{e}}_{\alpha } &=&(\ ^{\shortmid }%
\widetilde{\mathbf{e}}_{i}=\frac{\partial }{\partial x^{i}}-\ ^{\shortmid }%
\widetilde{N}_{ia}(x,p)\frac{\partial }{\partial p_{a}},\ ^{\shortmid }e^{b}=%
\frac{\partial }{\partial p_{b}});\ \ ^{\shortmid }\widetilde{\mathbf{e}}%
^{\alpha }=(\ ^{\shortmid }e^{i}=dx^{i},\ ^{\shortmid }\mathbf{e}%
_{a}=dp_{a}+\ ^{\shortmid }\widetilde{N}_{ia}(x,p)dx^{i}).  \notag
\end{eqnarray}

Such a canonical N-splitting $\ \widetilde{\mathbf{N}}:TTV=hTV\oplus vTV\ $ and
 $\ ^{\shortmid }\widetilde{\mathbf{N}}:TT^{\ast }V=hT^{\ast }V\oplus vT^{\ast }V$ is stated by respective generating Lagrange and/or Hamilton functions on any tangent and/or cotangent Lorentz bundle. The nonholonomic structure of phase spaces can be described in equivalent forms using canonical data $(\widetilde{L},\ \widetilde{\mathbf{N}};\widetilde{\mathbf{e}}_{\alpha },\widetilde{\mathbf{e}}^{\alpha }),$ with effective Largange density $\widetilde{L}$ (correspondingly, $(\widetilde{H},\ ^{\shortmid }%
\widetilde{\mathbf{N}};\ ^{\shortmid }\widetilde{\mathbf{e}}_{\alpha },\ ^{\shortmid }\widetilde{\mathbf{e}}^{\alpha }),$ with effective Hamilton density $\widetilde{H}$ ). We can consider a general N-splitting without effective Lagrangians (Hamiltonians), i.e. in terms of arbitrary geometric
data $(\mathbf{N};\mathbf{e}_{\alpha },\mathbf{e}^{\alpha })$ (correspondingly $(\ ^{\shortmid }\mathbf{N};\ ^{\shortmid }\mathbf{e}_{\alpha },\ ^{\shortmid }\mathbf{e}^{\alpha })$).\footnote{%
On nonholonomic (co) tangent bundles, we can consider d--vectors if they are written in a form adapted to a prescribed N--connection structure, for instance,
\begin{eqnarray*}
\mathbf{X} &=&\widetilde{\mathbf{X}}^{\alpha }\widetilde{\mathbf{e}}_{\alpha
}=\widetilde{\mathbf{X}}^{i}\widetilde{\mathbf{e}}_{i}+X^{b}e_{b}=\mathbf{X}%
^{\alpha }\mathbf{e}_{\alpha }=\mathbf{X}^{i}\mathbf{e}_{i}+X^{b}e_{b}\in T%
\mathbf{TV}, \\
\ ^{\shortmid }\mathbf{X} &=&\ ^{\shortmid }\widetilde{\mathbf{X}}^{\alpha }%
\widetilde{\mathbf{e}}_{\alpha }=\ ^{\shortmid }\widetilde{\mathbf{X}}^{i}\
^{\shortmid }\widetilde{\mathbf{e}}_{i}+\ ^{\shortmid }X_{b}\ ^{\shortmid
}e^{b}=\ ^{\shortmid }\mathbf{X}^{\alpha }\ ^{\shortmid }\mathbf{e}_{\alpha
}=\ ^{\shortmid }\mathbf{X}^{i}\ ^{\shortmid }\mathbf{e}_{i}+\ ^{\shortmid
}X_{b}\ ^{\shortmid }e^{b}\in T\mathbf{T}^{\ast }\mathbf{V.}
\end{eqnarray*}%
Such formulas can be written equivalently for decompositions with respect to
canonical, or arbitrary, N-adapted bases. In brief, the h-v and/or h-cv
decompositions can be written $\mathbf{X}^{\alpha }=\widetilde{\mathbf{X}}%
^{\alpha }=(\widetilde{\mathbf{X}}^{i},X^{b})=(\mathbf{X}^{i},X^{b}),\
^{\shortmid }\mathbf{X}^{\alpha }=\ ^{\shortmid }\widetilde{\mathbf{X}}%
^{\alpha }=(\ ^{\shortmid }\widetilde{\mathbf{X}}^{i},\ ^{\shortmid
}X_{b})=(\ ^{\shortmid }\mathbf{X}^{i},\ ^{\shortmid }X_{b}).$ Considering $%
\mathbf{X}$ and $\ ^{\shortmid }\mathbf{X}$ as 1-forms, we have
\begin{eqnarray*}
\mathbf{X} &=&\widetilde{\mathbf{X}}_{\alpha }\ \mathbf{e}^{\alpha }=X_{i}\
e^{i}+\widetilde{\mathbf{X}}^{a}\widetilde{\mathbf{e}}_{a}=\widetilde{%
\mathbf{X}}_{\alpha }\mathbf{e}^{\alpha }=X_{i}e^{i}+\mathbf{X}^{a}\mathbf{e}%
_{a}\ \in T^{\ast }\mathbf{TV} \\
\ ^{\shortmid }\mathbf{X} &=&\ ^{\shortmid }\widetilde{\mathbf{X}}_{\alpha
}\ ^{\shortmid }\mathbf{e}^{\alpha }=\ ^{\shortmid }X_{i}\ ^{\shortmid
}e^{i}+\ ^{\shortmid }\widetilde{\mathbf{X}}^{a}\ ^{\shortmid }\widetilde{%
\mathbf{e}}_{a}=\ ^{\shortmid }\widetilde{\mathbf{X}}_{\alpha }\ ^{\shortmid
}\mathbf{e}^{\alpha }=\ ^{\shortmid }X_{i}\ ^{\shortmid }e^{i}+\ ^{\shortmid
}\mathbf{X}^{a}\ ^{\shortmid }\mathbf{e}_{a}\ \in T^{\ast }\mathbf{T}^{\ast }%
\mathbf{V,}
\end{eqnarray*}%
or, in brief, $\mathbf{X}_{\alpha }=\widetilde{\mathbf{X}}_{\alpha }=(X_{i},%
\widetilde{\mathbf{X}}^{a})=(X_{i},\mathbf{X}^{a}),\ ^{\shortmid }\mathbf{%
X_{\alpha }=}\ ^{\shortmid }\widetilde{\mathbf{X}}_{\alpha }=(\ ^{\shortmid
}X_{i},\ ^{\shortmid }\widetilde{\mathbf{X}}^{a})=(\ ^{\shortmid }X_{i},\
^{\shortmid }\mathbf{X}^{a})$} Using tensor products of N-adapted (co)
frames on phase space, we can parameterize in N-adapted forms (canonical or
general ones) arbitrary tensors fields (d-tensors), connections and
d-connections and other types of geometric objects, d-objects.

\subsubsection{Canonical d-metric and almost complex structures}

There are canonical data $(\widetilde{L},\ \widetilde{\mathbf{N}};\widetilde{%
\mathbf{e}}_{\alpha },\widetilde{\mathbf{e}}^{\alpha };\widetilde{g}_{jk},%
\widetilde{g}^{jk})$ and/or $(\widetilde{H},\ ^{\shortmid }\widetilde{%
\mathbf{N}};\ ^{\shortmid }\widetilde{\mathbf{e}}_{\alpha },\ ^{\shortmid }%
\widetilde{\mathbf{e}}^{\alpha };\ \ ^{\shortmid }\widetilde{g}^{ab},\ \
^{\shortmid }\widetilde{g}_{ab})$ when the d-metrics are parameterized in
the Hessian form both for the h- and (c)v-components,%
\begin{eqnarray}
\widetilde{\mathbf{g}} &=&\widetilde{\mathbf{g}}_{\alpha \beta }(x,y)%
\widetilde{\mathbf{e}}^{\alpha }\mathbf{\otimes }\widetilde{\mathbf{e}}%
^{\beta }=\widetilde{g}_{ij}(x,y)e^{i}\otimes e^{j}+\widetilde{g}_{ab}(x,y)%
\widetilde{\mathbf{e}}^{a}\otimes \widetilde{\mathbf{e}}^{a}\mbox{
and/or }  \label{cdms} \\
\ ^{\shortmid }\widetilde{\mathbf{g}} &=&\ ^{\shortmid }\widetilde{\mathbf{g}%
}_{\alpha \beta }(x,p)\ ^{\shortmid }\widetilde{\mathbf{e}}^{\alpha }\mathbf{%
\otimes \ ^{\shortmid }}\widetilde{\mathbf{e}}^{\beta }=\ \ ^{\shortmid }%
\widetilde{g}_{ij}(x,p)e^{i}\otimes e^{j}+\ ^{\shortmid }\widetilde{g}%
^{ab}(x,p)\ ^{\shortmid }\widetilde{\mathbf{e}}_{a}\otimes \ ^{\shortmid }%
\widetilde{\mathbf{e}}_{b}.  \label{cdmds}
\end{eqnarray}

By frame transforms, the canonical d-metric structures (\ref{cdms}) and (\ref%
{cdmds}) [with tildes] can be written, respectively, in general d-metric
forms (\ref{dmt}) and (\ref{dmct}) [without tildes]. In explicit form, the
general vierbein transforms are written $e_{\alpha }=e_{\ \alpha }^{%
\underline{\alpha }}(u)\partial /\partial u^{\underline{\alpha }}$ and $%
e^{\beta }=e_{\ \underline{\beta }}^{\beta }(u)du^{\underline{\beta }}.$ We
underline the local coordinate indices in order to distinguish them from
arbitrary abstract ones. In such formulas, the matrix $e_{\ \underline{\beta
}}^{\beta }$ is inverse to $e_{\ \alpha }^{\underline{\alpha }}$ for
orthonormalized bases. \ For Hamilton like configurations on cotangent
bundles, we consider $\ ^{\shortmid }e_{\alpha }=\ ^{\shortmid }e_{\ \alpha
}^{\underline{\alpha }}(\ ^{\shortmid }u)\partial /\partial \ ^{\shortmid
}u^{\underline{\alpha }}$ and $\ ^{\shortmid }e^{\beta }=\ ^{\shortmid }e_{\
\underline{\beta }}^{\beta }(\ ^{\shortmid }u)d\ ^{\shortmid }u^{\underline{%
\beta }}.$ There are not used boldface symbols for such transforms because
they can be not adapted to a N--connection structure.

Using (\ref{cnddapb}), respectively, for (\ref{dmt}) and (\ref{dmct}) and
regrouping with respect to local coordinate bases, we prove that with
respect to local coordinate frames, any d--metric structures on $\mathbf{TV}$
and/or $\mathbf{T}^{\ast }\mathbf{V,}$%
\begin{equation*}
\mathbf{g}=\mathbf{g}_{\alpha \beta }(x,y)\mathbf{e}^{\alpha }\mathbf{%
\otimes e}^{\beta }=g_{\underline{\alpha }\underline{\beta }}(x,y)du^{%
\underline{\alpha }}\mathbf{\otimes }du^{\underline{\beta }}\mbox{
and/or }\ ^{\shortmid }\mathbf{g}=\ ^{\shortmid }\mathbf{g}_{\alpha \beta
}(x,p)\ ^{\shortmid }\mathbf{e}^{\alpha }\mathbf{\otimes \ ^{\shortmid }e}%
^{\beta }=\ ^{\shortmid }g_{\underline{\alpha }\underline{\beta }}(x,p)d\
^{\shortmid }u^{\underline{\alpha }}\mathbf{\otimes }d\ ^{\shortmid }u^{%
\underline{\beta }}.
\end{equation*}%
These formulas can be subjected to frame transforms, $\mathbf{g}_{\alpha
\beta }=e_{\ \alpha }^{\underline{\alpha }}e_{\ \beta }^{\underline{\beta }%
}g_{\underline{\alpha }\underline{\beta }}$ and $^{\shortmid }\mathbf{g}%
_{\alpha \beta }=\ ^{\shortmid }e_{\ \alpha }^{\underline{\alpha }}\
^{\shortmid }e_{\ \beta }^{\underline{\beta }}\ ^{\shortmid }g_{\underline{%
\alpha }\underline{\beta }},$ and written in equivalent off-diagonal forms:
\begin{eqnarray}
g_{\underline{\alpha }\underline{\beta }} &=&\left[
\begin{array}{cc}
g_{ij}(x)+g_{ab}(x,y)N_{i}^{a}(x,y)N_{j}^{b}(x,y) & g_{ae}(x,y)N_{j}^{e}(x,y)
\\
g_{be}(x,y)N_{i}^{e}(x,y) & g_{ab}(x,y)%
\end{array}%
\right] \mbox{
and/or }  \notag \\
\ ^{\shortmid }g_{\underline{\alpha }\underline{\beta }} &=&\left[
\begin{array}{cc}
\ ^{\shortmid }g_{ij}(x)+\ ^{\shortmid }g^{ab}(x,p)\ ^{\shortmid
}N_{ia}(x,p)\ ^{\shortmid }N_{jb}(x,p) & \ ^{\shortmid }g^{ae}\ ^{\shortmid
}N_{je}(x,p) \\
\ ^{\shortmid }g^{be}\ ^{\shortmid }N_{ie}(x,p) & \ ^{\shortmid
}g^{ab}(x,p)\
\end{array}%
\right] .  \label{offd}
\end{eqnarray}

Parameterizations of type (\ref{offd}) for metrics are considered, for
instance, in Kaluza--Klein theories on associated vector bundles. In our
cases, the constructions are on (co) tangent bundles for geometric mechanics
models. We conclude that if we fix a metric structure of type $\ ^{\shortmid
}\widetilde{\mathbf{g}}$ (\ref{cdmds}), we can elaborate equivalent models
with $\ ^{\shortmid }\mathbf{g}$ (\ref{dmct}) determined by certain classes
of nonholonomic frame transforms. Inversely, prescribing a d-metric $\
^{\shortmid }\mathbf{g,}$ we can define nonholonomic variables when this
metric structure can be represented as a $\ ^{\shortmid }\widetilde{\mathbf{g%
}},$ i.e. in mechanical like variables, when $\ ^{\shortmid }\mathbf{g=}\
^{\shortmid }\widetilde{\mathbf{g}}.$ In a more general context, we can
elaborate on bi-metric (and even multi-metric theories of gravity, geometric
mechanics and thermodynamics) if we consider that $\ ^{\shortmid }\widetilde{%
\mathbf{g}}$ and $\ ^{\shortmid }\mathbf{g}$ are related via certain
generalized nonholonomic transforms, see details an references in \cite%
{vacaru18,bubuianu18}.

The canonical N--connections $\widetilde{\mathbf{N}}$ and $\ ^{\shortmid }%
\widetilde{\mathbf{N}}$ define respectively certain canonical almost complex
structures $\widetilde{\mathbf{J}}\mathbf{,}$ on $\mathbf{TV},$ and $\
^{\shortmid }\widetilde{\mathbf{J}},$ on $\mathbf{T}^{\ast }\mathbf{V}.$
This follows, for instance, from such a construction on $\mathbf{T}^{\ast }%
\mathbf{V.}$ Let us consider a linear operator $\ ^{\shortmid }\widetilde{%
\mathbf{J}}$ acting on$\ ^{\shortmid }\mathbf{e}_{\alpha }=(\ ^{\shortmid }%
\mathbf{e}_{i},\ ^{\shortmid }e^{b})$ using formulas $\ ^{\shortmid }%
\widetilde{\mathbf{J}}(\ ^{\shortmid }\mathbf{e}_{i})=-\ ^{\shortmid
}e^{n+i} $ and $\ ^{\shortmid }\widetilde{\mathbf{J}}(\ ^{\shortmid
}e^{n+i})=\ ^{\shortmid }\mathbf{e}_{i}$. This $\ ^{\shortmid }\widetilde{%
\mathbf{J}}$ defines globally an almost complex structure ( $\ \ ^{\shortmid
}\widetilde{\mathbf{J}}\mathbf{\circ \ }\ ^{\shortmid }\widetilde{\mathbf{J}}%
=$ $-\mathbf{\ I,}$ where $\mathbf{I}$ is the unity matrix) on $\mathbf{T}%
^{\ast }\mathbf{V.}$ Such an operator is completely determined for Hamilton
spaces by a $\widetilde{H}(x,p).$

We note that $\widetilde{\mathbf{J}}$ and $\ ^{\shortmid }\widetilde{\mathbf{%
J}}$ are standard almost complex structures only for the Euclidean
signatures, respectively, on $\mathbf{TV}$ and $\mathbf{T}^{\ast }\mathbf{V}$%
. Contrary, we call them as pseudo almost complex structure. It is possible
to omit tildes and write $\mathbf{J}$ and $\ ^{\shortmid }\mathbf{J}$ for
arbitrary frame/ coordinate transforms.

The canonical Neijenhuis tensor fields determined by Lagrange and Hamilton
generating functions, for respective canonical almost complex structures $%
\widetilde{\mathbf{J}}$ on $\mathbf{TV}$ and/or $\ ^{\shortmid }\widetilde{%
\mathbf{J}}$ on $\mathbf{T}^{\ast }\mathbf{V},$ are introduced as curvatures
of respective N--connections
\begin{eqnarray}
\widetilde{\mathbf{\Omega }}\mathbf{(}\widetilde{\mathbf{X}}\mathbf{,}%
\widetilde{\mathbf{Y}}) &:=&\mathbf{-[\widetilde{\mathbf{X}}\mathbf{,}%
\widetilde{\mathbf{Y}}]+[\widetilde{\mathbf{J}}\widetilde{\mathbf{X}},%
\widetilde{\mathbf{J}}\widetilde{\mathbf{Y}}]-\widetilde{\mathbf{J}}[%
\widetilde{\mathbf{J}}\widetilde{\mathbf{X}},\widetilde{\mathbf{Y}}]-%
\widetilde{\mathbf{J}}[\widetilde{\mathbf{X}},\widetilde{\mathbf{J}}%
\widetilde{\mathbf{Y}}]}\mbox{ and/or }  \notag \\
\ ^{\shortmid }\widetilde{\mathbf{\Omega }}(\ ^{\shortmid }\widetilde{%
\mathbf{X}}\mathbf{,}\ ^{\shortmid }\widetilde{\mathbf{Y}}) &:=&%
\mathbf{-[\ ^{\shortmid }\widetilde{\mathbf{X}}\mathbf{,}\ ^{\shortmid }%
\widetilde{\mathbf{Y}}]+[\ ^{\shortmid }\widetilde{\mathbf{J}}\ ^{\shortmid }%
\widetilde{\mathbf{X}},\ ^{\shortmid }\widetilde{\mathbf{J}}\ ^{\shortmid }%
\widetilde{\mathbf{Y}}]-\ ^{\shortmid }\widetilde{\mathbf{J}}[\ ^{\shortmid }%
\widetilde{\mathbf{J}}\ ^{\shortmid }\widetilde{\mathbf{X}},\ ^{\shortmid }%
\widetilde{\mathbf{Y}}]-\ ^{\shortmid }\widetilde{\mathbf{J}}[\ ^{\shortmid }%
\widetilde{\mathbf{X}},\ ^{\shortmid }\widetilde{\mathbf{J}}\ ^{\shortmid }%
\widetilde{\mathbf{Y}}],}  \label{neijt}
\end{eqnarray}%
for any d--vectors $\mathbf{X,}$ $\mathbf{Y}$ and $\ ^{\shortmid }\mathbf{%
X,\ ^{\shortmid }Y.}$ Such formulas can be written in general form without
tilde values if there are considered arbitrary frame transforms. In local
form, a N--connection on $\mathbf{TV,}$ or $\mathbf{T}^{\ast }\mathbf{V,}$
is characterized by such coefficients of (\ref{neijt}) (i.e. the
N--connection curvature):
\begin{equation}
\Omega _{ij}^{a}=\frac{\partial N_{i}^{a}}{\partial x^{j}}-\frac{\partial
N_{j}^{a}}{\partial x^{i}}+N_{i}^{b}\frac{\partial N_{j}^{a}}{\partial y^{b}}%
-N_{j}^{b}\frac{\partial N_{i}^{a}}{\partial y^{b}},\mbox{\ or\ }\ \ \mathbf{%
\ ^{\shortmid }}\Omega _{ija}=\frac{\partial \mathbf{\ ^{\shortmid }}N_{ia}}{%
\partial x^{j}}-\frac{\partial \mathbf{\ ^{\shortmid }}N_{ja}}{\partial x^{i}%
}+\ \mathbf{^{\shortmid }}N_{ib}\frac{\partial \mathbf{\ ^{\shortmid }}N_{ja}%
}{\partial p_{b}}-\mathbf{\ ^{\shortmid }}N_{jb}\frac{\partial \mathbf{\
^{\shortmid }}N_{ia}}{\partial p_{b}}.  \label{neijtc}
\end{equation}

Almost complex structures $\mathbf{J}$ and $\ ^{\shortmid }\mathbf{J}$
transform into standard complex structures for Euclidean signatures if $%
\mathbf{\Omega }=0$ and/or $\ ^{\shortmid }\mathbf{\Omega }=0.$ For almost
complex canonical structures, we can consider canonical forms with "tilde"
values determined by $\widetilde{\mathbf{N}}=\{\widetilde{N}_{j}^{b}\}$ and $%
\ ^{\shortmid }\widetilde{\mathbf{N}}=\{\mathbf{\ ^{\shortmid }}\widetilde{N}%
_{ia}\}.$

Applying a straightforward N-adapted calculus using formulas $\widetilde{%
\mathbf{e}}_{\alpha }=(\widetilde{\mathbf{e}}_{i},e_{b})$ and $\ ^{\shortmid
}\widetilde{\mathbf{e}}_{\alpha }=(\ ^{\shortmid }\widetilde{\mathbf{e}}%
_{i},\ ^{\shortmid }e^{b}),$ see (\ref{cnddapb}) and (\ref{neijtc}), we
prove that the canonical nonholonomic frame structures on $\mathbf{TV}$
and/or $\mathbf{T}^{\ast }\mathbf{V}$ are characterized by corresponding
anholonomy relations
\begin{equation}
\lbrack \widetilde{\mathbf{e}}_{\alpha },\widetilde{\mathbf{e}}_{\beta }]=%
\widetilde{\mathbf{e}}_{\alpha }\widetilde{\mathbf{e}}_{\beta }-\widetilde{%
\mathbf{e}}_{\beta }\widetilde{\mathbf{e}}_{\alpha }=\widetilde{W}_{\alpha
\beta }^{\gamma }\widetilde{\mathbf{e}}_{\gamma }\mbox{ and }\lbrack \
^{\shortmid }\widetilde{\mathbf{e}}_{\alpha },\ ^{\shortmid }\widetilde{%
\mathbf{e}}_{\beta }]=\ ^{\shortmid }\widetilde{\mathbf{e}}_{\alpha }\
^{\shortmid }\widetilde{\mathbf{e}}_{\beta }-\ ^{\shortmid }\widetilde{%
\mathbf{e}}_{\beta }\ ^{\shortmid }\widetilde{\mathbf{e}}_{\alpha }=\
^{\shortmid }\widetilde{W}_{\alpha \beta }^{\gamma }\ ^{\shortmid }%
\widetilde{\mathbf{e}}_{\gamma }  \label{anhrelc}
\end{equation}%
with anholonomy coefficients $\widetilde{W}_{ia}^{b}=\partial _{a}\widetilde{%
N}_{i}^{b},$ $\widetilde{W}_{ji}^{a}=\widetilde{\Omega }_{ij}^{a},$ and $\
^{\shortmid }\widetilde{W}_{ib}^{a}=\partial \ ^{\shortmid }\widetilde{N}%
_{ib}/\partial p_{a}$ and $\ ^{\shortmid }\widetilde{W}_{jia}=\ \mathbf{\
^{\shortmid }}\widetilde{\Omega }_{ija}.$ We can define holonomic
(integrable) frame configurations if the respective anholonomy coefficients
in (\ref{anhrelc}) are zero.

In geometric mechanics, the canonical d-metric structures $\ \widetilde{%
\mathbf{g}}$ (\ref{cdms}) and $\ ^{\shortmid }\widetilde{\mathbf{g}}$ (\ref%
{cdmds}) are described by generic off--diagonal metrics (\ref{offd}) if
respective anholonomy coefficients (\ref{anhrelc}) are not trivial.

\subsection{Linear connections and curvatures for Lagrange--Hamilton spaces}

Elaborating on different type Lagrange-Hamilton models, we are not able to perform the constructions in N-adapted anholonomic form if we work only with generalized (Finsler like) metrics determined by nonlinear quadratic forms $L(x,y)$ and/or $H(x,p)$ (\ref{nqed}). The goal of this subsection is to analyze which classes of linear connections and respective covariant derivative operators can be generated canonically by fundamental generating functions.

\subsubsection{Distinguished connections, N-adapted distortions and curvatures}

We can define a linear connection $D$ on $\mathbf{TV}$ when a $\mathcal{L}$--duality between the tangent and corresponding cotangent bundles which can be defined by pull--back and push--forward maps. We omit geometric details on constructing such maps from/to base space to total space, considered, for
instance, in \cite{vacaru18,bubuianu18}. A linear connection $\ ^{\shortmid}D$ on
$\mathbf{T}^{\ast }\mathbf{V}$ is defined as follows:
$\ ^{\shortmid}D_{\ ^{\shortmid }\mathbf{X}}\ ^{\shortmid }\mathbf{Y}:=(D_{\mathbf{X}}\mathbf{Y})^{\ast }=\ ^{\shortmid }(D_{\mathbf{X}}\mathbf{Y}),$ for any vector fields $\ ^{\shortmid }\mathbf{X}$ and $\ ^{\shortmid }\mathbf{Y}$ on
$\mathbf{T}^{\ast }\mathbf{V}.$ Inversely, we can consider a linear connection $\ \ ^{\shortmid }D$ on $\mathbf{T}^{\ast }\mathbf{V}$ and then construct a linear connection $\ ^{\circ }D$ on $\mathbf{TV},$ following the rule $\ ^{\circ }D_{\mathbf{X}}\mathbf{Y}:=(\ ^{\shortmid }D_{\ ^{\shortmid }%
\mathbf{X}}\ ^{\shortmid }\mathbf{Y})^{\circ },$ for any vector fields $\mathbf{X}$ and $\mathbf{Y}$ on $\mathbf{TV}$.

A distinguished connection (d--connection) is a linear connection $\mathbf{D}$ on $\mathbf{TV}$ (or
$\ ^{\shortmid }\mathbf{D}$ on $\mathbf{T}^{\ast }\mathbf{V})$ which is compatible with the N--connection splitting (\ref{ncon}).

The coefficients of d--connections can be defined and computed in corresponding N-adapted forms,
\begin{equation*}
\mathbf{D}_{\mathbf{e}_{\beta }}\mathbf{e}_{\gamma }:=\mathbf{\Gamma }_{\
\beta \gamma }^{\alpha }\mathbf{e}_{\alpha }\mbox{ and }\mathbf{\
^{\shortmid }D}_{\mathbf{\ ^{\shortmid }e}_{\beta }}\ \mathbf{^{\shortmid }e}%
_{\gamma }:=\mathbf{\ ^{\shortmid }\Gamma }_{\ \beta \gamma }^{\alpha }%
\mathbf{\ ^{\shortmid }e}_{\alpha }.
\end{equation*}%
For a h-v splitting, $\ \mathbf{D}_{\mathbf{e}_{k}}\mathbf{e}_{j}:=L_{\
jk}^{i}\mathbf{e}_{i},\mathbf{D}_{\mathbf{e}_{k}}e_{b}:=\acute{L}_{\
bk}^{a}e_{a},\mathbf{D}_{e_{c}}\mathbf{e}_{j}:=\acute{C}_{\ jc}^{i}\mathbf{e}%
_{i},\mathbf{D}_{e_{c}}e_{b}:=C_{\ bc}^{a}e_{a}$ and a h-cv splitting, $\ \
^{\shortmid }\mathbf{D}_{\ ^{\shortmid }\mathbf{e}_{k}}\ ^{\shortmid }%
\mathbf{e}_{j}:=\ ^{\shortmid }L_{\ jk}^{i}\ ^{\shortmid }\mathbf{e}_{i},\
^{\shortmid }\mathbf{D}_{\mathbf{e}_{k}}\ ^{\shortmid }e^{b}:=-\ ^{\shortmid
}\acute{L}_{a\ k}^{\ b}\ ^{\shortmid }e^{a},\ ^{\shortmid }\mathbf{D}_{\
^{\shortmid }e^{c}}\ ^{\shortmid }\mathbf{e}_{j}:=\ ^{\shortmid }\acute{C}%
_{\ j}^{i\ c}\ ^{\shortmid }\mathbf{e}_{i},\ ^{\shortmid }\mathbf{D}_{\
^{\shortmid }e^{c}}\ ^{\shortmid }e^{b}:=-\ ^{\shortmid }C_{a}^{\ bc}\
^{\shortmid }e^{a}.$ In result, the N-adapted coefficients of d-connections
on (co) tangent Lorentz bundles can be parameterized (respectively) $\mathbf{%
\Gamma }_{\ \beta \gamma }^{\alpha }=\{L_{\ jk}^{i},\acute{L}_{\ bk}^{a},%
\acute{C}_{\ jc}^{i},C_{\ bc}^{a}\}$ and $\ ^{\shortmid }\mathbf{\Gamma }_{\
\beta \gamma }^{\alpha }=\{\ ^{\shortmid }L_{\ jk}^{i},\ ^{\shortmid }\acute{%
L}_{a\ k}^{\ b},\ ^{\shortmid }\acute{C}_{\ j}^{i\ c},\ ^{\shortmid
}C_{a}^{\ bc}\}.$ These coefficients can be used for explicit computations
of h-- and/or v--splitting, cv-splitting, of covariant derivatives
\begin{equation*}
\mathbf{D=}\left( _{h}\mathbf{D,\ }_{v}\mathbf{D}\right) \mbox{ and }
\ ^{\shortmid }\mathbf{D}=\left( \mathbf{\ }_{h}^{\shortmid }\mathbf{D }, \
_{v}^{\shortmid }\mathbf{D}\right) ,
\end{equation*}%
where $\ _{h}\mathbf{D}=\{L_{\ jk}^{i},\acute{L}_{\ bk}^{a}\},\ _{v}\mathbf{D%
}=\{\acute{C}_{\ jc}^{i},C_{\ bc}^{a}\}$ and $\ _{h}^{\shortmid }\mathbf{D}%
=\{\ ^{\shortmid }L_{\ jk}^{i},\ ^{\shortmid }\acute{L}_{a\ k}^{\ b}\},$ $\
_{v}^{\shortmid }\mathbf{D}=\{\ ^{\shortmid }\acute{C}_{\ j}^{i\ c},\
^{\shortmid }C_{a}^{\ bc}\}.$

We can consider a linear connection $\underline{D}$ (which is not obligatory a d-connection) and a d-connection $\mathbf{D}$ both defined on $\mathbf{TV}$. Such geometric objects are respectively denoted $\ ^{\shortmid }\underline{D}$ and $\ ^{\shortmid }\mathbf{D}$ on $\mathbf{T}^{\ast }\mathbf{V}$. For
(co)vector bundles, there are nonholonomic relations with respective distortion d-tensors
$\mathbf{Z:=D}-\underline{D}$ and $\ ^{\shortmid }\mathbf{Z:=\ ^{\shortmid }D}-\ ^{\shortmid }\underline{D}.$

Using similar definitions and theorems both for linear connections and d-connections, we prove that d--connection $\mathbf{D,}$ or $\ ^{\shortmid }\mathbf{D}$, is characterized by respective curvature $(\mathcal{R},$ or $\ ^{\shortmid }\mathcal{R}),$ torsion $(\mathcal{T},$ or $\ ^{\shortmid }%
\mathcal{T}),$ nonmetricity, $(\mathcal{Q},$ or $\ ^{\shortmid }\mathcal{Q})$, d-tensors,
{\small
\begin{eqnarray}
\mathcal{R}(\mathbf{X,Y})&:= &\mathbf{D}_{\mathbf{X}}\mathbf{D}_{\mathbf{Y}}-%
\mathbf{D}_{\mathbf{Y}}\mathbf{D}_{\mathbf{X}}-\mathbf{D}_{\mathbf{[X,Y]}},%
\mathcal{T}(\mathbf{X,Y}):=\mathbf{D}_{\mathbf{X}}\mathbf{Y}-\mathbf{D}_{%
\mathbf{Y}}\mathbf{X}-[\mathbf{X,Y}],\mathcal{Q}(\mathbf{X}):=\mathbf{D}_{%
\mathbf{X}}\mathbf{g},\mbox{ or }  \label{dcurvabstr} \\
\ ^{\shortmid }\mathcal{R}(\ ^{\shortmid }\mathbf{X,\ ^{\shortmid }Y}%
)&:=&^{\shortmid }\mathbf{D}_{\ ^{\shortmid }\mathbf{X}}\ ^{\shortmid }%
\mathbf{D}_{\ ^{\shortmid }\mathbf{Y}}-\ ^{\shortmid }\mathbf{D}_{\
^{\shortmid }\mathbf{Y}}\ ^{\shortmid }\mathbf{D}_{\ ^{\shortmid }\mathbf{X}%
}-\ ^{\shortmid }\mathbf{D}_{\mathbf{[\ ^{\shortmid }X,\ ^{\shortmid }Y]}},\
^{\shortmid }\mathcal{T}(\ ^{\shortmid }\mathbf{X,\ ^{\shortmid }Y}):=\
^{\shortmid }\mathbf{D}_{\ ^{\shortmid }\mathbf{X}}\ ^{\shortmid }\mathbf{Y}%
-\ ^{\shortmid }\mathbf{D}_{\ ^{\shortmid }\mathbf{Y}}\ ^{\shortmid }\mathbf{%
X}-[\ ^{\shortmid }\mathbf{X,\ ^{\shortmid }Y}],  \notag
\end{eqnarray}
} and $\ ^{\shortmid }\mathcal{Q}( \ ^{\shortmid }\mathbf{X}):=\ ^{\shortmid
}\mathbf{D}_{\ ^{\shortmid }\mathbf{X}}\ ^{\shortmid }\mathbf{g}$.
The N--adapted coefficients for the curvature, torsion and nonmetricity d-tensors are provided in Appendices to \cite{vacaru18,bubuianu18}, see also references therein. The geometric d-tensors (\ref{dcurvabstr}) are written, for instance, using tilde on symbols if such d-objects are defined and computed for Lagrange (or Hamilton) generating functions, see below.

\subsubsection{The Ricci and Einstein d--tensors on phase spaces and (co) vector bundles}

Respectively, the Ricci d--tensors are defined and computed as $Ric=\{%
\mathbf{R}_{\alpha \beta }:=\mathbf{R}_{\ \alpha \beta \tau }^{\tau }\},$
for a d-connection $\mathbf{D}$, and $\ ^{\shortmid }Ric=\{\ ^{\shortmid }%
\mathbf{R}_{\alpha \beta }:=\ ^{\shortmid }\mathbf{R}_{\ \alpha \beta \tau
}^{\tau }\},$ for a d-connection $\ ^{\shortmid }\mathbf{D,}$ see formulas (%
\ref{dcurvabstr}). In N-adapted form, we prove that the N-adapted
coefficients of the Ricci d--tensors of a d-connection $\mathbf{D}$ (or $\
^{\shortmid }\mathbf{D)}$ in respective phase spaces are parameterized in $h$%
- and/or $v$-, or $cv$-form, by formulas
\begin{eqnarray}
\mathbf{R}_{\alpha \beta } &=&\{R_{hj}:=R_{\ hji}^{i},\ \ R_{ja}:=-P_{\
jia}^{i},\ R_{bk}:=P_{\ bka}^{a},R_{\ bc}=S_{\ bca}^{a}\},\mbox{ or }
\label{dricci} \\
\ ^{\shortmid }\mathbf{R}_{\alpha \beta } &=&\{\ ^{\shortmid }R_{hj}:=\
^{\shortmid }R_{\ hji}^{i},\ ^{\shortmid }R_{j}^{\ a}:=-\ ^{\shortmid }P_{\
ji}^{i\ \ \ a},\ \ ^{\shortmid }R_{\ k}^{b}:=\ ^{\shortmid }P_{a\ k}^{\ b\ \
a},\ ^{\shortmid }R_{\ }^{bc}=\ ^{\shortmid }S_{a\ }^{\ bca}\}.
\label{driccid}
\end{eqnarray}

If a phase space is enabled both with a d-connection, $\mathbf{D}$ (or$\
^{\shortmid }\mathbf{D),}$ and d-metric, $\mathbf{g}$ (\ref{dmt}) (or$\
^{\shortmid }\mathbf{g}$ (\ref{dmct})) [in particular, we can consider
canonical "tilde" values with d-metrics $\widetilde{\mathbf{g}}$ (\ref{cdms}%
) and/or $\ ^{\shortmid }\widetilde{\mathbf{g}}$ (\ref{cdmds}), and their
frame transforms], we can define and compute nonholonomic Ricci scalars. In
result, we obtain that the scalar curvature of a d-connection $\mathbf{D,}$
or $\ ^{\shortmid }\mathbf{D,}$ can be defined and computed for the inverse
d-metric $\mathbf{g}^{\alpha \beta },$ or $\ ^{\shortmid }\mathbf{g}^{\alpha
\beta },$
\begin{equation*}
\ _{s}R:=\mathbf{g}^{\alpha \beta }\mathbf{R}_{\alpha \beta
}=g^{ij}R_{ij}+g^{ab}R_{ab}=R+S,\ \mbox{ or }\ _{s}^{\shortmid }R:=\
^{\shortmid }\mathbf{g}^{\alpha \beta }\ ^{\shortmid }\mathbf{R}_{\alpha
\beta }=\ ^{\shortmid }g^{ij}\ ^{\shortmid }R_{ij}+\ ^{\shortmid }g^{ab}\
^{\shortmid }R_{ab}=\ ^{\shortmid }R+\ ^{\shortmid }S,
\end{equation*}%
with respective h-- and v--components $R=g^{ij}R_{ij},S=g^{ab}S_{ab},$ or $\
^{\shortmid }R=\ ^{\shortmid }g^{ij}\ ^{\shortmid }R_{ij},\ ^{\shortmid }S=\
^{\shortmid }g_{ab}\ ^{\shortmid }S^{ab}.$

By constructions, the Einstein d-tensors on $\mathbf{TV}$ and/or $\mathbf{T}%
^{\ast }\mathbf{V}$ are defined:
\begin{equation*}
En=\{\mathbf{E}_{\alpha \beta }:=\mathbf{R}_{\alpha \beta }-\frac{1}{2}%
\mathbf{g}_{\alpha \beta }\ _{s}R\}\mbox{ and/or }\ ^{\shortmid }En=\{\
^{\shortmid }\mathbf{E}_{\alpha \beta }:=\ ^{\shortmid }\mathbf{R}_{\alpha
\beta }-\frac{1}{2}\ ^{\shortmid }\mathbf{g}_{\alpha \beta }\
_{s}^{\shortmid }R\}.
\end{equation*}
Such values can be used in MGTs and encoding geometric and physical models in quantum computing theories.

\subsubsection{Physically important d-connections for geometric mechanics}

The Lagrange and/or Hamilton phase spaces (with a possible $\mathcal{L}$%
--duality) can be endowed and characterized respectively by different type
geometric and physically important linear connections and canonical/ almost
symplectic connections, which are equivalent if respective distorting
relations are postulated. In our approaches to geometric mechanics and
classical and quantum field/ thermodynamic and gravity theories, we use such
linear connection structures:
\begin{eqnarray}
\lbrack \mathbf{g,N]} &\mathbf{\simeq }&\mathbf{[}\widetilde{\mathbf{g}},%
\widetilde{\mathbf{N}}]\mathbf{\simeq \lbrack }\widetilde{\theta }:=%
\widetilde{\mathbf{g}}(\widetilde{\mathbf{J}}\cdot ,\cdot ),\widetilde{%
\mathbf{P}}\mathbf{,}\widetilde{\mathbf{J}}\mathbf{,}\widetilde{\mathbb{J}}]
\label{canondcl} \\
&\Longrightarrow &\left\{
\begin{array}{ccccc}
\nabla : &  & \nabla \mathbf{g}=0;\ \mathbf{T[\nabla ]}=0, &  & %
\mbox{Lagrange LC--connection}; \\
\widehat{\mathbf{D}}: &  & \widehat{\mathbf{D}}\ \mathbf{g}=0;\ h\widehat{%
\mathbf{T}}=0,\ v\widehat{\mathbf{T}}=0. &  &
\mbox{canonical Lagrange
d-connection}; \\
\widetilde{\mathbf{D}}: &  & \widetilde{\mathbf{D}}\widetilde{\theta }=0,%
\widetilde{\mathbf{D}}\widetilde{\theta }=0 &  &
\mbox{almost sympl.
Lagrange d-connection};%
\end{array}%
\right.  \notag
\end{eqnarray}%
{\small \ }%
\begin{eqnarray}
\lbrack \ ^{\shortmid }\mathbf{g,\ ^{\shortmid }N]} &\simeq &[\ ^{\shortmid }%
\widetilde{\mathbf{g}},\ ^{\shortmid }\widetilde{\mathbf{N}}] \simeq \lbrack
\ ^{\shortmid }\widetilde{\theta }:=\ ^{\shortmid }\widetilde{\mathbf{g}}(\
^{\shortmid }\widetilde{\mathbf{J}}\cdot ,\cdot ),\ ^{\shortmid }\widetilde{%
\mathbf{P}},\ ^{\shortmid }\widetilde{\mathbf{J}},\ ^{\shortmid }\widetilde{%
\mathbb{J}}]  \label{canondch} \\
&\Longrightarrow &\left\{
\begin{array}{ccccc}
\ ^{\shortmid }\nabla : &  & \ ^{\shortmid }\nabla \ ^{\shortmid }\mathbf{g}%
=0;\ \ ^{\shortmid }\mathbf{T[\ ^{\shortmid }\nabla ]}=0, &  & %
\mbox{Hamilton LC-connection}; \\
\ ^{\shortmid }\widehat{\mathbf{D}}: &  & \ ^{\shortmid }\widehat{\mathbf{D}}%
\ \mathbf{g}=0;\ h\ ^{\shortmid }\widehat{\mathbf{T}}=0,\ cv\ ^{\shortmid }%
\widehat{\mathbf{T}}=0. &  & \mbox{canonical Hamilton d-connection}; \\
\ ^{\shortmid }\widetilde{\mathbf{D}}: &  & \ ^{\shortmid }\widetilde{%
\mathbf{D}}\ ^{\shortmid }\widetilde{\theta }=0,\ ^{\shortmid }\widetilde{%
\mathbf{D}}\ ^{\shortmid }\widetilde{\theta }=0 &  &
\mbox{almost sympl.
Hamilton d-connection.}%
\end{array}%
\right.  \notag
\end{eqnarray}

We can consider distortion relations
\begin{eqnarray}
\widehat{\mathbf{D}} &=&\nabla +\widehat{\mathbf{Z}},\widetilde{\mathbf{D}}%
=\nabla +\widetilde{\mathbf{Z}},\mbox{ and }\widehat{\mathbf{D}}=\widetilde{%
\mathbf{D}}+\mathbf{Z,}\mbox{  determined by }(\mathbf{g,N)};
\label{candistr} \\
\ ^{\shortmid }\widehat{\mathbf{D}} &=&\ ^{\shortmid }\nabla +\ ^{\shortmid }%
\widehat{\mathbf{Z}},\ ^{\shortmid }\widetilde{\mathbf{D}}=\ ^{\shortmid
}\nabla +\ ^{\shortmid }\widetilde{\mathbf{Z}},\mbox{ and }\ ^{\shortmid }%
\widehat{\mathbf{D}}=\ ^{\shortmid }\widetilde{\mathbf{D}}+\ ^{\shortmid }%
\mathbf{Z,}\mbox{ determined by }(\ ^{\shortmid }\mathbf{g,\ ^{\shortmid }N)}%
;  \notag
\end{eqnarray}%
with distortion d-tensors $\widehat{\mathbf{Z}},\widetilde{\mathbf{Z}},$ and
$\mathbf{Z,}$ on $T\mathbf{TV,}$ and $\ ^{\shortmid }\widehat{\mathbf{Z}},\
^{\shortmid }\widetilde{\mathbf{Z}},$ and $\ ^{\shortmid }\mathbf{Z,}$ on $T%
\mathbf{T}^{\ast }\mathbf{V.}$

Geometric mechanic models are characterized by respective canonical and/or
almost symplectic distortion d-tensors $\widehat{\mathbf{Z}}[\widetilde{%
\mathbf{g}},\widetilde{\mathbf{N}}],\widetilde{\mathbf{Z}}[\widetilde{%
\mathbf{g}},\widetilde{\mathbf{N}}],$ and $\mathbf{Z}[\widetilde{\mathbf{g}},%
\widetilde{\mathbf{N}}],$ for (almost symplectic) Lagrange models, and $\
^{\shortmid }\widehat{\mathbf{Z}}[\ ^{\shortmid }\widetilde{\mathbf{g}},\
^{\shortmid }\widetilde{\mathbf{N}}],\ ^{\shortmid }\widetilde{\mathbf{Z}}[\
^{\shortmid }\widetilde{\mathbf{g}},\ ^{\shortmid }\widetilde{\mathbf{N}}],$
and $\ ^{\shortmid }\mathbf{Z}[\ ^{\shortmid }\widetilde{\mathbf{g}},\
^{\shortmid }\widetilde{\mathbf{N}}],$ for (almost symplectic) Hamilton
models. Respective phase space geometries can be described in equivalent
forms by such data{\small
\begin{equation*}
\begin{array}{ccccc}
\mbox{ on }\mathbf{TV:} &  & (\mathbf{g,N,}\widehat{\mathbf{D}}%
)\leftrightarrows (L:\widetilde{\mathbf{g}}, \widetilde{\mathbf{N}},%
\widetilde{\mathbf{D}}) & \leftrightarrow (\widetilde{\theta },\widetilde{%
\mathbf{P}},\widetilde{\mathbf{J}},\widetilde{\mathbb{J}},\widetilde{\mathbf{%
D}}) & \leftrightarrow \lbrack (\mathbf{g[}N],\nabla )]; \\
&  & \updownarrow \mbox{ possible }\mathcal{L}\mbox{-duality }\& &  &
\updownarrow \mbox{ not N-adapted } \\
\mbox{on }\mathbf{T}^{\ast }\mathbf{V:} &  & (\ ^{\shortmid }\mathbf{g,\
^{\shortmid }N,}\ ^{\shortmid }\widehat{\mathbf{D}})\leftrightarrows (H:\
^{\shortmid }\widetilde{\mathbf{g}},\ ^{\shortmid }\widetilde{\mathbf{N}},\
^{\shortmid }\widetilde{\mathbf{D}}) & \leftrightarrow (\ ^{\shortmid }%
\widetilde{\theta },\ ^{\shortmid }\widetilde{\mathbf{P}},\ ^{\shortmid }%
\widetilde{\mathbf{J}},\ ^{\shortmid }\widetilde{\mathbb{J}},\ ^{\shortmid }%
\widetilde{\mathbf{D}}) & \leftrightarrow \lbrack (\ ^{\shortmid }\mathbf{g}%
[\ ^{\shortmid }N],\ ^{\shortmid }\nabla )].%
\end{array}%
\end{equation*}%
} We can work with canonical d-connection structures on (co) tangent
bundles, $\widehat{\mathbf{D}}$ and/or $\ ^{\shortmid }\widehat{\mathbf{D}}$
which allows us to decouple and integrate in most general exact and
parametric forms certain effective geometric flow and/or modified
gravitational field equations. Here we note that Lagrange--Finsler variables
can be introduced on 4-d, and higher dimension, (pseudo) Riemannian spaces
and in GR if nonholonomic fibered structures are considered on spacetime manifolds, see discussions and examples in Refs. \cite{vacaru18,bubuianu18,ruchin13,gheorghiu16,bubuianu19,vacaru19,vacaru19a,rajpoot17}.

An important example is that when imposing certain (in general,
nonholonomic) constraints of type $\widehat{\mathbf{Z}}=0,$ we obtain $%
\widehat{\mathbf{D}}_{|\widehat{\mathbf{Z}}=0}\simeq \nabla $ even $\widehat{%
\mathbf{D}}\neq \nabla .$ If such conditions are satisfied, we can extract
(pseudo) Riemannian or effective geometric mechanical (with tilde values)
LC-configurations from more (general) nonholonmic metric-affine structures.
For instance, we can obtain LC-configurations for geometric models with $%
\widehat{\mathbf{D}}$ and/or $\ ^{\shortmid }\widehat{\mathbf{D}}$ \ for
respective zero distortions, $\widehat{\mathbf{Z}}$ and/or $\ ^{\shortmid }%
\widehat{\mathbf{Z}}.$ Equivalently, one can be considered the zero torsion
conditions for $\widehat{\mathcal{T}}$ $=\{\widehat{\mathbf{T}}_{\ \alpha
\beta }^{\gamma }\}=0$ and/or $\ \ ^{\shortmid }\widehat{\mathcal{T}}$ $=\{\
^{\shortmid }\widehat{\mathbf{T}}_{\ \alpha \beta }^{\gamma }\}=0.$

Using distortions of linear connections, we can prove in abstract and
N-adapted forms that there are canonical distortion relations encoding
generating functions for respective Lagrange-Hamilton and equivalent
nonholonomic variables: For the curvature d-tensors, we compute%
\begin{equation*}
\widehat{\mathcal{R}}[\mathbf{g},\widehat{\mathbf{D}}=\nabla +\widehat{%
\mathbf{Z}}]=\mathcal{R}[\mathbf{g},\nabla ]+\widehat{\mathcal{Z}}[\mathbf{g}%
,\widehat{\mathbf{Z}}],\ ^{\shortmid }\widehat{\mathcal{R}}[\ ^{\shortmid }%
\mathbf{g},\ ^{\shortmid }\widehat{\mathbf{D}}=\ ^{\shortmid }\nabla +\
^{\shortmid }\widehat{\mathbf{Z}}]=\ ^{\shortmid }\mathcal{R}[\ ^{\shortmid }%
\mathbf{g},\ ^{\shortmid }\nabla ]+\ ^{\shortmid }\widehat{\mathcal{Z}}[\
^{\shortmid }\mathbf{g},\ ^{\shortmid }\widehat{\mathbf{Z}}],
\end{equation*}%
with respective distortion d-tensors $\ \widehat{\mathcal{Z}},$ on $\mathbf{%
TV,}$ and $\ ^{\shortmid }\widehat{\mathcal{Z}},$ on $\mathbf{T}^{\ast }%
\mathbf{V}.$ Similarly, we obtain for the Ricci d-tensors,%
\begin{equation*}
\widehat{R}ic[\mathbf{g},\widehat{\mathbf{D}}=\nabla +\widehat{\mathbf{Z}}%
]=Ric[\mathbf{g},\nabla ]+\widehat{Z}ic[\mathbf{g},\widehat{\mathbf{Z}}],\
^{\shortmid }\widehat{R}ic[\ ^{\shortmid }\mathbf{g},\ ^{\shortmid }\widehat{%
\mathbf{D}}=\ ^{\shortmid }\nabla +\ ^{\shortmid }\widehat{\mathbf{Z}}]=\
^{\shortmid }Ric[\ ^{\shortmid }\mathbf{g},\ ^{\shortmid }\nabla ]+\
^{\shortmid }\widehat{Z}ic[\ ^{\shortmid }\mathbf{g},\ ^{\shortmid }\widehat{%
\mathbf{Z}}],
\end{equation*}%
with respective distortion d-tensors $\ \widehat{Z}ic,$ on $\mathbf{TV,}$
and $\ \ ^{\shortmid }\widehat{Z}ic,$ on $\mathbf{T}^{\ast }\mathbf{V}$.
Finally, for the scalar curvature of canonical d-connection $\widehat{%
\mathbf{D}},$ or $\ ^{\shortmid }\widehat{\mathbf{D}},$%
\begin{equation*}
\ _{s}^{\shortmid }\widehat{R}[\mathbf{g},\widehat{\mathbf{D}}=\nabla +%
\widehat{\mathbf{Z}}]=\mathcal{R}[\mathbf{g},\nabla ]+\ _{s}\widehat{Z}[%
\mathbf{g},\widehat{\mathbf{Z}}],\ _{s}^{\shortmid }\widehat{R}[\
^{\shortmid }\mathbf{g},\ ^{\shortmid }\widehat{\mathbf{D}}=\ ^{\shortmid
}\nabla +\ ^{\shortmid }\widehat{\mathbf{Z}}]=\ _{s}^{\shortmid }R[\
^{\shortmid }\mathbf{g},\ ^{\shortmid }\nabla ]+\ _{s}^{\shortmid }\widehat{Z%
}[\ ^{\shortmid }\mathbf{g},\ ^{\shortmid }\widehat{\mathbf{Z}}],
\end{equation*}%
with respective distortion scalar functionals $\ _{s}\widehat{Z},$ on $\mathbf{TV,}$ and
$\ _{s}^{\shortmid }\widehat{Z},$ on $\mathbf{T}^{\ast }\mathbf{V}$.

Above formulas can be reformulated for distortions of the almost symplectic Lagrange, or Finsler, d-connections, for instance, considering
\begin{eqnarray*}
\widetilde{\mathcal{R}}[\widetilde{\mathbf{g}} &\simeq &\widetilde{\theta },%
\widetilde{\mathbf{D}}=\nabla +\widetilde{\mathbf{Z}}]=\mathcal{R}[%
\widetilde{\mathbf{g}}\simeq \widetilde{\theta },\nabla ]+\widetilde{%
\mathcal{Z}}[\widetilde{\mathbf{g}}\simeq \widetilde{\theta },\widetilde{%
\mathbf{Z}}], \\
\ ^{\shortmid }\widetilde{\mathcal{R}}[\ ^{\shortmid }\widetilde{\mathbf{g}}
&\simeq &\ ^{\shortmid }\widetilde{\theta },\ ^{\shortmid }\widetilde{%
\mathbf{D}}=\ ^{\shortmid }\nabla +\ ^{\shortmid }\widetilde{\mathbf{Z}}]=\
^{\shortmid }\mathcal{R}[\ ^{\shortmid }\widetilde{\mathbf{g}}\simeq \
^{\shortmid }\widetilde{\theta },\ ^{\shortmid }\nabla ]+\ ^{\shortmid }%
\widetilde{\mathcal{Z}}[\ ^{\shortmid }\mathbf{g}\simeq \ ^{\shortmid }%
\widetilde{\theta },\ ^{\shortmid }\widetilde{\mathbf{Z}}],
\end{eqnarray*}%
and any similar geometric objects with "tilde" symbols.

\section{Geometric flow evolution of classical mechanical systems}

\label{s3} The goal of this section is to formulate in canonical Hamilton
variables the theory of nonholonomic geometric flows of relativistic
mechanical systems. This is important for further developments in classical
and quantum information theories when the Hamilton variables are used in
explicit form. We shall present also the main results in canonical Lagrange
variables because such formulas are very important for investigating various
connections between quantum field theory, QFT, quantum gravity, QG, and
quantum information theory.\footnote{%
Such a research is related to author's project on geometric flows and
applications in physics which was elaborated in 2005 for a sabbatical
professor fellowship at CSIC, Madrid, in Spain, and further developments
supported by a project IDEI, in Romania; and related visiting projects at
CERN (Switzerland); M. Planck Institute, Munich, and A. Einstein Institute,
Postdam, (Germany) etc. Those projects were on applications of nonholonomic
geometric methods in classical and quantum mechanics and physics, with
various generalizations to deformation quantization, noncommutative geometry
etc. A sub-direction of former research was devoted to studies on flow evolution of
Lagrange-Hamilton systems geometrized on (co) tangent bundles, which
resulted in a series of works on the nonholonomic geometric evolution of
Finsler-Lagrange-Hamilton space spaces, see historical remarks and a
comprehensive bibliography in Appendix B.4.17 to Ref. \cite{vacaru18}. Here
we note that nonholonomic generalizations of G. Perelman functionals and R.
Hamilton geometric evolution equations were considered for Finsler--Lagrange
systems in Refs. \cite{vjmp08,vrmp09}, see further generalizations for
almost K\"{a}hler --Lagrange-Hamilton models on Lie algebroids ,
relativistic Lagrange-Hamilton mechanics etc. \cite%
{vacaru11,vacaru13,alexiou}. In principle, Finsler-Lagrange-Hamilton
variables can be introduced on any (non) commutative / (super) manifold,
which allows to re-write in effective (super/noncommutative) mechanic forms
all results on geometric flows of physical theories elaborated in \cite%
{vacaru09,rajpoot17,ruchin13,gheorghiu16,bubuianu19,vacaru19,vacaru19a}, see
also references therein.}

\subsection{Relativistic geometric flows and Perelman's thermodynamics for
phase spacetimes}

Let us consider families of nonholonomic 8--d tangent and cotangent Lorentz
bundles, $T\mathbf{V}(\tau )$ and $T^{\ast }\mathbf{V}(\tau )$ parameterized
by a positive parameter $\tau ,0\leq \tau \leq \tau _{0}.$ Such phase
spacetimes are enabled with corresponding sets of canonical d-metrics of
type (\ref{cdms}) and (\ref{cdmds}), $\widetilde{\mathbf{g}}(\tau )=%
\widetilde{\mathbf{g}}(\tau ,u)$ and $\ ^{\shortmid }\widetilde{\mathbf{g}}%
(\tau )=\ ^{\shortmid }\widetilde{\mathbf{g}}(\tau ,\ ^{\shortmid }u)$ and
canonical N--connections of type (\ref{cannc}), $\ ^{\shortmid }\widetilde{%
\mathbf{N}}(\tau )=\ ^{\shortmid }\widetilde{\mathbf{N}}(\tau ,\ ^{\shortmid
}u).$ Any relativistic nonholonomic phase spacetime $T\mathbf{V}$ $\subset T%
\mathbf{V}(\tau )$ and/or $T^{\ast }\mathbf{V}$ $\subset T^{\ast }\mathbf{V}%
(\tau )$ can be enabled with necessary types of double nonholonomic
(2+2)+(2+2) and (3+1)+(3+1) splitting, see details for such geometric
constructions in \cite{rajpoot17,ruchin13,gheorghiu16,bubuianu19,vacaru19}.%
\footnote{%
Additionally to coordinate and index conventions from footnote \ref%
{fncoordconv}, we label the local (3+1)+(3+1) coordinates in the form $\
^{\shortmid }u=\{\ ^{\shortmid }u^{\alpha }=\ ^{\shortmid }u^{\alpha
_{s}}=(x^{i_{1}},y^{a_{2}};p_{a_{3}},p_{a_{4}})=(x^{\grave{\imath}%
},u^{4}=y^{4}=t;p_{\grave{a}},p_{8}=E)\}$ for $i_{1},j_{1},k_{1},...=1,2;$ $%
a_{1},b_{1},c_{1},...=3,4;$ $a_{2},b_{2},c_{2},...=5,6;$ $%
a_{3},b_{3},c_{3},...=7,8;$ and $\grave{\imath},\grave{j},\grave{k}%
,...=1,2,3,$ respectively, $\grave{a},\grave{b},\grave{c},...=5,6,7$ can be
used for corresponding spacelike hyper surfaces on a base Lorentz manifold
and typical cofiber.} For instance, a nonholonomic (3+1)+(3+1) splitting on
a $T\mathbf{V}$ can be chosen in such a form that any open region on a base
Lorentz manifold, $U\subset $ $\mathbf{V,}$ is covered by a family of 3-d
spacelike hypersurfaces $\widehat{\Xi }_{t},$ or $\widetilde{\Xi }_{t},$
parameterized by a time like parameter $t.$ The parameterizations of
hypersurfaces can be labeled in certain forms which are adapted to the type
of canonical d-connection we use for our geometric constructions. In this
work, we prefer to use "tilde" labels/ values related to geometric
mechanics. On a typical cofiber of $T^{\ast }\mathbf{V,}$ we can consider a
3-d cofiber hypersurface $\ ^{\shortmid }\widetilde{\Xi }_{E}$, for
instance, of signature $(+++)$ with a label $E$ for parameterizations by an
energy type parameter. We can write correspondingly $\widetilde{\Xi }=(%
\widetilde{\Xi }_{t},\ \widetilde{\Xi }_{E})$ and $\ ^{\shortmid }\widetilde{%
\Xi }=(\widetilde{\Xi }_{t},\ ^{\shortmid }\widetilde{\Xi }_{E})$ $\ $for
nonholonomic distributions of base and fiber hypersurfaces with conventional
splitting 3+3 of signature (+++;+++) on total phase space. For additional
shell decompositions of type (2+2)+(2+2), we can use also a $s$-label, $\
_{s}^{\shortmid }\widehat{\Xi }=(\ _{s}\widehat{\Xi }_{t},\ _{s}^{\shortmid }%
\widehat{\Xi }_{E})\subset \ _{s}T^{\ast }\mathbf{V,}$ if we shall be
interested in constructing certain classes of exact or parametric solutions
of geometric flow equations. In general, we can elaborate on two generic
different types of geometric phase flow theories: The fist type is with a
conventional parameter $\tau (\chi )$ admitting re-parameterizations of a
temperature like parameter used for labeling 4-d Lorentz spacetimes and
their phase space configurations. The second type of models is with $\tau
(t) $ as a time like parameter when (3+3)-d spacelike phase configurations
evolve relativistically on a "redefined" time like coordinate. In this work,
we elaborate on theories of type 1.

\subsubsection{Perelman-Lagrange and Perelman-Hamilton functionals}

In \cite{vrmp09}, we studied geometric flows of Finsler-Lagrange theories using canonical data $(\mathbf{g}(\tau ),\widetilde{\mathbf{D}}(\tau ))$ when various generalizations and applications in MGTs were elaborated for the data $(\mathbf{g}(\tau ),\widehat{\mathbf{D}}(\tau ))$,
\cite{vjmp08,vacaru09,rajpoot17,ruchin13,gheorghiu16,bubuianu19,vacaru19,vacaru19a}. Those constructions were based on nonholonomic generalizations of Perelman's functionals \cite{perelman1} and distortion relations form the Levi-Civita configurations $(\mathbf{g}(\tau ),\nabla (\tau )).$ Let us
consider how Perelman's functionals can be generalized in relativistic form for geometric flow evolution of Lagrange--Hamilton spaces.

\paragraph{F- and W-functionals in canonical J. Lagrange variables:}

Considering canonical Lagrange data $(\widetilde{\mathbf{g}}(\tau ),%
\widetilde{\mathbf{D}}(\tau ))$ on tangent Lorentz bundles in order to
postulate the functionals:
\begin{eqnarray}
\widetilde{\mathcal{F}} &=&\widetilde{\int }e^{-\widetilde{f}}\sqrt{|%
\widetilde{\mathbf{g}}_{\alpha \beta }|}d^{8}u(\ _{s}\widetilde{R}+|%
\widetilde{\mathbf{D}}\widetilde{f}|^{2}) \mbox{ and }  \label{lffperelm} \\
\widetilde{\mathcal{W}} &=&\widetilde{\int }\widetilde{\mu }\sqrt{|%
\widetilde{\mathbf{g}}_{\alpha \beta }|}d^{8}u[\tau (\ \ _{s}\widetilde{R}%
+|\ \ _{h}\widetilde{\mathbf{D}}\ \widetilde{f}|+|\ \ _{v}\widetilde{\mathbf{%
D}}\ \widetilde{f}|)^{2}+\widetilde{f}-16],  \label{lwfperelm}
\end{eqnarray}%
where the normalizing function $\ \widetilde{f}(\tau ,u)$ satisfies the
conditions
\begin{equation*}
\widetilde{\int }\widetilde{\mu }\sqrt{|\widetilde{\mathbf{g}}_{\alpha \beta
}|}d^{8}u:=\int_{t_{1}}^{t_{2}}\int_{\widetilde{\Xi }_{t}}%
\int_{y_{(0)}^{8}}^{y^{8}}\int_{\ ^{\shortmid }\widetilde{\Xi }_{E}}%
\widetilde{\mu }\sqrt{|\widetilde{\mathbf{g}}_{\alpha \beta }|}d^{8}u=1
\end{equation*}
for $\widetilde{\mu }=\left( 4\pi \tau \right) ^{-8}e^{-\widetilde{f}},$
when the coefficients $16=2\times 8$ is for 8-d spaces. For 3-d hypersurface
LC-configurations with $\nabla ,$ such values transform into the standard G.
Perelman functionals. The Ricci scalar $\ _{s}\widetilde{R}$ is taken for
the Ricci d-tensor $\ \widetilde{\mathbf{R}}_{\alpha \beta }$ (\ref{dricci})
constructed for the canonical Lagrange data $(\widetilde{\mathbf{g}},%
\widetilde{\mathbf{D}})$. Re-defining the normalization functions and using
corresponding nonholonomic frame transforms and d-connection distortions, we
can re-write the functionals (\ref{lffperelm}) and (\ref{lwfperelm}) in
"hat" variables, $\widehat{\mathcal{F}}$ and $\widehat{\mathcal{W}},$ see
similar constructions in \cite{vjmp08,rajpoot17,ruchin13,gheorghiu16,bubuianu19,vacaru19}.

\paragraph{F- and W-functionals in canonical W. Hamilton variables:}

We use canonical data
$(\ ^{\shortmid }\widetilde{\mathbf{g}}(\tau ),\ ^{\shortmid }\widetilde{\mathbf{D}}(\tau ))$ on cotangent Lorentz bundles and postulate the functionals:
\begin{eqnarray}
\ ^{\shortmid }\widetilde{\mathcal{F}} &=&\ ^{\shortmid }\widetilde{\int }%
e^{-\ ^{\shortmid }\widetilde{f}}\sqrt{|\ ^{\shortmid }\widetilde{\mathbf{g}}%
_{\alpha \beta }|}d^{8}\ ^{\shortmid }u(\ _{s}^{\shortmid }\widetilde{R}+|\
^{\shortmid }\widetilde{\mathbf{D}}\ ^{\shortmid }\widetilde{f}|^{2}) \mbox{
and }  \label{ffperelmctl} \\
\ ^{\shortmid }\widetilde{\mathcal{W}} &=&\ ^{\shortmid }\widetilde{\int }\
^{\shortmid }\widetilde{\mu }\sqrt{|\ ^{\shortmid }\widetilde{\mathbf{g}}%
_{\alpha \beta }|}d^{8}\ ^{\shortmid }u[\tau (\ \ _{s}^{\shortmid }%
\widetilde{R}+|\ \ _{h}^{\shortmid }\widetilde{\mathbf{D}}\ ^{\shortmid }%
\widetilde{f}|+|\ \ _{v}^{\shortmid }\widetilde{\mathbf{D}}\ \ ^{\shortmid }%
\widetilde{f}|)^{2}+\ \ ^{\shortmid }\widetilde{f}-16],  \label{wfperelmctl}
\end{eqnarray}%
where the normalizing function $\ \ ^{\shortmid }\widetilde{f}(\tau ,\
^{\shortmid }u)$ satisfies
\begin{equation*}
\ ^{\shortmid }\widetilde{\int }\ ^{\shortmid }\widetilde{\mu }\sqrt{|\
^{\shortmid }\widetilde{\mathbf{g}}_{\alpha \beta }|}d^{8}\ ^{\shortmid
}u:=\int_{t_{1}}^{t_{2}}\int_{\widetilde{\Xi }_{t}}\int_{E_{1}}^{E_{2}}\int_{%
\ ^{\shortmid }\widetilde{\Xi }_{E}}\ ^{\shortmid }\widetilde{\mu }\sqrt{|\
^{\shortmid }\widetilde{\mathbf{g}}_{\alpha \beta }|}d^{8}\ ^{\shortmid }u=1
\end{equation*}
for $\ ^{\shortmid }\widetilde{\mu }=\left( 4\pi \tau \right) ^{-8}e^{-\
^{\shortmid }\widetilde{f}}$, when the coefficient $16=2\times 8$ is taken
for 8-d spaces. The Ricci scalar $\ _{s}^{\shortmid }\widetilde{R}$ is taken
for the Ricci d-tensor $\ ^{\shortmid }\widetilde{\mathbf{R}}_{\alpha \beta
} $ (\ref{driccid}) constructed using the canonical Hamilton data $(\
^{\shortmid }\widetilde{\mathbf{g}},\ ^{\shortmid }\widetilde{\mathbf{D}}).$

Similar functionals can be postulated for nonholonomic geometric flows on $%
T^{\ast }\mathbf{V}$ using data $(\ ^{\shortmid }\mathbf{g}(\tau ),\ \
^{\shortmid }\widehat{\mathbf{D}}(\tau ))$ and redefined integration
measures and normalizing functions on respective hypersurfaces. Considering
LC-configurations with $\ ^{\shortmid }\widetilde{\mathbf{D}}_{\mid \
^{\shortmid }\widetilde{\mathbf{T}}=0}=\ ^{\shortmid }\nabla $ and/or$\
^{\shortmid }\widehat{\mathbf{D}}_{\mid \ ^{\shortmid }\widehat{\mathbf{T}}%
=0}=\ ^{\shortmid }\nabla ,$ the values (\ref{ffperelmctl}) and (\ref%
{wfperelmctl}) transform respectively into 8-d phase space versions of the
so called Perelman's F-entropy and W-entropy. It should be noted that $\
^{\shortmid }\widetilde{\mathcal{W}}$ and/or $\ ^{\shortmid }\widehat{%
\mathcal{W}}$ \ do not have a character of entropy for pseudo--Riemannian
metrics but can be treated as a value characterizing relativistic geometric
hydrodynamic phase space flows.

\paragraph{Nonholonomic lapse and shift variables:}

Using N--adapted diadic shell and/or double (2+2)+(2+2) and (3+1)+(3+1)
frame and coordinate transforms of metrics with additional dependence on a
flow parameter, we can introduce various parameterizations of geometric
objects on phase spacetimes. To define thermodynamic like variables for
geometric flow evolution of stationary configurations on $T^{\ast }\mathbf{V,%
}$ we take {\small
\begin{eqnarray*}
&&\ ^{\shortmid }\mathbf{g}=\ ^{\shortmid }\mathbf{g}_{\alpha ^{\prime
}\beta ^{\prime }}(\tau ,\ ^{\shortmid }u)d\ ^{\shortmid }\mathbf{e}^{\alpha
^{\prime }}\otimes d\ ^{\shortmid }\mathbf{e}^{\beta ^{\prime }}=q_{i}(\tau
,x^{k})dx^{i}\otimes dx^{i}+q_{3}(\tau ,x^{k},y^{3})\mathbf{e}^{3}\otimes
\mathbf{e}^{3}-[\breve{N}(\tau ,x^{k},y^{3})]^{2}\mathbf{e}^{4}\otimes
\mathbf{e}^{4}+ \\
&&\ ^{\shortmid }q^{a_{2}}(\tau ,x^{k},y^{3},p_{b_{2}})\ ^{\shortmid }%
\mathbf{e}_{a_{2}}\otimes \ ^{\shortmid }\mathbf{e}_{a_{2}}+\ ^{\shortmid
}q^{7}(\tau ,x^{k},y^{3},p_{b_{2}},p_{b_{3}})\ ^{\shortmid }\mathbf{e}%
_{7}\otimes \ ^{\shortmid }\mathbf{e}_{7}-[\ ^{\shortmid }\check{N}(\tau
,x^{k},y^{3},p_{b_{2}},p_{b_{3}})]^{2}\ ^{\shortmid }\mathbf{e}_{8}\otimes \
^{\shortmid }\mathbf{e}_{8},
\end{eqnarray*}%
} where, for instance, $\ ^{\shortmid }e^{\alpha _{s}}$ are N-adapted bases
on total space of respective cotangent Lorentz bundles. This ansatz for
metrics is a general N-adapted one for a 8--d phase space metric which can
be written as an extension of a couple of 3--d metrics, $q_{ij}=diag(q_{%
\grave{\imath}})=(q_{i},q_{3})$ on a hypersurface $\widetilde{\Xi }_{t},$
and $\ ^{\shortmid }q^{\grave{a}\grave{b}}=diag(\ ^{\shortmid }q^{\grave{a}%
})=(\ ^{\shortmid }q^{a_{2}},\ ^{\shortmid }q^{7})$ on a hypersurface $\
^{\shortmid }\widetilde{\Xi }_{E},$ \ if
\begin{equation}
q_{3}=g_{3},\breve{N}^{2}=-g_{4}\mbox{ and }\ ^{\shortmid }q^{7}=\
^{\shortmid }g^{7},\ ^{\shortmid }\check{N}^{2}=-\ ^{\shortmid }g^{8},
\label{shift1}
\end{equation}%
where $\breve{N}$ is the lapse function on the base and $\ ^{\shortmid }%
\check{N}$ is the lapse function in the co-fiber (here we note that "the
inverse hat" labels are a bit different for the 4-th and 8-th coordinate).

On $T\mathbf{V},$ the nonholonomic lapse and shift variables are introduced
in a similar way, which results in d--matric parameterizations {\small
\begin{eqnarray}
&&\mathbf{g}=\mathbf{g}_{\alpha ^{\prime }\beta ^{\prime }}(\tau ,\ u)d\
\mathbf{e}^{\alpha ^{\prime }}\otimes d\mathbf{e}^{\beta ^{\prime
}}=q_{i}(\tau ,x^{k})dx^{i}\otimes dx^{i}+q_{3}(\tau ,x^{k},y^{3})\mathbf{e}%
^{3}\otimes \mathbf{e}^{3}-[\breve{N}(\tau ,x^{k},y^{3})]^{2}\mathbf{e}%
^{4}\otimes \mathbf{e}^{4}+  \notag \\
&&q^{a_{2}}(\tau ,x^{k},y^{3},y^{b_{2}})\mathbf{e}_{a_{2}}\otimes \mathbf{e}%
_{a_{2}}+q^{7}(\tau ,x^{k},y^{3},y^{b_{2}},y^{b_{3}})\mathbf{e}_{7}\otimes
\mathbf{e}_{7}-[\check{N}(\tau ,x^{k},y^{3},y^{b_{2}},y^{b_{3}})]^{2}\mathbf{%
e}_{8}\otimes \mathbf{e}_{8}.  \label{3amd1}
\end{eqnarray}%
} We consider respective hypersurface formulas, $q_{ij}=diag(q_{\grave{\imath%
}})=(q_{i},q_{3})$ on a hypersurface $\widetilde{\Xi }_{t},$ and $\ q_{%
\grave{a}\grave{b}}=diag(q_{\grave{a}})=(q_{a_{2}},q_{7})$ on a hypersurface
$\widetilde{\Xi }_{E},$ \ if $q_{3}=g_{3},\breve{N}^{2}=-g_{4}$ and$\
q_{7}=g_{7},\ \check{N}^{2}=-g_{8},$where $\breve{N}$ is the lapse function
on the base and $\ ^{\shortmid }\check{N}$ is the lapse function in the fiber

\subsubsection{Relativistic thermodynamic models for Lagrange--Hamilton
geometric evolution}

G. Perelman's very original idea was that the geometric flows of Riemannian metrics can be characterized by an analogous thermodynamic model \cite{perelman1}. In this work, we consider relativistic mechanical
generalizations related to geometric flow approaches to classical mechanics \cite{vjmp08,vrmp09}.

\paragraph{Some basic concepts from statistical thermodynamics:}

\label{ssbasicth}To elaborate analogous thermodynamical models we can
consider a partition function $Z=\int \exp (-\beta E)d\omega (E)$ for the
canonical ensemble at temperature $\beta ^{-1}=T$ (one should not confuse
this $T$ for thermodynamics with standard tensor notations with $T$
containing indices for respective for energy-momentum tensors and/or torsion
in MGTs) being defined by the measure taken to be the density of states $%
\omega (E).$ The thermodynamical values are computed in standard form for
the average energy, $\mathcal{E=}\ \left\langle E\right\rangle :=-\partial
\log Z/\partial \beta ,$ the entropy $\mathcal{S}:=\beta \left\langle
E\right\rangle +\log Z$ and the fluctuation $\eta :=\left\langle \left(
E-\left\langle E\right\rangle \right) ^{2}\right\rangle =\partial ^{2}\log
Z/\partial \beta ^{2}.$ Using $Z,$ we can define a conventional state
density (generalized as a density matrix, it is important for elaborations in
geometric flow thermodynamics and information theory, see next sections)%
\begin{equation*}
\sigma (\beta ,E)=Z^{-1}e^{-\beta E}.
\end{equation*}%
Considering $\log \sigma =-\beta \mathcal{E-}\log Z,$ we define the relative
entropy between any state density $\rho $ and $\sigma ,$%
\begin{equation*}
\mathcal{S}(\rho \shortparallel \sigma ):=-\mathcal{S}(\rho )+\int (\beta
\mathcal{E+}\log Z)\rho d\omega (E)=\beta \lbrack \mathcal{E}(\rho )-T%
\mathcal{S}(\rho )]+\log Z,
\end{equation*}%
where the average energy computed in the density matrix $\rho $ is
$\mathcal{E}(\rho )=\int \mathcal{E}\rho d\omega (E).$ The free energy corresponding to $\rho $ is
\begin{equation}
\mathcal{F}(\rho ):=\mathcal{E}(\rho )-T\mathcal{S}(\rho ).  \label{fren}
\end{equation}%
We note that if $\log Z$ is independent on $\rho $ (as we consider in above
formulas) we have $\mathcal{S}(\sigma \shortparallel \sigma )=0.$ This
allows us to write%
\begin{equation}
\mathcal{S}(\rho \shortparallel \sigma )=\beta \lbrack \mathcal{F}(\rho )-%
\mathcal{F}(\sigma )].  \label{relentr}
\end{equation}

In this work, we study the geometric flow evolution of thermodynamics systems that preserves the thermal equilibrium at temperature $\beta $ but maps $\rho \rightarrow \rho ^{\prime }$ (such density states are different ones) keeping the same density state $\sigma .$ We can provide a realistic physical interpretation for such systems if
\begin{equation}
\mathcal{S}(\rho \shortparallel \sigma )\geq \mathcal{S}(\rho ^{\prime
}\shortparallel \sigma ),\mbox{ i.e. }\mathcal{F}(\rho )\geq \mathcal{F}%
(\rho ^{\prime }).  \label{secondthlaw}
\end{equation}%
So, we should elaborate on thermodynamic geometric flows that preserve the thermal equilibrium and can only reduce the free energy. These aspects connect mechanical flow models to the second low of thermodynamics.\footnote{%
It should be noted here that G. Perelman treated $\tau =\beta ^{-1}$ as a temperature parameter and that he introduced the concept of W--entropy following an analogy to formulas for the entropy in statistical mechanics. We reproduce here the Remark 5.3 and next paragraph, just before section 6 in \cite{perelman1}:\ "An entropy formula for the Ricci flow in dimension two was found by Chow [C]; there seems to be no relation between his formula and ours. .... The interplay of statistical physics and (pseudo)-riemannian
geometry occurs in the subject of Black Hole Thermodynamics, developed by Hawking et al. Unfortunately, this subject is beyond my understanding at the moment." It should be also emphasized that G. Perelman had not specified what type of underlying microstates and their energy should be taken in order to explain the geometric flows corresponding to certain thermodynamical and gravity models. In this work, we are interested in geometric mechanics and the classical and quantum information theory developing our approaches elaborated in \cite{vjmp08,vacaru09,rajpoot17,ruchin13,gheorghiu16,bubuianu19,vacaru19,vacaru19a}.}

\paragraph{Thermodynamic values for relativistic Lagrange-Hamilton flows:}

For relativistic geometric flows of mechanical systems, we introduce respective thermodynamic generating functions%
\begin{eqnarray}
\widetilde{\mathcal{Z}}[\widetilde{\mathbf{g}}(\tau )] &=&\widetilde{\int }%
e^{-\widetilde{f}}\sqrt{|\widetilde{\mathbf{g}}_{\alpha \beta }|}d^{8}u(-%
\widetilde{f}+16),\mbox{ for }T\mathbf{V;}  \label{thgenerfunct} \\
\ ^{\shortmid }\widetilde{\mathcal{Z}}[\ ^{\shortmid }\widetilde{\mathbf{g}}%
(\tau )] &=&\ ^{\shortmid }\widetilde{\int }e^{-\ ^{\shortmid }\widetilde{f}}%
\sqrt{|\ ^{\shortmid }\widetilde{\mathbf{g}}_{\alpha \beta }|}d^{8}\
^{\shortmid }u(-\ ^{\shortmid }\widetilde{f}+16),\mbox{ for }T^{\ast }%
\mathbf{V,}  \notag
\end{eqnarray}%
where the respective functional dependence is given by $[\widetilde{\mathbf{g%
}}(\tau )]$ and $[\ ^{\shortmid }\widetilde{\mathbf{g}}(\tau )]$ (we shall
not write such dependencies if that will not result in ambiguities). For a
thermodynamic analogous interpretation we can consider that a density state $%
\sigma $ is associated to $\widetilde{\mathbf{g}}_{\alpha \beta },$ we can
write in functional form $\sigma \lbrack \widetilde{\mathbf{g}}],$ but the
geometric evolution may involve densities $\rho \lbrack \ _{1}\widetilde{%
\mathbf{g}}]$ and $\rho ^{\prime }[\ _{1}\widetilde{\mathbf{g}}],$ where the
left label 1 is used in order to distinguish two d-metrics $\widetilde{%
\mathbf{g}}$ and $\ _{1}\widetilde{\mathbf{g}}.$ On cotangent bundles, such
values are written respectively $\ ^{\shortmid }\sigma \lbrack \ ^{\shortmid
}\widetilde{\mathbf{g}}],\ ^{\shortmid }\rho \lbrack \ _{1}^{\shortmid }%
\widetilde{\mathbf{g}}]$ and $\rho ^{\prime }[\ _{1}^{\shortmid }\widetilde{%
\mathbf{g}}].$

Generalizing for nonholonomic deformations of metrics and d-connections
respective formulas related to respective entropy like functionals (\ref%
{lffperelm}), (\ref{lwfperelm}) and (\ref{ffperelmctl}), (\ref{wfperelmctl}%
), we can define and compute such relativisitic thermodynamic values for
geometric evolution flows of Lagrange mechanical systems,%
\begin{eqnarray}
\mbox{average  flow energy:  }\widetilde{\mathcal{E}}\  &=&-\tau ^{2}%
\widetilde{\int }e^{-\widetilde{f}}\sqrt{|q_{1}q_{2}q_{3}\breve{N}%
q_{5}q_{6}q_{7}\check{N}|}\delta ^{8}u(\ _{s}\widetilde{R}+|\widetilde{%
\mathbf{D}}\widetilde{f}|^{2}\mathbf{\ }-\frac{8}{\tau }),  \label{8rdthvls}
\\
\mbox{ flow entropy: }\widetilde{\mathcal{S}}\  &=&-\widetilde{\int }e^{-%
\widetilde{f}}\sqrt{|q_{1}q_{2}q_{3}\breve{N}q_{5}q_{6}q_{7}\check{N}|}%
\delta ^{8}u\left[ \tau \left( \ _{s}\widetilde{R}+|\widetilde{\mathbf{D}}%
\widetilde{f}|^{2}\right) +\tilde{f}-16\right] ,  \notag \\
\mbox{ flow fluctuation: }\widetilde{\eta }\  &=&-\widetilde{\int }e^{-%
\widetilde{f}}\sqrt{|q_{1}q_{2}q_{3}\breve{N}q_{5}q_{6}q_{7}\check{N}|}%
\delta ^{8}u[|\ \widetilde{\mathbf{R}}_{\alpha \beta }+\widetilde{\mathbf{D}}%
_{\alpha }\ \widetilde{\mathbf{D}}_{\beta }\tilde{f}-\frac{1}{2\tau }\mathbf{%
g}_{\alpha \beta }|^{2}],  \notag
\end{eqnarray}%
where $\delta ^{8}u$ contains N-elongated differentials of type (\ref{nadapb}%
) (when we compute such integrals in N-adapted form). Using such values, we
can compute the respective free energy (\ref{fren}) and relative entropy (%
\ref{relentr}),%
\begin{equation*}
\widetilde{\mathcal{F}}(\ _{1}\widetilde{\mathbf{g}})=\widetilde{\mathcal{E}}%
(\ _{1}\widetilde{\mathbf{g}})-\beta ^{-1}\widetilde{\mathcal{S}}(\ _{1}%
\widetilde{\mathbf{g}})\mbox{ and }\widetilde{\mathcal{S}}(\ _{1}\widetilde{%
\mathbf{g}}\shortparallel \sigma )=\beta \lbrack \widetilde{\mathcal{F}}(\
_{1}\widetilde{\mathbf{g}})-\widetilde{\mathcal{F}}(\widetilde{\mathbf{g}})],%
\mbox{ where }
\end{equation*}%
\begin{eqnarray*}
\widetilde{\mathcal{E}}(\ _{1}\widetilde{\mathbf{g}}) &=&-\tau ^{2}%
\widetilde{\int }e^{-\widetilde{f}}\sqrt{|q_{1}q_{2}q_{3}\breve{N}%
q_{5}q_{6}q_{7}\check{N}|}\delta ^{8}u[\ _{s}\widetilde{R}(\ _{1}\widetilde{%
\mathbf{g}})+|\widetilde{\mathbf{D}}(\ _{1}\widetilde{\mathbf{g}})\widetilde{%
f}(\tau ,u)|^{2}\mathbf{\ }-\frac{8}{\tau }], \\
\widetilde{\mathcal{S}}(\ _{1}\widetilde{\mathbf{g}}) &=&-\widetilde{\int }%
e^{-\widetilde{f}}\sqrt{|q_{1}q_{2}q_{3}\breve{N}q_{5}q_{6}q_{7}\check{N}|}%
\delta ^{8}u\left[ \tau \left( \ _{s}\widetilde{R}(\ _{1}\widetilde{\mathbf{g%
}})+|\widetilde{\mathbf{D}}(\ _{1}\widetilde{\mathbf{g}})\widetilde{f}(\tau
,u)|^{2}\right) +\tilde{f}(\tau ,u)-16\right] .
\end{eqnarray*}%
Such values are in relativistic thermodynamic relation if the second
thermodynamic law (\ref{secondthlaw}) is satisfied. This impose certain
constraints on the class of normalizing and generating functions we consider
for the termodynamic description of such relativistic Lagrange systems.

For geometric evolution flows of Hamilton mechanical systems, the
relativistic thermodynamic values are%
\begin{eqnarray}
\ ^{\shortmid }\widetilde{\mathcal{E}}\ &=&-\tau ^{2}\ ^{\shortmid }%
\widetilde{\int }e^{-\ ^{\shortmid }\widetilde{f}}\sqrt{|q_{1}q_{2}q_{3}%
\breve{N}\ ^{\shortmid }q_{5}\ ^{\shortmid }q_{6}\ ^{\shortmid }q_{7}\
^{\shortmid }\check{N}|}\delta ^{8}\ ^{\shortmid }u(\ \ _{s}^{\shortmid }%
\widetilde{R}+|\ ^{\shortmid }\widetilde{\mathbf{D}}\ ^{\shortmid }%
\widetilde{f}|^{2}\mathbf{\ }-\frac{8}{\tau }),  \label{8rdthvhs} \\
\ ^{\shortmid }\widetilde{\mathcal{S}}\ &=&-\ ^{\shortmid }\widetilde{\int }%
e^{-\ ^{\shortmid }\widetilde{f}}\sqrt{|q_{1}q_{2}q_{3}\breve{N}\
^{\shortmid }q_{5}\ ^{\shortmid }q_{6}\ ^{\shortmid }q_{7}\ ^{\shortmid }%
\check{N}|}\delta ^{8}\ ^{\shortmid }u\left[ \tau \left( \ _{s}^{\shortmid }%
\widetilde{R}+|\ ^{\shortmid }\widetilde{\mathbf{D}}\ ^{\shortmid }%
\widetilde{f}|^{2}\right) +\ ^{\shortmid }\tilde{f}-16\right] ,  \notag \\
\ ^{\shortmid }\widetilde{\eta }\ &=&-\ ^{\shortmid }\widetilde{\int }e^{-\
^{\shortmid }\widetilde{f}}\sqrt{|q_{1}q_{2}q_{3}\breve{N}\ ^{\shortmid
}q_{5}\ ^{\shortmid }q_{6}\ ^{\shortmid }q_{7}\ ^{\shortmid }\check{N}|}%
\delta ^{8}\ ^{\shortmid }u[|\ \ ^{\shortmid }\widetilde{\mathbf{R}}_{\alpha
\beta }+\ ^{\shortmid }\widetilde{\mathbf{D}}_{\alpha }\ \ ^{\shortmid }%
\widetilde{\mathbf{D}}_{\beta }\ ^{\shortmid }\tilde{f}-\frac{1}{2\tau }\
^{\shortmid }\mathbf{g}_{\alpha \beta }|^{2}].  \notag
\end{eqnarray}%
Other thermodynamic values and conditions can be computed by analogy to
above relativistic Lagrange thermodynamic configurations and formulas (\ref%
{fren}), (\ref{relentr}) and (\ref{secondthlaw}).

Finally we note that above formulas can be written respectively and
equivalently in terms of the canonical d--connections $\widehat{\mathbf{D}}$
and $\ _{\shortmid }\widehat{\mathbf{D}}$ if we consider nonholonomic
deformations to certain systems of nonlinear partial differential equations
with general decoupling.

\subsubsection{Curved spaces emerging from relativistic phase space
geometric evolution}

The geometric flow evolution of 4-d (pseudo) Riemannian configurations is
described by nonholonomically modified Perelman's functionals integrated on
(co) fiber variables (\ref{lffperelm}), (\ref{lwfperelm}) and/or (\ref%
{ffperelmctl}), (\ref{wfperelmctl}). A subclass of such relativistic flows
are generated for parameterizations with d-metrics (\ref{lqe}) and (\ref%
{lqed})). Re-defining the normalizing functions, $\widetilde{f}\rightarrow
\widehat{f}(x^{1},x^{2},y^{3},y^{4})$ and/or $\widetilde{f}\rightarrow \
^{\shortmid }\widehat{f},$ for general frame transforms on a base Lorentz
manifold, we obtain such functionals:
\begin{eqnarray}
\widehat{\mathcal{F}} &=&\int_{t_{1}}^{t_{2}}\int_{\widehat{\Xi }_{t}}e^{-%
\widehat{f}}\sqrt{|q_{1}q_{2}q_{3}\breve{N}|}\delta ^{4}u(\ _{s}\widehat{R}+|%
\widehat{\mathbf{D}}\widehat{f}|^{2})\mbox{ and }  \label{perem4drelat} \\
\widehat{\mathcal{W}} &=&\int_{t_{1}}^{t_{2}}\int_{\widehat{\Xi }_{t}}\left(
4\pi \tau \right) ^{-4}e^{-f}\sqrt{|q_{1}q_{2}q_{3}\breve{N}|}\delta
^{4}u[\tau (\ \ _{s}\widehat{R}+|\ \ _{h}\widehat{\mathbf{D}}\ \widehat{f}%
|+|\ \ _{v}\widehat{\mathbf{D}}\ \widehat{f}|)^{2}+\widehat{f}-8].  \notag
\end{eqnarray}%
In these formulas, geometric fllows of $\ _{s}\widehat{R}$ are for
respective $\widehat{\mathbf{D}}=(\ _{h}\widehat{\mathbf{D}},\ _{v}\widehat{%
\mathbf{D}})$ on a family of bases $\mathbf{V}(\tau ),$where the normalizing
function $\ \widehat{f}(\tau ,u)$ satisfies the conditions $%
\int_{t_{1}}^{t_{2}}\int_{\widetilde{\Xi }_{t}}\widehat{\mu }\sqrt{%
|q_{1}q_{2}q_{3}\breve{N}|}\delta ^{4}u=1$ for $\widehat{\mu }=\left( 4\pi
\tau \right) ^{-4}e^{-\widehat{f}},$ when the coefficient $8=2\times 4$ is
taken for 4-d manifolds.

Using formulas for distortions of connections (\ref{candistr}) re-defined
for 4-d nonholonomic manifolds, the functionals (\ref{perem4drelat}) can
re-written using geometric data $(\widetilde{\mathbf{g}},\widetilde{\mathbf{D%
}})$ and/or $(\mathbf{g},\nabla ).$ Such F- and W--functionals define
nonholonomic geometric evolution flows of vacuum gravitational fields in
MGTs and GR, see details in Refs. \cite{ruchin13,gheorghiu16,bubuianu19,vacaru19}. We can consider that, in principle, (modified) gravitational interactions are induced as certain emergent fields from geometric evolution flows of mechanical Lagrange/ Hamilton systems.

The thermodynamic generating function corresponding to (\ref{perem4drelat}) can be defined in the form
\begin{equation*}
\widehat{\mathcal{Z}}=\int_{t_{1}}^{t_{2}}\int_{\widehat{\Xi }_{t}}e^{-%
\widehat{f}}\sqrt{|q_{1}q_{2}q_{3}\breve{N}|}\delta ^{4}u(-\widehat{f}+8),%
\mbox{
for }\mathbf{V.}
\end{equation*}%
In result, we can characterize emergent (pseudo) Riemannian geometries by
such relativistic thermodynamic values,%
\begin{eqnarray}
\widehat{\mathcal{E}}\ &=&-\tau ^{2}\int_{t_{1}}^{t_{2}}\int_{\widehat{\Xi }%
_{t}}e^{-\widehat{f}}\sqrt{|q_{1}q_{2}q_{3}\breve{N}|}\delta ^{4}u(\ _{s}%
\widehat{R}+|\widehat{\mathbf{D}}\widehat{f}|^{2}\mathbf{\ }-\frac{4}{\tau }%
),  \label{4dthermodval} \\
\widehat{\mathcal{S}}\ &=&-\int_{t_{1}}^{t_{2}}\int_{\widehat{\Xi }_{t}}e^{-%
\widehat{f}}\sqrt{|q_{1}q_{2}q_{3}\breve{N}|}\delta ^{4}u\left[ \tau \left(
\ _{s}\widehat{R}+|\widehat{\mathbf{D}}\widehat{f}|^{2}\right) +\widehat{f}-8%
\right] ,  \notag \\
\widehat{\eta }\ &=&-\int_{t_{1}}^{t_{2}}\int_{\widehat{\Xi }_{t}}e^{-%
\widehat{f}}\sqrt{|q_{1}q_{2}q_{3}\breve{N}|}\delta ^{4}u[|\ \widehat{%
\mathbf{R}}_{\alpha \beta }+\widehat{\mathbf{D}}_{\alpha }\ \widehat{\mathbf{%
D}}_{\beta }\tilde{f}-\frac{1}{2\tau }\mathbf{g}_{\alpha \beta }|^{2}],
\notag
\end{eqnarray}%
where all geometric objects and indices are for 4-d base manifolds. Up to
nonholonomic frame transforms and deformations of connections, such vaules
encode explicit information (integrated on fiber variables and/or projected
on base spacetime manifolds) on certain total space Lagrange/ Hamilton
generating functions.

There are different approaches for elaborating models of 3--d Ricci flow
evolution of mechanical systems and (emergent of prescribe) 4--d spacetimes
with pseudo--Euclidean signature. In principle, there are two general
possibilities. In the first case, is to approach the problem as in the
theories of stochastic / diffusion and kinetic processes with local
anisotropy, fractional geometric evolution etc.  For such models, one
elaborates on thermofield models of Ricci flow evolution on imaginary time $%
\tau =-it(0\leq \tau \leq 1/\kappa T,$ where $\kappa $ is Boltzmann's
constant. In corresponding formulas, the symbol $T$ \ is used for the
temperature (such a letter with respective indices for torsion and
energy-momentum tensors is also used in gravity theories). In result, the
pseudo--Riemannian spacetime is transformed into a Riemannian configuration
space as one elaborates in thermal and/or finite temperature quantum field
theory. The second class consists from theories modelled on 3-d
hypersurfaces and evolving relativistically, for instance, on a 4-d Ricci
soliton configuration. In such cases, the evolution parameter $\tau \sim t$
is a time like coordinate. In this work, we study evolution of relativistic
mechanics systems on a temperature like parameter $\tau \sim T.$

\subsubsection{Effective nonholonomic 3-d space like hypersurface F- and
W-functionals}

Lagrange and Hamilton mechanical systems on Lorentz manifolds can be also
characterized by 3-d space like hypersurface functionals. Such values can be
defined respectively for (\ref{perem4drelat}) and (\ref{4dthermodval}) for
any 3+1 splitting with 3-d closed hypersurface fibrations $\widehat{\Xi }%
_{t}.$

We denote by$\ _{\circ }\widehat{\mathbf{D}}=\widehat{\mathbf{D}}_{\mid
\widehat{\Xi }_{t}}$ the canonical d--connection $\widehat{\mathbf{D}}$
defined on a 3-d hypersurface $\widehat{\Xi }_{t}.$ In a similar form, there
are defined hypersurface "tilde" variables with$\ _{\circ }\widetilde{%
\mathbf{D}}=\widetilde{\mathbf{D}}_{\mid \widetilde{\Xi }_{t}}$ determined
as a projection of 8-d canonical Lagrange-Hamilton d--connection defined on
a 3-d hypersurface $\widetilde{\Xi }_{t}.$ For geometric flow evolution, all
such values depend on a temperature like parameter $\tau (\tau ^{\prime })$
with possible scale re-definitions for another parameter $\tau ^{\prime }$
etc. We define also $\ _{\circ }^{s}\widehat{R}:=$ $\ ^{s}\widehat{R}_{\mid
\widehat{\Xi }_{t}}$ and $\ _{\circ }^{s}\widetilde{R}:=$ $\ ^{s}\widetilde{R%
}_{\mid \widetilde{\Xi }_{t}}.$ Using $q_{\grave{\imath}}(\tau )=[q_{i}(\tau
),q_{3}(\tau )]$ in a family of d-metrics (see, for instance, (\ref{3amd1}%
)), we define 3-d F- and W-functionals parameterized in N--adapted form for
the canonical d-connection,
\begin{eqnarray}
\ _{\circ }\widehat{\mathcal{F}} &=&\int_{\widehat{\Xi }_{t}}e^{-\ \ _{\circ
}\widehat{f}}\sqrt{|q_{1}q_{2}q_{3}|}\delta \grave{x}^{3}\left[ (\ \ _{\circ
}^{s}\widehat{R}\mathcal{+}|\ _{\circ }\widehat{\mathbf{D}}\ \ _{\circ }%
\widehat{f}|^{2})\right] ,\mbox{ and }  \label{perelm3f} \\
\ \ _{\circ }\widehat{\mathcal{W}} &=&\int_{\widehat{\Xi }_{t}}\ \ _{\circ }%
\widehat{\mu }\sqrt{|q_{1}q_{2}q_{3}|}\delta \grave{x}^{3}\left[ \tau \left(
\ _{\circ }^{s}\widehat{R}+|\ _{\circ }^{h}\widehat{\mathbf{D}}\ _{\circ }%
\widehat{f}|+|\ _{\circ }^{v}\widehat{\mathbf{D}}\ _{\circ }\widehat{f}%
|\right) ^{2}+\ \ _{\circ }\widehat{f}-6\right] .  \label{perelm3w}
\end{eqnarray}%
These functionals are for a redefined normalization function $\ _{\circ }%
\widehat{f}.$ We can always chose a necessary type scaling function $\
_{\circ }\widehat{f}$ which satisfies normalization conditions $\int_{%
\widehat{\Xi }_{t}}\ \ _{\circ }\widehat{\mu }\sqrt{|q_{1}q_{2}q_{3}|}\delta
\grave{x}^{3}=1$ for $\ \ _{\circ }\widehat{\mu }=\left( 4\pi \tau \right)
^{-3}e^{-\ _{\circ }\widehat{f}}$ $.$ For topological considerations, the
type of normalization is not important. Such conditions can be imposed as
via frame/coordinate transforms and deformations of linear connections which
allows to solve derived geometric flow evolution equations in explicit form.
For certain applications, we can consider $_{\circ }\widehat{f}$ as an
undetermined scalar function which can be related to certain conformal
transforms or re-parameterizations.

Using $\ _{\circ }\widehat{\mathcal{F}}$ (\ref{perelm3f}) and the
thermodynamic generating function  $\ _{\circ }\widehat{\mathcal{Z}}=\exp
[\int_{\widehat{\Xi }_{t}}\ _{\circ }\widehat{\mu }\sqrt{|q_{1}q_{2}q_{3}|}%
\delta \grave{x}^{3}(-\ _{\circ }\widehat{f}+6)]$, we can define and compute
such 3-d hypersurface thermodynamic values:
\begin{eqnarray}
\ \ _{\circ }\widehat{\mathcal{E}}\ &=&-\tau ^{2}\int_{\widehat{\Xi }_{t}}\
_{\circ }\widehat{\mu }\sqrt{|q_{1}q_{2}q_{3}|}\delta \grave{x}^{3}\left(
\mathbf{\ }\ \ _{\circ }^{s}\widehat{R}+|\ _{\circ }\widehat{\mathbf{D}}\ \
_{\circ }\widehat{f}|^{2}-\frac{3}{\tau }\right) ,  \label{3hsthermod} \\
\ \ _{\circ }\widehat{\mathcal{S}} &=&-\int_{\widehat{\Xi }_{t}}\ _{\circ }%
\widehat{\mu }\sqrt{|q_{1}q_{2}q_{3}|}\delta \grave{x}^{3}\left[ \tau \left(
\ \ \ _{\circ }^{s}\widehat{R}\mathcal{+}|\ _{\circ }\widehat{\mathbf{D}}\ \
_{\circ }\widehat{f}|^{2}\right) +\tilde{f}-6\right] ,  \notag \\
\ \ _{\circ }\widehat{\eta } &=&2\ \tau ^{4}\int_{\widehat{\Xi }_{t}}\
_{\circ }\widehat{\mu }\sqrt{|q_{1}q_{2}q_{3}|}\delta \grave{x}^{3}[|\ \
_{\circ }\widehat{\mathbf{R}}_{\grave{\imath}\grave{j}}+\ \ _{\circ }%
\widehat{\mathbf{D}}_{\grave{\imath}}\ \ _{\circ }\widehat{\mathbf{D}}_{%
\grave{j}}\tilde{f}-\frac{1}{2\tau }q_{\grave{\imath}\grave{j}}|^{2}].
\notag
\end{eqnarray}%
These formulas can be considered for 4--d configurations (\ref{4dthermodval}%
) taking the lapse function $\breve{N}=1$ for N-adapted Gaussian
coordinates. We can also write such formulas in equivalent form using
geometric data $(\widetilde{\mathbf{q}},\ _{\circ }\widetilde{\mathbf{D}})$
and/or $(\mathbf{q},\ _{\circ }\nabla )$ for respectively re-defined
normalizing functions. For LC-configurations, the 3-d hypersurface formulas (%
\ref{perelm3f}), (\ref{perelm3w}) and (\ref{3hsthermod}) transform into the
standard ones from G. Perelman's preprint \cite{perelman1}. The main
difference is that in our approach such Riemannian hypersufrace flow
evolution scenarios are determined by Lagrange-Hamilton mechanical systems.

\subsection{Generalized R. Hamilton flow evolution equations and geometric
mechanics}

In this section we show that Lagrange and/or Hamilton mechanical systems are characterized not only by dynamical equations (which is well-known from classical mechanics \cite{abraham,arnold,deleon85}) but
also by certain classes of geometric flow evolution equations \cite{vjmp08,vrmp09}. Relativistic variants of such systems of nonlinear PDEs can be proven by applying a variational N-adapted calculus for respective F- and W-functionals as in  \cite{vjmp08,vacaru09,rajpoot17,ruchin13,gheorghiu16,bubuianu19,vacaru19,vacaru19a}.  For holonomic Riemannian manifolds, such proofs can be found in \cite{perelman1,monogrrf1,monogrrf2,monogrrf3}.

\subsubsection{Riemannian geometric flows on 3-d spacelike hypersurface}

Applying a N--adapted variational procedure on a 3-d hypersurface to a
functional (\ref{perelm3f}) or (\ref{perelm3w}) defined by data $(\widetilde{%
g}_{\grave{\imath}\grave{j}},\widetilde{\nabla }),$ we obtain such equations
in the form
\begin{equation}
\frac{\partial \widetilde{g}_{\grave{\imath}\grave{j}}}{\partial \tau }=-2\
\widetilde{R}_{\grave{\imath}\grave{j}},  \label{heq1a}
\end{equation}%
where $\tau $ is an evolution real parameter. There are used local coordinates $u^{\grave{\imath}}$ with indices $\grave{\imath},\grave{j}=1,2,3 $ and Ricci tensor $\widetilde{R}_{\grave{\imath}\grave{j}}$ for a
3-d Riemannian manifold (in this work constructed as an emergent from geometric mechanics curve space). These equations are equivalent to the (non-relativistic) Ricci flow evolution equations postulated heuristically
by R. Hamilton \cite{hamilt1,hamilt2,hamilt3}. G. Perelman proved such equations using his F- and W-functionals.

The equations (\ref{heq1a}) describe a nonlinear diffusion process for
geometric flow evolution of relativistic mechanical systems encoded up to
frame transforms into 3-d Riemannian metrics (we can omit tilde and write $%
{g}_{\grave{\imath}\grave{j}}$ in certain general covariant form). For models with small
deformations of a 3--d Euclidean metric $g_{\grave{\imath}\grave{j}}\approx
\delta _{\grave{\imath}\grave{j}}+$ $h_{\grave{\imath}\grave{j}},$ with $%
\delta _{\grave{\imath}\grave{j}}=diag[1,1,1]$ and $h_{\grave{\imath}\grave{j%
}}|\ll 1$, the Ricci tensor approximates the 3-d Laplace operator $\Delta =%
\frac{\partial ^{2}}{(\partial u^{1})^{2}}+\frac{\partial ^{2}}{(\partial
u^{2})^{2}}+\frac{\partial ^{2}}{(\partial u^{3})^{2}}.$ On 3-d
hypersurfaces and "slow" evolution, the geometric flows of mechanical
systems are described by a linear diffusion equation with $R_{\grave{\imath}%
\grave{j}}\sim \Delta h_{\grave{\imath}\grave{j}}.$ For relativistic models,
we have to elaborate on hydrodynamic anisotropic like transports of entropic
fields and derived geometric objects \cite{rajpoot17,ruchin13,gheorghiu16}.

\subsubsection{Geometric flow equations for relativistic Lagrange--Hamilton
systems}

Applying a N-adapted variational procedure with a corresponding
re-definition of normalizing function for $\widetilde{\mathcal{F}}$ (\ref%
{lffperelm}) determined by geometric data $(\widetilde{\mathbf{g}}\mathbf{=\{%
}\widetilde{\mathbf{g}}_{\mu \nu }=[\widetilde{g}_{ij},\widetilde{g}_{ab}]\},%
\widetilde{\mathbf{N}}\mathbf{=\{}\widetilde{N}_{i}^{a}\},\widetilde{\mathbf{%
D}}),$ we obtain a system of nonlinear PDEs generalizing the R. Hamilton
equations for geometric flow evolution of relativistic Lagrange systems,
\begin{eqnarray}
\partial _{\tau }\widetilde{g}_{ij} &=&-2\widetilde{\mathbf{R}}_{ij};\
\partial _{\tau }\widetilde{g}_{ab} = -2\widetilde{\mathbf{R}}_{ab};
\label{ricciflowr2} \\
\widetilde{\mathbf{R}}_{ia} &=&\widetilde{\mathbf{R}}_{ai}=0;\widetilde{%
\mathbf{R}}_{ij}=\widetilde{\mathbf{R}}_{ji};\widetilde{\mathbf{R}}_{ab}=%
\widetilde{\mathbf{R}}_{ba};  \notag \\
\partial _{\tau }\widetilde{f} &=&-\widetilde{\square }\widetilde{f}%
+\left\vert \widetilde{\mathbf{D}}\widetilde{f}\right\vert ^{2}-\ \ _{s}%
\widetilde{R}).  \notag
\end{eqnarray}%
In these formulas, $\widetilde{\square }(\tau )=\widetilde{\mathbf{D}}%
^{\alpha }(\tau )\widetilde{\mathbf{D}}_{\alpha }(\tau )$ and the conditions
$\widetilde{\mathbf{R}}_{ia}=0$ and $\widetilde{\mathbf{R}}_{ai}=0$ for the
Ricci tensor $Ric[\widetilde{\mathbf{D}}]=\{\widetilde{\mathbf{R}}_{\alpha
\beta }=[\widetilde{\mathbf{R}}_{ij},\widetilde{\mathbf{R}}_{ia},\widetilde{%
\mathbf{R}}_{ai},\widetilde{\mathbf{R}}_{ab}]\}$ are imposed in order to
keep a symmetric metric evolution.

For the geometric flow evolution of relativisitic Hamilton mechanical systems, the analogs of (\ref{ricciflowr2}) can be written (in principle, such equations can be proven in abstract form  dualizing  geometric objects from the tangent Lorentz bundles to respective cotangent bundles and
functional $\ ^{\shortmid }\widetilde{\mathcal{F}}$ (\ref{ffperelmctl})) for
the geometric data $(\ ^{\shortmid }\widetilde{\mathbf{g}}=\{ \ ^{\shortmid }\widetilde{\mathbf{g}}_{\mu \nu }=[\ ^{\shortmid }\widetilde{g}_{ij},\ ^{\shortmid }\widetilde{g}_{ab}]\},\ ^{\shortmid }\widetilde{\mathbf{N}}\mathbf{=\{}\ ^{\shortmid }\widetilde{N}_{i}^{a}\},\ ^{\shortmid }%
\widetilde{\mathbf{D}}),$
\begin{eqnarray}
\partial _{\tau }\ ^{\shortmid }\widetilde{g}_{ij} &=&-2\ ^{\shortmid }%
\widetilde{\mathbf{R}}_{ij};\ \partial _{\tau }\ ^{\shortmid }\widetilde{g}%
_{ab} = -2\ ^{\shortmid }\widetilde{\mathbf{R}}_{ab};  \label{ricciflowr2a}
\\
\ ^{\shortmid }\widetilde{\mathbf{R}}_{ia} &=&\ ^{\shortmid }\widetilde{%
\mathbf{R}}_{ai}=0;\ ^{\shortmid }\widetilde{\mathbf{R}}_{ij}=\ ^{\shortmid }%
\widetilde{\mathbf{R}}_{ji};\ ^{\shortmid }\widetilde{\mathbf{R}}_{ab}=\
^{\shortmid }\widetilde{\mathbf{R}}_{ba};  \notag \\
\partial _{\tau }\ ^{\shortmid }\widetilde{f} &=&-\ ^{\shortmid }\widetilde{%
\square }\widetilde{f}+\left\vert \ ^{\shortmid }\widetilde{\mathbf{D}}\
^{\shortmid }\widetilde{f}\right\vert ^{2}-\ \ _{s}^{\shortmid }\widetilde{R}%
),  \notag
\end{eqnarray}%
where $\ ^{\shortmid }\widetilde{\square }(\tau )=\ ^{\shortmid }\widetilde{%
\mathbf{D}}^{\alpha }(\tau )\ ^{\shortmid }\widetilde{\mathbf{D}}_{\alpha
}(\tau ).$

Using nonholonomic deformations of d-connections (\ref{candistr}),
respective frame transforms and re-definition of normalizing functions, the
geometric flow evolution equations can be written in "hat" variables or for
LC-configurations. Imposing corresponding classes of nonholonomic
constraints, we may drive the flows of geometric objects in a "pure"
mechanical form, or mix the frames and indices and generate new classes of
nonholonomic phase spacetimes.

\subsubsection{Nonholonomic Ricci solitons, emergent gravity, and geometric
mechanics}

For self-similar configurations in a fixed point $\tau =\tau _{0},$ the
geometric flows (\ref{heq1a}) are described by nonholonomic Ricci soliton
equations
\begin{equation}
\widetilde{R}_{\grave{\imath}\grave{j}}-\lambda \widetilde{g}_{\grave{\imath}%
\grave{j}}=\widetilde{\nabla }_{\grave{\imath}}v_{\grave{j}}+\widetilde{%
\nabla }_{\grave{j}}v_{\grave{\imath}},  \label{friedsoliton}
\end{equation}%
for $\lambda =\pm 1,0$ and a vector field $v_{\grave{j}}.$ In these
formulas, $\lambda $ is taken for a corresponding normalization function,
which defines a 3-d hypersurface version of the Einstein equations with
cosmological constant. We keep tilde on symbols in order to emphasize that
the geometric objects are determined by certain Lagrange or Hamilton
generating function on a 8-d (co) tangent bundle.

In a similar form, we can consider self-similar point $\tau =\tau _{0}$
configurations for the systems of nonlinear PDEs (\ref{ricciflowr2}) and/or (%
\ref{ricciflowr2a}), when $\partial _{\tau }\widetilde{\mathbf{g}}_{\mu \nu
}=0$ and/or $\partial _{\tau }\widetilde{\mathbf{g}}_{\mu \nu }=0,$ with a
corresponding choice of the normalizing geometric flow functions (for
simplicity, we can take a zero vector field $v_{\alpha }=0),$ the equations (%
\ref{ricciflowr2}) transform into relativistic nonholonomic Ricci soliton
equations%
\begin{eqnarray}
\widetilde{\mathbf{R}}_{ij} &=&\lambda \widetilde{g}_{\grave{\imath}\grave{j}%
},\ \widetilde{\mathbf{R}}_{ab}=\lambda \widetilde{g}_{ab},\ \widetilde{%
\mathbf{R}}_{ia}=\widetilde{\mathbf{R}}_{ai}=0,\mbox{ on }T\mathbf{V};
\label{entrsolit} \\
\ ^{\shortmid }\widetilde{\mathbf{R}}_{ij} &=&\lambda \ ^{\shortmid }%
\widetilde{g}_{\grave{\imath}\grave{j}},\ \ ^{\shortmid }\widetilde{\mathbf{R%
}}_{ab}=\lambda \ ^{\shortmid }\widetilde{g}_{ab},\ \ ^{\shortmid }%
\widetilde{\mathbf{R}}_{ia}=\ ^{\shortmid }\widetilde{\mathbf{R}}_{ai}=0,%
\mbox{ on }T^{\ast }\mathbf{V}.  \notag
\end{eqnarray}%
Such equation can be written in hat and/or LC-variables using nonholonomic
deformations of d-connections (\ref{candistr}) and frame transforms.
Projecting (\ref{entrsolit}) on a base 4-d Lorentz manifold $\mathbf{V,}$ we
obtain nonholonomically deformed vacuum Einstein equations with cosmological
constant $\lambda .$

In this work, we do not study gravitational and matter field geometric field interactions. Nevertheless, we note that in our nonholonomic geometric flow approach to investigating the evolution of Lagrange-Hamilton systems, the gravitational field equations emerge from geometric flows of mechanical systems being characterized by a W-entropy (\ref{perem4drelat}) and respective thermodynamical values (\ref{4dthermodval}). The gravitational constant can be introduced for identifications with respective spherical symmetric solutions with an additional assumption that at long distances the standard Newton gravitational potential is generated. In certain sense, for such theories, a W-entropy acts as an entropic force for the E. Verlinde model \cite{verlinde10,verlinde16}, see proofs in \cite{bubuianu19,vacaru19,vacaru19a}.

\section{Classical \& quantum mechanical geometric information flow theories}

\label{s4} This section is a short introduction to basic aspects of
classical and quantum geometric information flow (respectively, GIF and
QGIF) models and related subjects from the theory of geometric evolution of
relativistic mechanical systems (elaborated in previous sections). Using
modified G. Perelman entropy functionals and the nonholonomically
adapted von Neumann entropy for quantum density matrices, there are defined
quantum conditional entropy, relative entropy, and mutual information values
as basic ingredients of the QGIF theory.

\subsection{Geometric information flow theory of classical mechanical systems}

Classical information theory is based on fundamental concepts of Shannon,
conditional and relative entropies \cite%
{preskill,witten18,nielsen,cover,wilde}. To elaborate on classical aspects
of geometric information flow, GIF, models we have to define analogous
values determined by (modified) Perelman entropy functionals and associated thermodynamical models.

\subsubsection{Shannon entropy and geometric flow entropy in information
theories}

Let us remember the general definition of the Shannon entropy $S_{B}$ of a
probability distribution for a random variable $B$ taking certain values $b_{%
\underline{1}},b_{\underline{2}},...,b_{\underline{k}}$ (for instance, to
send a long message $\underline{N} \gg 1$ with $k$ letters) with respective
probabilities to observe such values $p_{\underline{1}},p_{\underline{2}%
},...,p_{\underline{k}}.$\footnote{%
In this section, we should not confuse symbols for probabilities $p_{%
\underline{i}}$ with similar notations for cofiber coordinates; and a number
 $\underline{N}$ is different from the symbol $\mathbf{N}$ used for the
N-connections. Here we note that it is almost impossible and not optimal to
elaborate an unified system of notations with completely different symbols
in an article involving different directions in differential geometry,
geometric mechanics, probability and diffusion, classical and quantum
information theory. We try to keep traditional notations for different
directions in mathematics or physics but (if necessary) underly symbols and
provide respective remarks allowing to avoid notation ambiguities.} By
definition,%
\begin{equation*}
S_{B}:=-\sum\limits_{\underline{j}=1}^{\underline{k}}p_{\underline{j}}\log
p_{\underline{j}}\geq 0\mbox{ with }\sum\limits_{\underline{j}=1}^{%
\underline{k}}p_{\underline{j}}=1.
\end{equation*}%
This is for the probability theory with random variables. In classical
information models, $\underline{N}S_{B}$ is the number of bits of
information which can be extracted from a message with $\underline{N}$
symbols which are randomly generated. For engineering applications, $%
\underline{N}S_{B}$ is the number of bits to which a message with $%
\underline{N}$ letters can be compressed. Typically, such messages are not
randomly generated but contain certain information. To encode certain real
messages with correlations between letters (for instance, words for grammar
and syntax) and loose less modifications is a more complex random process.
In the ideal gaze limit (ignoring correlations), we can consider that the
entropy of a long message is just $\underline{N}S$, when $S$ is the entropy
of a message consisting of only one letter. We can formalize the
constructions as models of statistical mechanics if we introduce a classical
Hamiltonian $H$ determining the probability of a $i$-th symbol $b_{i}$ in a
statistical thermodynamical model via formula $p_{\underline{i}}=2^{-H(b_{%
\underline{i}})}.$

The theory of geometric flows is different from the standard theory of
random processes, classical information models and "simple" engineering
applications. The flow evolution is characterized by the W-entropy and
(which is important for our further developments) additional assumptions on
associated statistical thermodynamic values like mean energy, entropy and
fluctuation. For classical mechanical systems, such values are canonically
determined by generating functions $\widetilde{L}$ and $\widetilde{H},$ see
formulas $\widetilde{\mathcal{W}}$ (\ref{lwfperelm}) and $\ ^{\shortmid }%
\widetilde{\mathcal{W}}$ (\ref{wfperelmctl}), and, respectively, for flow
evolution of Hessian metrics, by $\left[ \widetilde{\mathcal{E}},\widetilde{%
\mathcal{S}},\widetilde{\eta }\ \right]$ (\ref{8rdthvls}) and $\left[ \
^{\shortmid }\widetilde{\mathcal{E}},\ ^{\shortmid }\widetilde{\mathcal{S}}%
,\ ^{\shortmid }\widetilde{\eta }\right] $ (\ref{8rdthvhs}). On a discrete
network with random variables, we can introduce probabilities, for instance,
$\widetilde{p}_{\underline{n}}=2^{-\widetilde{H}(b_{\underline{n}})}$ and $\
^{\shortmid }\widetilde{p}_{\underline{n}}=2^{-\ ^{\shortmid }\widetilde{H}%
(b_{\underline{n}})},$ or, for statistical ansambles, $\widetilde{p}_{%
\underline{n}}=2^{-\widetilde{\mathcal{E}}(b_{\underline{n}})}$ and $\
^{\shortmid }\widetilde{p}_{\underline{n}}=2^{-\ ^{\shortmid }\widetilde{%
\mathcal{E}}(b_{\underline{n}})}.$ In result, it is possible to elaborate
classical information theories determined by effective Hamiltonians $%
\widetilde{H},$ or energy functionals $\widetilde{\mathcal{E}}$ and $\
^{\shortmid }\widetilde{\mathcal{E}}.$ This is for certain discrete versions
with probability models and correlations encoding information on geometric
flows of mechanical systems.

In this subsection, we  elaborate on
continuous information flow models encoding geometric evolution of
mechanical systems using the thermodynamic entropies $\widetilde{\mathcal{S}}%
[\widetilde{\mathbf{g}}(\tau )]$ and $\ ^{\shortmid }\widetilde{\mathcal{S}}%
[\ ^{\shortmid }\widetilde{\mathbf{g}}(\tau )]$ without involving in the
constructions probability distributions which appear for random variables.
Geometric flows can be described by $\widetilde{\mathcal{S}}[\widetilde{%
\mathbf{g}}(\tau )]$ and $\ ^{\shortmid }\widetilde{\mathcal{S}}[\
^{\shortmid }\widetilde{\mathbf{g}}(\tau )]$. We can elaborate equivalent
constructions for W-entropies $\widetilde{\mathcal{W}}[\widetilde{\mathbf{g}}%
(\tau )]$ and $\ ^{\shortmid }\widetilde{\mathcal{W}}[\ ^{\shortmid }%
\widetilde{\mathbf{g}}(\tau )])$. Systems under geometric flows are denoted
as $\widetilde{B}=\widetilde{B}[\widetilde{\mathbf{g}}(\tau )]$ and $\
^{\shortmid }\widetilde{B}=\ ^{\shortmid }\widetilde{B}[\ ^{\shortmid }%
\widetilde{\mathbf{g}}(\tau )] $\ determined by corresponding canonical
d-metrics on phase spacetimes.

\subsubsection{Conditional entropy and geometric information flows GIF}

In information theory, there are studied various conventional models with
communicating humans called, for instance, Alice and Bob, see \cite%
{preskill,witten18}. Let us suppose that Alice sends a message via a noisy
telephone connection with many letters (any letter is a random variable $X$
taking possible values $x_{\underline{1}},...,x_{\underline{k}}$). Bob
receives instead of $X$ a random variable $Y$ consisting from possible
letters $y_{\underline{1}},...,y_{\underline{r}}$. In classical information
theory, one computes how many bits of information does Bob receives form
Alice's message with $N$ letters? Traditionally, the random variables are
denoted as $X,Y,Z$ etc. For one variable, the probability to observe $X=x_{%
\underline{i}}$ is denoted $P_{X}(x_{\underline{i}})$ subjected to the
condition that $\sum_{\underline{i}}$ $P_{X}(x_{\underline{i}})=1.$ The
communication between Alice and Bob is a random process of two variables
defined by a joint distribution $P_{X,Y}(x_{\underline{i}},y_{\underline{j}})
$ as the probability that Alice sends $X=x_{\underline{i}}$ and Bob hears $%
Y=y_{\underline{j}}.$ It is considered that the value $\ P_{Y}(y_{\underline{%
j}})=\sum_{\underline{i}}P_{X,Y}(x_{\underline{i}},y_{\underline{j}})$ is
the probability that Bob hears $Y=y_{\underline{j}}$ (summation is over all
choices of what Alice could send). The \textit{conditional probability}
\begin{equation*}
P_{X|Y}(x_{\underline{i}}|y_{\underline{j}}):=\frac{P_{X,Y}(x_{\underline{i}%
},y_{\underline{j}})}{P_{Y}(y_{\underline{j}})}
\end{equation*}%
is by definition a value characterizing that if Bob hear $Y=y_{\underline{j}%
},$ he can estimate the probability that Alice sent $x_{i}.$ We can write
for Alice's messages $P_{X}(x_{\underline{i}})=\sum_{\underline{j}}$ $%
P_{X,Y}(x_{\underline{i}},y_{\underline{j}}),$ or consider $P_{X}(x_{%
\underline{i}})$ as an independent probability density. Using these formulas,
one defines such important values:
\begin{eqnarray}
&&S_{X|Y=y_{j}}:= -\sum_{\underline{i}}P_{X|Y}(x_{\underline{i}}|y_{%
\underline{j}})\log P_{X|Y}(x_{\underline{i}}|y_{\underline{j}}),%
\mbox{ the
Shannon entropy of the conditional probability};  \notag \\
&&S_{XY}:= -\sum_{\underline{i},\underline{j}}P_{X,Y}(x_{\underline{i}},y_{%
\underline{j}})\log P_{X,Y}(x_{\underline{i}},y_{\underline{j}}),%
\mbox{
the entropy of joint distribution };  \notag \\
&&S_{Y}:= -\sum_{\underline{i},\underline{j}}P_{X,Y}(x_{\underline{i}},y_{%
\underline{j}})\log P_{Y}(y_{\underline{j}}),\mbox{ the total information
content received by Bob };  \notag \\
&& S_{X}:= -\sum_{\underline{i},\underline{j}}P_{X,Y}(x_{\underline{i}},y_{%
\underline{j}})\log P_{X}(x_{\underline{i}}),%
\mbox{  the total information
content in  Alice's message };  \label{probentropies} \\
&& S_{X|Y}:= S(X|Y)=\sum_{\underline{j}}P_{Y}(y_{\underline{j}%
})S_{X|Y=y_{j}},\mbox{  the
conditional entropy }.  \notag
\end{eqnarray}%
Using such formulas, one prove that (this can be violated by quantum
systems)
\begin{equation}
S(X|Y)=S_{XY}-S_{Y}\geq 0  \label{condentr}
\end{equation}%
and the \textit{mutual information} between $X$ and $Y$ (a measure of how
much we learn about $X$ observing $Y)$%
\begin{equation}
I(X;Y):=S_{X}-S_{XY}+S_{Y}\geq 0.  \label{mutualinf}
\end{equation}

Now, let us analyse another type of communications between Alice and Bob. We
suppose that they are research scientists and know advanced differential
geometry, classical mechanics, information theory, and theory of geometric
flows. Alice sends to Bob not only simple messages consisting from letters
and density probabilities but messages encoding that (in her world ) she
study geometric flow evolution processes of a mechanical system of type $%
\widetilde{A}=\widetilde{A}[\widetilde{\mathbf{g}}(\tau )],$ or $\
^{\shortmid }\widetilde{A}=\ ^{\shortmid }\widetilde{A}[\ ^{\shortmid }%
\widetilde{\mathbf{g}}(\tau )]$, determined by flows of Hessian metrics. Bob
will receive Alice's message (it may be a short letter) and knows that Alice
plays a game with geometric flow modeling. We denote Bob's geometric
evolution systems as $\widetilde{B}=\widetilde{B}[\ _{1}\widetilde{\mathbf{g}%
}(\tau )],$ or $\ ^{\shortmid }\widetilde{B}=\ ^{\shortmid }\widetilde{B}[\
_{1}^{\shortmid }\widetilde{\mathbf{g}}(\tau )].$ In elaborating such GIF
models, Alice and Bob could work or not with probability densities. In
principle, the thermodynamic generating functions $\widetilde{\mathcal{Z}}[%
\widetilde{\mathbf{g}}(\tau )]$ and/or $\ ^{\shortmid }\widetilde{\mathcal{Z}%
}[\ ^{\shortmid }\widetilde{\mathbf{g}}(\tau )]$ from (\ref{thgenerfunct})
can be considered as geometric flow analogs of probability densities but
they may use directly the W-entropy $\widetilde{\mathcal{W}}$ (\ref%
{lwfperelm}), or $\ ^{\shortmid }\widetilde{\mathcal{W}}$ (\ref{wfperelmctl}%
), and, respectively, for ansambles of Hessian metrics, by $\left[
\widetilde{\mathcal{E}},\widetilde{\mathcal{S}},\widetilde{\eta }\ \right] $
(\ref{8rdthvls}), or \ $\left[ \ ^{\shortmid }\widetilde{\mathcal{E}},\
^{\shortmid }\widetilde{\mathcal{S}},\ ^{\shortmid }\widetilde{\eta }\ %
\right] $ (\ref{8rdthvhs}). For simplicity, we analyze here how they may
GIF-communicate using instead of messages with random letters certain
geometric flow transfers of information encoding concepts of mechanical dual
phase spacetimes for Lorentz cotangent bundles. In such a case, we have to
use the geometric flow thermodynamic entropy $\ ^{\shortmid }\widetilde{%
\mathcal{S}}[\ ^{\shortmid }\widetilde{\mathbf{g}}(\tau )]$ associated to \
W-entropy $\ ^{\shortmid }\widetilde{\mathcal{W}}[\ ^{\shortmid }\widetilde{%
\mathbf{g}}(\tau )]$ and formulas considered in subsection \ref{ssbasicth}.
We shall use also geometric flow models on $T^{\ast }\mathbf{V}\otimes $\ $%
T^{\ast }\mathbf{V}$ with one cotangent bundle for Alice and another one for
Bob. The local coordinates on such  products of cotangent bundles are
labeled $(\ ^{\shortmid }u,\ _{1}^{\shortmid }u)$ and the normalizing
functions are of type $\ _{AB}^{\shortmid }\widetilde{f}(\ ^{\shortmid }u,\
_{1}^{\shortmid }u).$ The canonical d-metric structure on such tensor
products of phase spacetimes is of type%
\begin{equation*}
\ _{AB}^{\shortmid }\widetilde{\mathbf{g}}=\{\ ^{\shortmid }\widetilde{%
\mathbf{g}}=[q_{1},q_{2},q_{3},\breve{N},\ ^{\shortmid }q_{5},\ ^{\shortmid
}q_{6},\ ^{\shortmid }q_{7},\ ^{\shortmid }\check{N}],\ _{1}^{\shortmid }%
\widetilde{\mathbf{g}}=[\ _{1}q_{1},\ _{1}q_{2},\ _{1}q_{3},\ _{1}\breve{N}%
,\ _{1}^{\shortmid }q_{5},\ _{1}^{\shortmid }q_{6},\ _{1}^{\shortmid
}q_{7},\ _{1}^{\shortmid }\check{N}]\}.
\end{equation*}%
Respectively, we consider a canonical d--connection $\ _{AB}^{\shortmid }%
\widetilde{ \mathbf{D}}=\ ^{\shortmid }\widetilde{\mathbf{D}}+\
_{B}^{\shortmid }\widetilde{\mathbf{D}}$ and respective scalar curvature $\
_{sAB}^{\shortmid }\widetilde{R}=\ _{s}^{\shortmid }\widetilde{R}+\
_{s1}^{\shortmid }\widetilde{R}.$

We work with $\ ^{\shortmid }\widetilde{\mathcal{S}}[\widetilde{A}]$ and $\
^{\shortmid }\widetilde{\mathcal{S}}[\widetilde{B}]$ defined by respective
formulas for $\ ^{\shortmid }\widetilde{\mathbf{g}}(\tau )$ and $\
_{1}^{\shortmid }\widetilde{\mathbf{g}}(\tau )$ as in (\ref{8rdthvhs}). They
are analogs of $S_{X}$ and $S_{Y}$ in above formulas. As an analog of $%
S_{XY} $ for GIF, we consider the thermodynamic generating function (as a
generalization of (\ref{thgenerfunct}))
\begin{equation*}
\ _{AB}^{\shortmid }\widetilde{\mathcal{Z}}[\ ^{\shortmid }\widetilde{%
\mathbf{g}}(\tau ),\ _{1}^{\shortmid }\widetilde{\mathbf{g}}(\tau )]=\
^{\shortmid }\widetilde{\int \ }\ _{1}^{\shortmid }\widetilde{\int }e^{-\
_{AB}^{\shortmid }\widetilde{f}}\sqrt{|\ ^{\shortmid }\widetilde{\mathbf{g}}%
_{\alpha \beta }|}\sqrt{|\ _{1}^{\shortmid }\widetilde{\mathbf{g}}_{\alpha
\beta }|}d^{8}\ ^{\shortmid }u\ d^{8}\ _{1}^{\shortmid }u(-\
_{AB}^{\shortmid }\widetilde{f}+32),\mbox{ for }T^{\ast }\mathbf{V\otimes \
T^{\ast }\mathbf{V},}
\end{equation*}%
and resulting entropy function
\begin{eqnarray*}
&&\ _{AB}^{\shortmid }\widetilde{\mathcal{S}}=\ ^{\shortmid }\widetilde{%
\mathcal{S}}\ [\widetilde{A},\widetilde{B}]=-\ ^{\shortmid }\widetilde{\int }%
\ _{1}^{\shortmid }\widetilde{\int }e^{-\ _{AB}^{\shortmid }\widetilde{f}}%
\sqrt{|q_{1}q_{2}q_{3}\breve{N}\ ^{\shortmid }q_{5}\ ^{\shortmid }q_{6}\
^{\shortmid }q_{7}\ ^{\shortmid }\check{N}|} \\
&&\sqrt{|\ _{1}q_{1}\ _{1}q_{2}\ _{1}q_{3}\ _{1}\breve{N}\ _{1}^{\shortmid
}q_{5}\ _{1}^{\shortmid }q_{6}\ _{1}^{\shortmid }q_{7}\ _{1}^{\shortmid }%
\check{N}|}\delta ^{8}\ ^{\shortmid }u\ d^{8}\ _{1}^{\shortmid }u\ \left[
\tau \left( \ _{s}^{\shortmid }\widetilde{R}+\ _{s1}^{\shortmid }\widetilde{R%
}+|\ ^{\shortmid }\widetilde{\mathbf{D}}\ _{AB}^{\shortmid }\widetilde{f}+\
_{1}^{\shortmid }\widetilde{\mathbf{D}}\ _{AB}^{\shortmid }\widetilde{f}%
|^{2}\right) +\ _{AB}^{\shortmid }\tilde{f}-32\right] .
\end{eqnarray*}%
Using such formulas, we claim that for GIFs the formulas for the conditional
entropy (\ref{8rdthvhs}) and mutual information (\ref{8rdthvhs}) are
respectively generalized
\begin{eqnarray}
\ ^{\shortmid }\widetilde{\mathcal{S}}\ [\widetilde{A}|\widetilde{B}]&:=&\
_{AB}^{\shortmid }\widetilde{\mathcal{S}}-\ ^{\shortmid }\widetilde{\mathcal{%
S}}[\widetilde{B}]\geq 0\mbox{ and }  \label{condgfentropy} \\
\ ^{\shortmid }\widetilde{\mathcal{J}}\ [\widetilde{A};\widetilde{B}]&:=&\
^{\shortmid }\widetilde{\mathcal{S}}[\widetilde{A}]-\ _{AB}^{\shortmid }%
\widetilde{\mathcal{S}}+\ ^{\shortmid }\widetilde{\mathcal{S}}[\widetilde{B}%
]\geq 0.  \label{conditgfmutinf}
\end{eqnarray}

Similar claims can be formulated if we use the W-entropy $\ ^{\shortmid }%
\widetilde{\mathcal{W}}$ (\ref{wfperelmctl}):
\begin{equation*}
\ ^{\shortmid }\widetilde{\mathcal{W}}\ [\widetilde{A}|\widetilde{B}]:=\
_{AB}^{\shortmid }\widetilde{\mathcal{W}}-\ ^{\shortmid }\widetilde{\mathcal{%
W}}[\widetilde{B}]\geq 0\mbox{ and }\ ^{\shortmid }\widetilde{\mathcal{J}}\ [%
\widetilde{A};\widetilde{B}]:=\ ^{\shortmid }\widetilde{\mathcal{W}}[%
\widetilde{A}]-\ _{AB}^{\shortmid }\widetilde{\mathcal{W}}+\ ^{\shortmid }%
\widetilde{\mathcal{W}}[\widetilde{B}]\geq 0,
\end{equation*}%
with respective formulas computed for the W--entropy instead of the
S-entropy in the standard probability theory. For relativistic information
flows, such formulas can be applied without additional assumptions on
formulating associated statistical thermodynamic models.\footnote{\label%
{fnproofs} Let us explain why we use the word "claim" for these formulas. In
principle, the conditions of non--negativity of respective values can be
violated if Alice sends to Bob GIFs as solutions, for instance, of
generalized R. Hamilton geometric flow equations (\ref{ricciflowr2a}). For
such variants, we use the claims (\ref{condgfentropy}) and (\ref%
{conditgfmutinf}) as criteria for selecting physically realistic and viable
solutions for the information theory of geometric flow evolution of W.
Hamilton mechanical systems. Nevertheless, working on cotangent Lorentz
bundles, such claims can be transformed into theorems and proven if we
consider a causal axiomatic approach to Finsler-Lagrange-Hamilton theories
elaborated in \cite{vacaru18,bubuianu18}. Here we sketch the idea and key
steps for proving such formulas. For physicists, such formulas seem to be
natural ones; rigorous mathematical proofs require hundreds of pages and
application  of a corresponding interference of methods outlined in \cite%
{perelman1,monogrrf1,monogrrf2,monogrrf3} together with \cite%
{preskill,witten18,nielsen,cover,wilde} and, for nonholonomic
configurations, in our works \cite%
{vjmp08,vacaru09,rajpoot17,ruchin13,gheorghiu16,bubuianu19,vacaru19,vacaru19a}%
. The W-entropy and respective thermodynamic values can be defined on a 3-d
hypersurface as in (\ref{perelm3f}), (\ref{perelm3w}) and (\ref{3hsthermod}%
), and then extended for evolution on a time like curve to formulas (\ref%
{perem4drelat}) and (\ref{4dthermodval}). Then the formulas are dualized to
momentum type local coordinates on some open regions on $T^{\ast }\mathbf{%
V\otimes \ T^{\ast }\mathbf{V.}}$ Such causal curves can be defined to cover
a subspace on respective phase spacetimes, their tensor products, and
projections on lower dimensions. Here we note that in any point of a causal
curve in $T^{\ast }\mathbf{V}$ and related tensor products/ projection
spaces and subspaces we can define entopies of type (\ref{probentropies}).
This way, the geometric flow information values can be completed with
certain random variables. Alice's letters to Bob will encode not only GIFs
but also random bit information processes. We can associate entropies of
type $\ ^{\shortmid }\widetilde{\mathcal{W}}$ and/or $\ ^{\shortmid }%
\widetilde{\mathcal{S}}$ to probabilistic entropies.}

Finally, we note that above formulas can be defined and proven respectively,
and in similar forms, on $T\mathbf{V,}T\mathbf{V\otimes \ T\mathbf{V,}}$ and
other tensor products and lower dimension projections involving Lagrange
generating functions. For instance,
\begin{eqnarray*}
&&\widetilde{\mathcal{S}}\ [\widetilde{A}|\widetilde{B}]:= \ _{AB}\widetilde{%
\mathcal{S}}-\ \widetilde{\mathcal{S}}[\widetilde{B}]\geq 0\mbox{ and }\
\widetilde{\mathcal{J}}\ [\widetilde{A};\widetilde{B}]:=\ \widetilde{%
\mathcal{S}}[\widetilde{A}]-\ _{AB}\widetilde{\mathcal{S}}+\ \widetilde{%
\mathcal{S}}[\widetilde{B}]\geq 0; \\
&& \widetilde{\mathcal{W}}\ [\widetilde{A}|\widetilde{B}]:= \ _{AB}%
\widetilde{\mathcal{W}}-\ \widetilde{\mathcal{W}}[\widetilde{B}]\geq 0%
\mbox{
and } \widetilde{\mathcal{J}}_{\ \widetilde{\mathcal{W}}}\ [\widetilde{A};%
\widetilde{B}]:=\ \widetilde{\mathcal{W}}[\widetilde{A}]-\ _{AB}\widetilde{%
\mathcal{W}}+\ \widetilde{\mathcal{W}}[\widetilde{B}]\geq 0.
\end{eqnarray*}%
Such values can satisfy certain Legendre conditions and duality conditions
to respective formulas (\ref{condgfentropy}) and (\ref{condgfentropy}) and
W-analogs. The models for cotangent bundles are important for elaborating
quantum mechanical theories of GIFs with Hamilton generating functions. In
their turn, the GIF models on tangent bundles are important for encoding
quantum field theories formulated using the Lagrange formalism.

\subsubsection{Relative GIF entropy and monotonicity}

In the standard probability theory, the concept of relative entropy is
introduced if (for a random variable $X$) there are considered two
probability distributions $P_{X}$ and $Q_{X},$ where for $X=x_{\underline{i}%
},$ labeled by \underline{$i$}$=\{1,2,...s\},$ one obtains $p_{\underline{i}%
}=P_{X}(x_{\underline{i}})$ and $q_{\underline{i}}=Q_{X}(x_{\underline{i}}),$
let say, for some long messages with $\underline{N}$ letters. The key issue
is to decide which distribution describe a random process more
realistically. The relative entropy per observation (or Kullback--Liebler
divergence) is defined $S(P_{X}||Q_{X}):=\sum_{\underline{i}}p_{\underline{i}%
}(\log p_{\underline{i}}-\log q_{\underline{i}})\geq 1$ under assumption
that $\underline{N}S(P_{X}||Q_{X})\gg 1.$ This is an asymmetric value on $%
P_{X}$ and $Q_{X}$ and measure the difference between these two probability
distributions when we consider that $P_{X}$ is a correct answer and $Q_{X}$
is an initial hypothesis.

Let us study a pair of random variables $X$ and $Y$ for which we consider
two probability distributions. The fist one is a possible correlated joint
distribution
\begin{equation}
P_{X,Y}(x_{\underline{i}},y_{\underline{j}})\mbox{ and }P_{X}(x_{\underline{i%
}}):=\sum_{\underline{j}}P_{X,Y}(x_{\underline{i}},y_{\underline{j}%
}),P_{Y}(y_{\underline{j}}):=\sum_{\underline{i}}P_{X,Y}(x_{\underline{i}%
},y_{\underline{j}}).  \label{classprob}
\end{equation}%
A second probability distribution $Q_{X,Y}(x_{\underline{i}},y_{\underline{j}%
})=P_{X}(x_{\underline{i}})$ $P_{Y}(y_{\underline{j}})$ can be defined to
ignore correlations between $X$ and $Y.$  In a general context, $%
Q_{X,Y}(x_{\underline{i}},y_{\underline{j}})$ can be with correlations when $%
Q_{X}(x_{\underline{i}}):=\sum_{\underline{j}}Q_{X,Y}(x_{\underline{i}},y_{%
\underline{j}}).$ For more general constructions, we can introduce three
random variables $X,Y,Z$ described by a joint probability distribution and
related values:%
\begin{equation*}
P_{X,Y,Z}(x_{\underline{i}},y_{\underline{j}},z_{\underline{k}})\mbox{ and }%
P_{X}(x_{\underline{i}}):=\sum_{\underline{j},\underline{k}}P_{X,Y,Z}(x_{%
\underline{i}},y_{\underline{j}},z_{\underline{k}}),P_{Y,Z}(y_{\underline{j}%
},z_{\underline{k}}):=\sum_{\underline{i}}P_{X,Y,Z}(x_{\underline{i}},y_{%
\underline{j}},z_{\underline{k}}).
\end{equation*}%
If we forget the correlations between $X$ and $YZ,$ we define $Q_{X,Y,Z}(x_{%
\underline{i}},y_{\underline{j}},z_{\underline{k}}):=P_{X}(x_{\underline{i}%
})P_{Y,Z}(y_{\underline{j}},z_{\underline{k}}).$ Other type values can be
defined if we observe the subsystem $XY,$ when
\begin{equation*}
P_{X,Y}(x_{\underline{i}},y_{\underline{j}}):=\sum_{\underline{k}%
}P_{X,Y,Z}(x_{\underline{i}},y_{\underline{j}},z_{\underline{k}}),Q_{X,Y}(x_{%
\underline{i}},y_{\underline{j}}):=\sum_{\underline{k}}Q_{X,Y,Z}(x_{%
\underline{i}},y_{\underline{j}},z_{\underline{k}})=P_{X}(x_{\underline{i}%
})P_{Y}(y_{\underline{j}}).
\end{equation*}

Now, we can calculate the relative entropy $S$ and mutual information $I$ between two distributions
\begin{eqnarray*}
S(P_{X}||Q_{X}):=\sum_{i,j}P_{X,Y}(x_{\underline{i}},y_{\underline{j}})[\log
P_{X,Y}(x_{\underline{i}},y_{\underline{j}})-\log (P_{X}(x_{\underline{i}%
})P_{Y}(y_{\underline{j}}))] =S_{X}-S_{XY}+S_{Y}&=&I(X;Y); \\
S(P_{X,Y}||Q_{X,Y}):=S_{X}-S_{XY}+S_{Y}&=&I(X;Y); \\
S(P_{X,Y,Z}||Q_{X,Y,Z}):=S_{XY}-S_{XYZ}-S_{YZ}&=&I(X;YZ).
\end{eqnarray*}%
In result, one proves by explicit calculations such properties%
\begin{eqnarray*}
&&I(X;Y):= S_{X}+S_{Y}-S_{XY}\geq 0,\mbox{ subadditivity of entropy }; \\
&&S(P_{X,Y}||Q_{X,Y}) \geq S(P_{X}||Q_{X}),S(P_{X,Y,Z}||Q_{X,Y,Z})\geq
S(P_{X,Y}||Q_{X,Y}),\mbox{ monotonicity of relative entropy}.
\end{eqnarray*}%
There is also the condition of strong subadditivity
\begin{equation*}
S_{X}-S_{XYZ}-S_{YZ}\geq S_{X}-S_{XY}+S_{Y},\mbox{ or }S_{XY}+S_{YZ}\geq
S_{Y}+S_{XYZ},
\end{equation*}%
which is equivalent for the condition of monotonity of mutual information $%
I(X;YZ)\geq I(X;Y).$

Above formulas for $S$ and $I$ can be generalized for respective relative
entropy and mutual information of geometric flows of mechanical systems (for
simplicity, we consider formulas generated by certain relativistic Hamilton
generating functions $H(x,p)$). For such evolution systems, there are
considered $\ _{A}^{\shortmid }\widetilde{\mathcal{Z}}:=$ $\ ^{\shortmid }%
\widetilde{\mathcal{Z}}[\ ^{\shortmid }\widetilde{\mathbf{g}}(\tau )]$ and $%
\ _{B}^{\shortmid }\widetilde{\mathcal{Z}}:=\ _{1}^{\shortmid }\widetilde{%
\mathcal{Z}}[\ _{1}^{\shortmid }\widetilde{\mathbf{g}}(\tau )],$ see (\ref%
{thgenerfunct}), as analogs of $p_{i}=P_{X}(x_{i})$ and $q_{i}=Q_{X}(x_{i}),$
see also formulas in the previous subsection. In general, there are
considered three evolution flow canonical mechanical systems $\widetilde{A},%
\widetilde{B},\widetilde{C}.$ In result, we claim (and can prove following
the method sketched in footnote \ref{fnproofs}) by explicit integral
N-adapted calculations on $T^{\ast }\mathbf{V\otimes \ T^{\ast }\mathbf{V}}$
$\mathbf{\otimes \ T^{\ast }\mathbf{V}}$ such properties%
\begin{eqnarray*}
\ ^{\shortmid }\widetilde{\mathcal{J}}\ [\widetilde{A};\widetilde{B}] &:=&\
^{\shortmid }\widetilde{\mathcal{S}}[\widetilde{A}]-\ _{AB}^{\shortmid }%
\widetilde{\mathcal{S}}+\ ^{\shortmid }\widetilde{\mathcal{S}}[\widetilde{B}%
]\geq 0,\mbox{ subadditivity of entropy}; \\
\ ^{\shortmid }\widetilde{\mathcal{S}}\ [\ _{AB}^{\shortmid }\widetilde{%
\mathcal{Z}}||\ _{AB}^{\shortmid }\widetilde{\mathcal{Z}}] &\geq &\
^{\shortmid }\widetilde{\mathcal{S}}\ [\ _{A}^{\shortmid }\widetilde{%
\mathcal{Z}}||\ _{A}^{\shortmid }\widetilde{\mathcal{Z}}],\ ^{\shortmid }%
\widetilde{\mathcal{S}}\ [\ _{ABC}^{\shortmid }\widetilde{\mathcal{Z}}||\
_{ABC}^{\shortmid }\widetilde{\mathcal{Z}}]\geq \ ^{\shortmid }\widetilde{%
\mathcal{S}}\ [\ _{AB}^{\shortmid }\widetilde{\mathcal{Z}}||\
_{AB}^{\shortmid }\widetilde{\mathcal{Z}}], \\
&&\mbox{ monotonicity of relative entropy }.
\end{eqnarray*}%
The conditions of strong subadditivity for GIF entropies are claimed
\begin{equation*}
\ _{A}^{\shortmid }\widetilde{\mathcal{S}}-\ _{ABC}^{\shortmid }\widetilde{%
\mathcal{S}}-\ _{BC}^{\shortmid }\widetilde{\mathcal{S}} \geq \
_{A}^{\shortmid }\widetilde{\mathcal{S}}-\ _{AB}^{\shortmid }\widetilde{%
\mathcal{S}}+\ _{B}^{\shortmid }\widetilde{\mathcal{S}},\mbox{ or } \
_{AB}^{\shortmid }\widetilde{\mathcal{S}}+\ _{BC}^{\shortmid }\widetilde{%
\mathcal{S}} \geq \ _{B}^{\shortmid }\widetilde{\mathcal{S}}+\
_{ABC}^{\shortmid }\widetilde{\mathcal{S}}.
\end{equation*}%
In equivalent form, these formulas can be written as the condition of
monotonicity of GIFs mutual information,
\begin{equation*}
\ ^{\shortmid }\widetilde{\mathcal{J}}\ [\widetilde{A};\widetilde{B}%
\widetilde{C}]\geq \ ^{\shortmid }\widetilde{\mathcal{J}}\ [\widetilde{A};%
\widetilde{B}].
\end{equation*}

Above formulas involve, for instance, the thermodynamic generating function
(as a generalization of (\ref{thgenerfunct}))
\begin{eqnarray*}
\ _{ABC}^{\shortmid }\widetilde{\mathcal{Z}}[\ ^{\shortmid }\widetilde{%
\mathbf{g}}(\tau ),\ _{1}^{\shortmid }\widetilde{\mathbf{g}}(\tau ),\
_{2}^{\shortmid }\widetilde{\mathbf{g}}(\tau )] &=&\ ^{\shortmid }\widetilde{%
\int \ }\ _{1}^{\shortmid }\widetilde{\int }\ _{2}^{\shortmid }\widetilde{%
\int }e^{-\ _{ABC}^{\shortmid }\widetilde{f}}\sqrt{|\ ^{\shortmid }%
\widetilde{\mathbf{g}}_{\alpha \beta }|}\sqrt{|\ _{1}^{\shortmid }\widetilde{%
\mathbf{g}}_{\alpha \beta }|}\sqrt{|\ _{2}^{\shortmid }\widetilde{\mathbf{g}}%
_{\alpha \beta }|}d^{8}\ ^{\shortmid }u\ d^{8}\ _{1}^{\shortmid }u\ d^{8}\
_{2}^{\shortmid }u \\
&&(-\ _{ABC}^{\shortmid }\widetilde{f}+48),\mbox{ for }T^{\ast }\mathbf{%
V\otimes \ T^{\ast }\mathbf{V}\otimes \ T^{\ast }\mathbf{V},}
\end{eqnarray*}%
with a normalizing function $\ _{ABC}^{\shortmid }\widetilde{f}(\
^{\shortmid }u,\ _{1}^{\shortmid }u,\ _{2}^{\shortmid }u),$ when the local
coordinates on such such products of cotangent bundles are labeled $(\
^{\shortmid }u,\ _{1}^{\shortmid }u,\ _{2}^{\shortmid }u).$ The canonical
d-metric structure on such tensor products of phase spacetimes is of type%
\begin{eqnarray*}
\ _{ABC}^{\shortmid }\widetilde{\mathbf{g}} &=&\{\ ^{\shortmid }\widetilde{%
\mathbf{g}}=[q_{1},q_{2},q_{3},\breve{N},\ ^{\shortmid }q_{5},\ ^{\shortmid
}q_{6},\ ^{\shortmid }q_{7},\ ^{\shortmid }\check{N}],\ _{1}^{\shortmid }%
\widetilde{\mathbf{g}}=[\ _{1}q_{1},\ _{1}q_{2},\ _{1}q_{3},\ _{1}\breve{N}%
,\ _{1}^{\shortmid }q_{5},\ _{1}^{\shortmid }q_{6},\ _{1}^{\shortmid
}q_{7},\ _{1}^{\shortmid }\check{N}], \\
\ _{2}^{\shortmid }\widetilde{\mathbf{g}} &=&[\ _{2}q_{1},\ _{2}q_{2},\
_{2}q_{3},\ _{2}\breve{N},\ _{2}^{\shortmid }q_{5},\ _{2}^{\shortmid
}q_{6},\ _{2}^{\shortmid }q_{7},\ _{2}^{\shortmid }\check{N}]\}.
\end{eqnarray*}%
We can consider a canonical d--connection $\ _{ABC}^{\shortmid }\widetilde{%
\mathbf{D}}=\ ^{\shortmid }\widetilde{\mathbf{D}}+\ _{B}^{\shortmid }%
\widetilde{\mathbf{D}}$ $+\ _{C}^{\shortmid }\widetilde{\mathbf{D}}$ and
respective scalar curvature $\ _{sABC}^{\shortmid }\widetilde{R}=\
_{s}^{\shortmid }\widetilde{R}+\ _{s1}^{\shortmid }\widetilde{R}+\
_{s2}^{\shortmid }\widetilde{R}.$ The resulting entropy function
\begin{eqnarray*}
\ _{ABC}^{\shortmid }\widetilde{\mathcal{S}} &=&\ ^{\shortmid }\widetilde{%
\mathcal{S}}\ [\widetilde{A},\widetilde{B},\widetilde{C}]=-\ ^{\shortmid }%
\widetilde{\int }\ _{1}^{\shortmid }\widetilde{\int }\ _{2}^{\shortmid }%
\widetilde{\int }e^{-\ _{ABC}^{\shortmid }\widetilde{f}}\sqrt{%
|q_{1}q_{2}q_{3}\breve{N}\ ^{\shortmid }q_{5}\ ^{\shortmid }q_{6}\
^{\shortmid }q_{7}\ ^{\shortmid }\check{N}|} \\
&&\sqrt{|\ _{1}q_{1}\ _{1}q_{2}\ _{1}q_{3}\ _{1}\breve{N}\ _{1}^{\shortmid
}q_{5}\ _{1}^{\shortmid }q_{6}\ _{1}^{\shortmid }q_{7}\ _{1}^{\shortmid }%
\check{N}|}\sqrt{|\ _{2}q_{1}\ _{2}q_{2}\ _{2}q_{3}\ _{2}\breve{N}\
_{2}^{\shortmid }q_{5}\ _{2}^{\shortmid }q_{6}\ _{2}^{\shortmid }q_{7}\
_{2}^{\shortmid }\check{N}|}\delta ^{8}\ ^{\shortmid }u\ d^{8}\
_{1}^{\shortmid }u\ d^{8}\ _{2}^{\shortmid }u \\
&&\left[ \tau \left( \ _{s}^{\shortmid }\widetilde{R}+\ _{s1}^{\shortmid }%
\widetilde{R}++\ _{s2}^{\shortmid }\widetilde{R}+|\ ^{\shortmid }\widetilde{%
\mathbf{D}}\ _{ABC}^{\shortmid }\widetilde{f}+\ _{1}^{\shortmid }\widetilde{%
\mathbf{D}}\ _{ABC}^{\shortmid }\widetilde{f}+\ _{2}^{\shortmid }\widetilde{%
\mathbf{D}}\ _{ABC}^{\shortmid }\widetilde{f}|^{2}\right) +\
_{ABC}^{\shortmid }\tilde{f}-48\right] .
\end{eqnarray*}%
Similar formulas can be derived for W-entropies and for Lagrange GIFs on $T%
\mathbf{V\otimes \ T\mathbf{V}}$ $\mathbf{\otimes \ T\mathbf{V.}}$

We conclude this introduction to the GIF theory of canonical classical
mechanical systems with two remarks: First, such constructions can be
generalized for stochastic maps and nonholonomic flow evolution and kinetic
processes of Lagrange-Hamilton systems as we studied in \cite%
{vacaru2000,vacaru2012,vacaru2013}. Here, we shall analyse a QGIF analog
when the quantum relative entropy is monotonic in any quantum channel,
including those associated to evolution of Hamiltonian quantum mechanical
systems.

Second, we shown that we are able both in the probability theory and for
geometric flow models to define conditional on some observation entropies.
There is not a good analog of the probability conditional distribution in the
quantum mechanical case. Nevertheless, there is a miracle that many
conclusions have quantum analogs \cite{witten18}. For GIFs of mechanical
Hamilton systems with a $H(\tau ,x,p)$, this is not a miracle because the
flow evolution of Hessian Hamilton metrics $\ ^{\shortmid }\widetilde{g}%
^{ab}(\tau ,x,p):=\frac{1}{2}\partial ^{2}H/\partial p_{a}\partial p_{b}$ (%
\ref{hesshs}) and respective canonical d-metrics $\ ^{\shortmid }\widetilde{%
\mathbf{g}}(\tau )$ (\ref{cdms}) are characterized by well--defined concepts
of W-entropy $\ ^{\shortmid }\widetilde{\mathcal{W}}$ (\ref{lwfperelm}) and
respective thermodynamical variables $\left[ \ ^{\shortmid }\widetilde{%
\mathcal{E}},\ ^{\shortmid }\widetilde{\mathcal{S}},\ ^{\shortmid }%
\widetilde{\eta }\right] $ (\ref{8rdthvhs}). In result, we can introduce GIF
formulas for conditional entropy and mutual entropy and their W-analogs. For
quantum developments in next subsection, we shall speculate on strong
subadditivity of quantum entropy which holds also for quantum analogs of
mechanical Hamilton systems.

\subsection{Basic ingredients of the quantum geometric information flow
theory}

The goal of this subsection is to analyze how the main concepts and formulas
for GIFs of mechanical systems can be extended to quantum theory and
formulate an approach to the theory of QGIFs. We note that a noncommutative
version of geometric flow theory was elaborated in \cite{vacaru09}. Those
results can be extended for elaborating noncommutative models of quantum
information theory. In a more simplified approach, we can consider quantum
mechanical models, and respective quantum geometric flows, by quantizing
certain relativistic mechanical Hamiltonians $H(\tau ,x,p),$ when in the
quasi-classical limits the geometric mechanics theory with Hessian metrics $%
\ ^{\shortmid }\widetilde{g}^{ab}(\tau ,x,p)$ emerges. In this work, the
main goal is to elaborate on quantum information theory for geometric flows
of mechanical systems characterized by geometric thermodynamical data $\left[
\ ^{\shortmid }\widetilde{\mathcal{W}};\ ^{\shortmid }\widetilde{\mathcal{E}}%
,\ ^{\shortmid }\widetilde{\mathcal{S}},\ ^{\shortmid }\widetilde{\eta }%
\right] ,$ see (\ref{lwfperelm}) and (\ref{8rdthvhs}).

\subsubsection{Density matrices and properties of quantum entropies for GIFs}

\paragraph{Statistical density matrix for relativistic mechanical Hamilton
flows:}

The thermodynamic generating function $\ ^{\shortmid }\widetilde{\mathcal{Z}}%
[\ ^{\shortmid }\widetilde{\mathbf{g}}(\tau )]$ (\ref{thgenerfunct}) with
canonical geometric objects determined by a Hamilton function $\widetilde{H}%
, $ see also subsection \ref{ssbasicth}, can be used for defining the state
density
\begin{equation}
\ ^{\shortmid }\widetilde{\sigma }(\beta ,\widetilde{H},\ ^{\shortmid }%
\widetilde{\mathbf{g}})=\ ^{\shortmid }\widetilde{\mathcal{Z}}^{-1}e^{-\beta
\widetilde{H}},  \label{statedens}
\end{equation}%
with $\beta =1/T,$ $\tau =T,$ as a classical analog of the density matrix in
quantum mechanics. The relative entropy between any state density$\
^{\shortmid }\widetilde{\rho }(\beta ,\widetilde{H},\ _{1}^{\shortmid }%
\widetilde{\mathbf{g}})$ and $\ ^{\shortmid }\widetilde{\sigma }(\beta ,%
\widetilde{H},\ ^{\shortmid }\widetilde{\mathbf{g}})$ is computed for a
prescribed measure $\omega (\widetilde{H}),$ for instance, on a cotangent
Lorentz bundle with $E$ considered as a thermodynamical energy parameter.

Using formulas (\ref{relentr}) and (\ref{fren}), we define for the
conditional entropy for geometric flows of Hamilton mechanical systems%
\begin{equation}
\ ^{\shortmid }\widetilde{\mathcal{S}}(\ ^{\shortmid }\widetilde{\rho }%
\shortparallel \ ^{\shortmid }\widetilde{\sigma })=\beta \lbrack \
^{\shortmid }\widetilde{\mathcal{F}}(\ ^{\shortmid }\widetilde{\rho })-\
^{\shortmid }\widetilde{\mathcal{F}}(\ ^{\shortmid }\widetilde{\sigma })],
\label{condhamentr}
\end{equation}%
where the free energy corresponding to $\ ^{\shortmid }\widetilde{\rho }$ is
$\ ^{\shortmid }\widetilde{\mathcal{F}}(\ ^{\shortmid }\widetilde{\rho }):=\
^{\shortmid }\widetilde{\mathcal{E}}(\ ^{\shortmid }\widetilde{\rho })-T\
^{\shortmid }\widetilde{\mathcal{S}}(\ ^{\shortmid }\widetilde{\rho })$. In
these formulas, the average energy is computed $\ ^{\shortmid }\widetilde{%
\mathcal{E}}(\ ^{\shortmid }\widetilde{\rho })=\int \ ^{\shortmid }%
\widetilde{\rho }\widetilde{H}d\omega (\widetilde{H})$ (i.e. using the
density matrix $\ ^{\shortmid }\widetilde{\rho })$ and the thermodynamic
entropy is $\ ^{\shortmid }\widetilde{\mathcal{S}}(\ ^{\shortmid }\widetilde{%
\rho }):=\beta \ ^{\shortmid }\widetilde{\mathcal{E}}(\ ^{\shortmid }%
\widetilde{\rho })+\log \ ^{\shortmid }\widetilde{\mathcal{Z}}(\ ^{\shortmid
}\widetilde{\rho }).$ Both values $\ ^{\shortmid }\widetilde{\mathcal{E}}(\
^{\shortmid }\widetilde{\rho })$ and \ $\ ^{\shortmid }\widetilde{\mathcal{S}%
}(\ ^{\shortmid }\widetilde{\rho })$ can be written equivalently to (\ref%
{8rdthvhs}). We note that if $\log \ ^{\shortmid }\widetilde{\mathcal{Z}}$
is independent on $\ ^{\shortmid }\widetilde{\rho }$ (as we consider in
above formulas) we have $\ ^{\shortmid }\widetilde{\mathcal{S}}(\
^{\shortmid }\widetilde{\sigma }\shortparallel \ ^{\shortmid }\widetilde{%
\sigma })=0.$

In this subsection, we elaborate on how GIFs of classical mechanical systems
can be generalized to QGIFs using basic concepts of quantum mechanics, QM,
and information theory. QM involves probabilities not as classical
probability distributions for a quantum state but, in general, as densities
matrices. Certain special QM systems can be described by pure states.
Nevertheless, to study quantum models of GIFs systems is necessary to
consider density matrices as quantum analogs of state densities of type $\
^{\shortmid }\widetilde{\sigma }$ (\ref{statedens}).

\paragraph{Density matrix for quantum information theory and associated
Hamilton mechanical systems:}

In an idealized case, a Hamiltonian GIF system $\widetilde{A}=\left[ \
^{\shortmid }\widetilde{\mathcal{E}},\ ^{\shortmid }\widetilde{\mathcal{S}}%
,\ ^{\shortmid }\widetilde{\eta }\right] $ (\ref{8rdthvhs}) can be described
by a Hilbert space $\widetilde{\mathcal{H}}_{A}.$ A state vector $\widetilde{%
\psi }_{A}\in \widetilde{\mathcal{H}}_{A}$ can be defined as infinite
dimensional complex vector solving the Schr\"{o}dinger equation with a
Hamiltonian $\widehat{H}$ taken as a well-defined quantum version of a
canonical Hamiltonian $\widetilde{H}.$ In the quasi-classical limit, from a
quantum mechanical model with $\widehat{H},$ we obtain a relativistic $%
\widetilde{H}$ and respective Hessian $\ ^{\shortmid }\widetilde{g}%
^{ab}(x,p) $ (\ref{hesshs}) and canonical d-metric $\ ^{\shortmid }%
\widetilde{\mathbf{g}}$ (\ref{cdmds}) (from which "non-tilde" d-metrics $\
^{\shortmid }\mathbf{g} $ (\ref{dmct}) emerge for general frame and
coordinate transforms on a $T\mathbf{V)}.$ We can consider unitary
transforms of type $\widetilde{\psi }_{A}\rightarrow U\psi _{A}$ and
describe the system $\widetilde{A}$ in an abstract Hilbert space $\mathcal{H}%
_{\widetilde{A}}$ (we put tilde on certain symbols if it is necessary to
emphasize that the constructions are related to quantization of a canonical
mechanical Hamiltonian system). For applications in the information theory,
a Hilbert space is approximated to a complex vector space of dimension $%
\underline{N}$ with Hermitian product, see details in \cite%
{preskill,witten18}.

We can consider a complementary system $B$ (we write $\widetilde{B}$ if it
is a quantum mechanical analog of a classical Hamilton mechanics) with an
associate Hilbert space $\mathcal{H}_{B},$ or $\mathcal{H}_{\widetilde{B}},$
with state vectors of type $\psi _{B}\in \mathcal{H}_{B}$ and/or unitary
transforms of type $\widetilde{\psi }_{B}\rightarrow \psi _{B}V\in \mathcal{H%
}_{\widetilde{B}}$. The combined Hilbert space is defined as a tensor
product, $\mathcal{H}_{A}\otimes \mathcal{H}_{B}$ and/or $\mathcal{H}_{%
\widetilde{A}}\otimes \mathcal{H}_{\widetilde{B}}.$ The state vectors for
the combined system are of type
\begin{equation*}
\psi _{AB}=\psi _{A}\otimes \psi _{B}\in \mathcal{H}_{AB}=\mathcal{H}%
_{A}\otimes \mathcal{H}_{B},
\end{equation*}%
where, for instance, $\psi _{B}=1_{B}$ is considered as the unity state
vector. For such products, the predictions about a system $\widetilde{A}$
can be made using the state vector $\widetilde{\psi }_{A}$ and forgetting
about the system $B$. In general, a generic pure state $\psi _{AB}\in
\mathcal{H}_{AB}$ is not a tensor product vector but is "entangled". This
means that if the respective dimension $\dim \mathcal{H}_{A}=\underline{N}$
and $\dim \mathcal{H}_{B}=\underline{M}$ then a generic state $\psi _{AB}$
is described by an $\underline{N}\times \underline{M}$ matrix. In quantum
information theory, it is considered that any pure state can be written as a
Schmidt decomposition%
\begin{equation}
\psi _{AB}=\sum_{\underline{i}}\sqrt{p_{\underline{i}}}\psi _{A}^{\underline{%
i}}\otimes \psi _{B}^{\underline{i}}\mbox{ or }\widetilde{\psi }_{AB}=\sum_{%
\underline{i}}\sqrt{p_{\underline{i}}}\widetilde{\psi }_{A}^{\underline{i}%
}\otimes \widetilde{\psi }_{B}^{\underline{i}}.  \label{bipartsyst}
\end{equation}%
In such formulas, the state vectors are orthonormal: for instance, $<\psi
_{A}^{\underline{i}},\psi _{A}^{\underline{j}}>=<\psi _{B}^{\underline{i}%
},\psi _{B}^{\underline{j}}>=\delta ^{\underline{i}\underline{j}},$ where $%
\delta ^{\underline{i}\underline{j}}$ is the Kronecker symbol. If $p_{%
\underline{i}}>0$ and $\sum_{\underline{i}}p_{\underline{i}}=1$ (this is
equivalent to the condition that, for instance, $\psi _{AB}$ is a unit
vector), we can treat $p_{\underline{i}}$ as probabilities. Here we note
that $\psi _{A}^{\underline{i}},$ or $\psi _{B}^{\underline{i}},$ may not be
bases of $\mathcal{H}_{A},$ or $\mathcal{H}_{B}$ (in principle, they may be
not enough for such bases).

The quantum density matrix for a system $A,$ or $\widetilde{A},$ is defined
\begin{equation*}
\rho _{A}:=\sum_{\underline{i}}p_{\underline{i}}|\psi _{A}^{\underline{i}%
}><\otimes \psi _{A}^{\underline{i}}|\mbox{ or }\rho _{\widetilde{A}}:=\sum_{%
\underline{i}}p_{\underline{i}}|\psi _{\widetilde{A}}^{\underline{i}%
}><\otimes \psi _{\widetilde{A}}^{\underline{i}}|.
\end{equation*}%
This operator is Hermitian and positive semi-definite, with trace $Tr_{%
\mathcal{H}_{A}}\rho _{A}=Tr_{\mathcal{H}_{\widetilde{A}}}\rho _{\widetilde{A%
}}=1.$ Using $\rho _{A},$ or $\rho _{\widetilde{A}},$ we can compute the
expectation value of any operator $\mathcal{O}_{A},$ or $\mathcal{O}_{%
\widetilde{A}},$ following, for instance, the rules%
\begin{eqnarray}
<\mathcal{O}>_{AB} &=&<\psi _{AB}|\mathcal{O}_{A}\otimes 1_{B}|\psi
_{AB}>=\sum_{\underline{i}}p_{\underline{i}}<\psi _{A}^{\underline{i}}|%
\mathcal{O}_{A}|\psi _{A}^{\underline{i}}><\psi _{B}^{\underline{i}%
}|1_{B}|\psi _{B}^{\underline{i}}>=  \notag \\
<\mathcal{O}>_{A} &=&\sum_{\underline{i}}p_{\underline{i}}<\psi _{A}^{%
\underline{i}}|\mathcal{O}_{A}|\psi _{A}^{\underline{i}}>=Tr_{\mathcal{H}%
_{A}}\rho _{A}\mathcal{O}_{A}.  \label{expectvalues}
\end{eqnarray}

In above formulas, we considered a bipartite system $AB,$ or $\widetilde{A}%
\widetilde{B}.$ Such systems are described in general form by quantum
denstity matrices of type $\rho _{AB},$ or $\rho _{\widetilde{A}\widetilde{B}%
}.$ Here we note that in the classical probability theory a bipartite system
$XY$ is described by a joint probability distribution $P_{X,Y}(x_{\underline{%
i}},y_{\underline{j}}),$ where $P_{X}(x_{\underline{i}}):=\sum_{\underline{j}%
}P_{X,Y}(x_{\underline{i}},y_{\underline{j}}),$ see (\ref{classprob}). For $%
AB$ as a bipartite quantum system with Hilbert space $\mathcal{H}_{A}\otimes
\mathcal{H}_{B},$ the density matrix $\rho _{AB}$ is defined in standard
quantum mechanical form: Let us consider $|\underline{i}>_{A},$ $\underline{i%
}=1,2,...,\underline{n}$ as an orthonormal basis of $\mathcal{H}_{A}$ and $%
|\underline{b}>_{B},$ $\underline{b}=1,2,..., \underline{m}$ as an
orthonormal basis of $\mathcal{H}_{B}.$ We write%
\begin{equation*}
\rho _{AB}=\sum_{\underline{i},\underline{i}^{\prime },\underline{b},%
\underline{b}^{\prime }}\rho _{\underline{i}\underline{i}^{\prime }%
\underline{b}\underline{b}^{\prime }}|\underline{i}>_{A}\otimes |\underline{b%
}>_{B}\ _{A}<\underline{i}^{\prime }|\otimes \ _{B}<\underline{b}^{\prime }|.
\end{equation*}%
For measurements of the system $A,$ it is considered the reduced density
matrix obtained by respective contracting of indices,
\begin{equation*}
\rho _{A}=Tr_{\mathcal{H}_{B}}\rho _{AB}=\sum_{\underline{i},\underline{i}%
^{\prime },\underline{b},\underline{b}}\rho _{\underline{i}\underline{i}%
^{\prime }\underline{b}\underline{b}}|\underline{i}>_{A}\ _{A}<\underline{i}%
^{\prime }|,\mbox{ for }|\underline{b}>_{B}\ _{B}<\underline{b}|=1.
\end{equation*}%
In a similar form, it is defined $\rho _{B}=Tr_{\mathcal{H}_{A}}\rho _{AB}.$
Using such formulas, we can elaborate on quantum information theory (see
reviews \cite{preskill,witten18}) and develop the approach for QGIFs.

\paragraph{Quantum density matrix for GIFs of mechanical Hamilton systems:}

Using formulas (\ref{expectvalues}), we can compute expectation values of a
state density $\ ^{\shortmid }\widetilde{\sigma }$ (\ref{statedens}) and
define a respective quantum density%
\begin{eqnarray}
\ ^{\shortmid }\widetilde{\sigma }_{AB} &=&<\ ^{\shortmid }\widetilde{\sigma
}>_{AB}=<\psi _{AB}|\ ^{\shortmid }\widetilde{\sigma }\otimes 1_{B}|\psi
_{AB}>=\sum_{\underline{i}}p_{\underline{i}}<\psi _{A}^{\underline{i}}|\
^{\shortmid }\widetilde{\sigma }|\psi _{A}^{\underline{i}}><\psi _{B}^{%
\underline{i}}|1_{B}|\psi _{B}^{\underline{i}}>=  \notag \\
\ ^{\shortmid }\widetilde{\sigma }_{A} &=&<\ ^{\shortmid }\widetilde{\sigma }%
>_{A}=\sum_{\underline{i}}p_{\underline{i}}<\psi _{A}^{\underline{i}}|\
^{\shortmid }\widetilde{\sigma }|\psi _{A}^{\underline{i}}>=Tr_{\mathcal{H}%
_{A}}\rho _{A}\ ^{\shortmid }\widetilde{\sigma }.  \label{aux01}
\end{eqnarray}%
Here the density matrix $\rho _{A}$ is taken for computing the QGIF density
matrix $\ ^{\shortmid }\widetilde{\sigma }_{A}$ determined by a state
density of the thermodynamical model for GIFs of a classical mechanical
Hamiltonian system $\ ^{\shortmid }\widetilde{\sigma }.$ For such systems,
we can work directly with quantum density matrices $\ ^{\shortmid }%
\widetilde{\sigma }_{AB}$ and $\ ^{\shortmid }\widetilde{\sigma }_{A}$ and
respective partial traces%
\begin{equation}
\ ^{\shortmid }\widetilde{\sigma }_{A}=Tr_{\mathcal{H}_{B}}\ ^{\shortmid }%
\widetilde{\sigma }_{AB}\mbox{ and }\ ^{\shortmid }\widetilde{\sigma }%
_{B}=Tr_{\mathcal{H}_{A}}\ ^{\shortmid }\widetilde{\sigma }_{AB}.
\label{aux02}
\end{equation}%
In coefficient form, we obtain such formulas
\begin{equation*}
\ ^{\shortmid }\widetilde{\sigma }_{AB}=\sum_{\underline{i},\underline{i}%
^{\prime },\underline{b},\underline{b}^{\prime }}\ ^{\shortmid }\widetilde{%
\sigma }_{\underline{i}\underline{i}^{\prime }\underline{b}\underline{b}%
^{\prime }}|\underline{i}>_{A}\otimes |\underline{b}>_{B}\ _{A}<\underline{i}%
^{\prime }|\otimes \ _{B}<\underline{b}^{\prime }|\mbox{ and }\ ^{\shortmid }%
\widetilde{\sigma }_{A}=\sum_{\underline{i},\underline{i}^{\prime },%
\underline{b},\underline{b}}\ ^{\shortmid }\widetilde{\sigma }_{\underline{i}%
\underline{i}^{\prime }\underline{b}\underline{b}}|\underline{i}>_{A}\ _{A}<%
\underline{i}^{\prime }|.
\end{equation*}

Let us discuss a concrete example with  density matrices. Consider an
isolated classical mechanical Hamitonian systems for which a QM model can be
constructed. To describe thermodynamically the geometric flow evolution of
both classical and quantum models we need respective state density and
quantum density matrix. In a pure state formalism, the mathematical
machinery gets bigger and bigger involving differential geometric concepts,
quantum mechanics and probability theories. This can be organized as quantum
information flow evolution model. Using a density matrix encoding the data
for Hamilton mechanical system, we can compute respective thermodynamical
values.

\subsubsection{Properties of entropies for QGIFs}

\paragraph{The von Neumann entropy of density matrix for QGIFs of mechanical
systems:}

Using $\ ^{\shortmid }\widetilde{\sigma }_{A},$ we can describe QGIF in a
standard QM form when the respective von Neumann entropy is used instead of
the Shannon entropy for a probability distribution,%
\begin{equation}
\ _{q}^{\shortmid }\widetilde{\mathcal{S}}(\ ^{\shortmid }\widetilde{\sigma }%
_{A}):=Tr\ ^{\shortmid }\widetilde{\sigma }_{A}\log \ ^{\shortmid }%
\widetilde{\sigma }_{A},  \label{neumgfentr}
\end{equation}%
where the trace is written in a simplified form without a label for the
corresponding Hilbert space. We use a left label $q$ as "quantum" and
emphasize that such an entropy is a quantum analog of $\ ^{\shortmid }%
\widetilde{\mathcal{S}}$ used in the thermodynamic model for geometric flow
evolution of Hamilton mechanical systems. The QGIF entropy $\
_{q}^{\shortmid }\widetilde{\mathcal{S}}(\ ^{\shortmid }\widetilde{\sigma }%
_{A})\geq 0$ and is manifestly invariant under a unitary transformation $\
^{\shortmid }\widetilde{\sigma }_{A}\rightarrow U\ ^{\shortmid }\widetilde{%
\sigma }_{A}U^{-1}.$

The quantum value $\ _{q}^{\shortmid }\widetilde{\mathcal{S}}(\ ^{\shortmid }%
\widetilde{\sigma }_{A})$ has a purifying property which is typical for
quantum information theory and does not have a classical analog. For a
bipartite system $\widetilde{\psi }_{AB}=\sum_{\underline{i}}\sqrt{p_{%
\underline{i}}}\widetilde{\psi }_{A}^{\underline{i}}\otimes \widetilde{\psi }%
_{B}^{\underline{i}}$ (\ref{bipartsyst}) and $\rho _{A}:=\sum_{\underline{i}%
}p_{\underline{i}}|\psi _{A}^{\underline{i}}>\otimes <\psi _{A}^{\underline{i%
}}|$, we write
\begin{eqnarray}
\ ^{\shortmid }\widetilde{\sigma }_{A}&:= &\sum_{\underline{i},\underline{i}%
^{\prime },\underline{b},\underline{b}}\ \sum_{\underline{k}}^{\shortmid }%
\widetilde{\sigma }_{\underline{i}\underline{i}^{\prime }\underline{b}%
\underline{b}}p_{\underline{k}}\ _{A}<\underline{i}^{\prime }||\psi _{A}^{%
\underline{k}}><\otimes \psi _{A}^{\underline{k}}||\underline{i}>_{A},
\label{aux03} \\
\ ^{\shortmid }\widetilde{\sigma }_{B}&:=&\sum_{\underline{j},\underline{j}%
^{\prime },\underline{b},\underline{b}}\ \sum_{\underline{k}}^{\shortmid }%
\widetilde{\sigma }_{\underline{j}\underline{j}^{\prime }\underline{b}%
\underline{b}}p_{\underline{k}}\ _{B}<\underline{j}^{\prime }||\psi _{B}^{%
\underline{k}}><\otimes \psi _{B}^{\underline{k}}||\underline{j}>_{B}.
\notag
\end{eqnarray}%
In both these formulas, we have the sam probabilities $p_{\underline{k}}\ $%
even the matrices and bases are different. So, it is clear that $\
_{q}^{\shortmid }\widetilde{\mathcal{S}}(\ ^{\shortmid }\widetilde{\sigma }%
_{A})=\ _{q}^{\shortmid }\widetilde{\mathcal{S}}(\ ^{\shortmid }\widetilde{%
\sigma }_{B}),$ which proves that a system $A$ and a purifying system $B$
always have the same QGIF von Neumann entropy. This holds true if $%
\widetilde{A}$ is taken for GIFs of a mechanical Hamilton system.

Because $\ _{q}^{\shortmid }\widetilde{\mathcal{S}}(\ ^{\shortmid }%
\widetilde{\sigma })$ is a typical von Neumann entropy, it has another very
important \textit{concavity }property. Let explain this for QGIFs because
there are involved certain important features induced by geometric flow
evolution. This mean that for any two density mechanical matrices $%
^{\shortmid }\widetilde{\sigma }_{1}$ and $^{\shortmid }\widetilde{\sigma }%
_{2}$ we can introduce $^{\shortmid }\widetilde{\sigma }(\lambda )=\lambda $
$^{\shortmid }\widetilde{\sigma }_{1}+(1-\lambda )$ $^{\shortmid }\widetilde{%
\sigma }_{2},$ for $0\leq \lambda \leq 1,$ and prove that $d^{2}$ $\
_{q}^{\shortmid }\widetilde{\mathcal{S}}(\ ^{\shortmid }\widetilde{\sigma }%
)/d\lambda ^{2}\leq 0.$ In result, one obtains $\ _{q}^{\shortmid }%
\widetilde{\mathcal{S}}(\ ^{\shortmid }\widetilde{\sigma }_{D})\geq $ $\
_{q}^{\shortmid }\widetilde{\mathcal{S}}(\ ^{\shortmid }\widetilde{\sigma }%
), $ here D is from diagonal, which means that dropping the off-diagonal
part of density matrix (this holds in any basis) results in entropy
increasing. $\ $This has important implications, for instance, in gravity
models emerging from (quantum) mechanical evolution theories. Pure diagonal
configurations have higher entropy than the generic off-diagonal ones.

\paragraph{Quantum generalizations of W- and thermodynamic entropy of mechanical systems:}

QGIFs can characterized not only by a von Neumann entropy of type (\ref{neumgfentr}) but also by quantum analogs of entropy values used for classical geometric flows (associated thermodynamics entropy and W-entropy). Such values can be introduced and computed in explcity form using respective formulas (\ref{aux01}), (\ref{aux02}), (\ref{aux03}) for classical conditional and mutual entropy used in formulas (\ref{condgfentropy}) and (\ref{conditgfmutinf}). The quantum formulas introduced in this subsection
can be considered for geometric flows of arbitrary systems and not only for mechanical ones. So, we write $A,B,...$ instead of $\widetilde{A},\widetilde{B},...$ and define%
\begin{equation*}
\ _{q}^{\shortmid }\widetilde{\mathcal{S}}_{AB}=Tr_{\mathcal{H}_{AB}}[(\
^{\shortmid }\widetilde{\sigma }_{AB})(\ _{AB}^{\shortmid }\widetilde{%
\mathcal{S}})]\mbox{ and }\ _{q}^{\shortmid }\widetilde{\mathcal{S}}_{A}=Tr_{%
\mathcal{H}_{A}}[(\ ^{\shortmid }\widetilde{\sigma }_{A})(\ _{A}^{\shortmid }%
\widetilde{\mathcal{S}})],\ _{q}^{\shortmid }\widetilde{\mathcal{S}}_{B}=Tr_{%
\mathcal{H}_{B}}[(\ ^{\shortmid }\widetilde{\sigma }_{B})(\ _{B}^{\shortmid }%
\widetilde{\mathcal{S}})].
\end{equation*}%
Similar formulas can be provided for the quantum version of W-entropy,%
\begin{equation*}
\ _{q}^{\shortmid }\widetilde{\mathcal{W}}_{AB}=Tr_{\mathcal{H}_{AB}}[(\
^{\shortmid }\widetilde{\sigma }_{AB})(\ _{AB}^{\shortmid }\widetilde{%
\mathcal{W}})]\mbox{ and }\ _{q}^{\shortmid }\widetilde{\mathcal{W}}_{A}=Tr_{%
\mathcal{H}_{A}}[(\ ^{\shortmid }\widetilde{\sigma }_{A})(\ _{A}^{\shortmid }%
\widetilde{\mathcal{W}})],\ _{q}^{\shortmid }\widetilde{\mathcal{W}}_{B}=Tr_{%
\mathcal{H}_{B}}[(\ ^{\shortmid }\widetilde{\sigma }_{B})(\ _{B}^{\shortmid }%
\widetilde{\mathcal{W}})].
\end{equation*}%
Such values describe QGIFs of Hamiltonian (quantum) mechanical systems.

The quantum probabilistic characteristics are described by the von Neumann entropy $\ _{q}^{\shortmid }\widetilde{\mathcal{S}}(\ ^{\shortmid } \widetilde{\sigma }_{A})$ (\ref{neumgfentr}) and corresponding
generalizations for $AB$ and $B$ systems
\begin{equation*}
\ _{q}^{\shortmid }\widetilde{\mathcal{S}}(\ ^{\shortmid }\widetilde{\sigma }%
_{AB}):=Tr\ ^{\shortmid }\widetilde{\sigma }_{AB}\log \ ^{\shortmid }%
\widetilde{\sigma }_{AB}\mbox{ and }\ _{q}^{\shortmid }\widetilde{\mathcal{S}%
}(\ ^{\shortmid }\widetilde{\sigma }_{A}):=Tr\ ^{\shortmid }\widetilde{%
\sigma }_{A}\log \ ^{\shortmid }\widetilde{\sigma }_{A},\ _{q}^{\shortmid }%
\widetilde{\mathcal{S}}(\ ^{\shortmid }\widetilde{\sigma }_{B}):=Tr\
^{\shortmid }\widetilde{\sigma }_{B}\log \ ^{\shortmid }\widetilde{\sigma }%
_{B}.
\end{equation*}%
Finally, we note that the entropies
$\ _{q}^{\shortmid }\widetilde{\mathcal{S}}_{A},\ _{q}^{\shortmid }\widetilde{\mathcal{W}}_{A},$ and $\ _{q}^{\shortmid }\widetilde{\mathcal{S}}(\ ^{\shortmid }\widetilde{\sigma }_{A})$ characterize respectively different thermodynamic, geometric flow and probabilistic properties of QGIFs of geometric mechanical Hamilton flows. In a similar form, we can omit the label "$\ ^{\shortmid }$" and derive
respective formulas for quantum flows of Lagrange systems. Such a formalism is more sophisticate mathematically because the Lagrange generating functions can not be used directly for constructing base vectors for respective Hilbert spaces.

\paragraph{Conditional and relative quantum entropy for QGIFs of mechanical systems:}

For QGIFs, we can imitate formally many classical definitions for GIFs. As it is stated in section 3.4 of \cite{witten18}, the quantum versions are potentially misleading or not good or usual notions. This is not surprising in the case of geometric flows because they are characterized not only by certain probabilistic quantum entropies but also by G. Perelman W-entropy and geometric thermodynamic entropy. Let us outline the main equations for respective von Neumann and conditional and relative entropy of quantum mechanical geometric flows.

Using quantum matrix computations with formulas of type (\ref{aux01}), (\ref{aux02}), (\ref{aux03}), we prove such quantum properties of entropies for QGIFs:%
\begin{equation}
\ _{q}^{\shortmid }\widetilde{\mathcal{S}}[A|B]=\ _{q}^{\shortmid }%
\widetilde{\mathcal{S}}_{AB}-\ _{q}^{\shortmid }\widetilde{\mathcal{S}}_{B}%
\mbox{ and }\ _{q}^{\shortmid }\widetilde{\mathcal{J}}[A;B]=\
_{q}^{\shortmid }\widetilde{\mathcal{S}}_{A}+\ _{q}^{\shortmid }\widetilde{%
\mathcal{S}}_{AB}+\ _{q}^{\shortmid }\widetilde{\mathcal{S}}_{B}\geq 0.
\label{condqentr}
\end{equation}%
Similar claims can be formulated (from small quantum perturbations, we can prove respective theorems) for the Neumann (\ref{neumgfentr}) and quantum W-entropy (\ref{wfperelmctl}),%
\begin{eqnarray*}
\ _{q}^{\shortmid }\widetilde{\mathcal{S}}(\ ^{\shortmid }\widetilde{\sigma }%
_{A|B}) &:=&\ _{q}^{\shortmid }\widetilde{\mathcal{S}}(\ ^{\shortmid }%
\widetilde{\sigma }_{AB})-\ _{q}^{\shortmid }\widetilde{\mathcal{S}}(\
^{\shortmid }\widetilde{\sigma }_{B})\mbox{ and }\ _{q}^{\shortmid }%
\widetilde{\mathcal{J}}(\ ^{\shortmid }\widetilde{\sigma }_{A;B}):=\
_{q}^{\shortmid }\widetilde{\mathcal{S}}(\ ^{\shortmid }\widetilde{\sigma }%
_{A})-\ _{q}^{\shortmid }\widetilde{\mathcal{S}}(\ ^{\shortmid }\widetilde{%
\sigma }_{AB})+\ _{q}^{\shortmid }\widetilde{\mathcal{S}}(\ ^{\shortmid }%
\widetilde{\sigma }_{B}); \\
\ _{q}^{\shortmid }\widetilde{\mathcal{W}}[A|B] &=&\ _{q}^{\shortmid }%
\widetilde{\mathcal{W}}_{AB}-\ _{q}^{\shortmid }\widetilde{\mathcal{W}}_{B}%
\mbox{ and }\ _{q}^{\shortmid }\widetilde{\mathcal{J}}_{\ \widetilde{%
\mathcal{W}}}[A;B]=\ _{q}^{\shortmid }\widetilde{\mathcal{W}}_{A}+\
_{q}^{\shortmid }\widetilde{\mathcal{W}}_{AB}+\ _{q}^{\shortmid }\widetilde{%
\mathcal{W}}_{B}\geq 0.
\end{eqnarray*}

It should be noted that different entropies and related mutual information values characterize different properties of the QGIFs of mechanical Hamilton systems. The von Neumann type values $\ _{q}^{\shortmid }\widetilde{\mathcal{S}}(\ ^{\shortmid }\widetilde{\sigma }_{A|B})$ and $_{q}^{\shortmid }%
\widetilde{\mathcal{J}}(\ ^{\shortmid }\widetilde{\sigma }_{A;B})$ can be used for proofs of entanglement and purifcation properties of such systems following standard methods of quantum information theory. Unlike the classical case, the quantum conditional entropy is not conditional on certain classical or quantum processes. But for QGIFs, the systems are with nonholonomic structure encoding classical and/or quantum mechanical systems. The conditional properties of such systems are encoded in $\ _{q}^{\shortmid
}\widetilde{\mathcal{S}}_{A}$ and $\ _{q}^{\shortmid }\widetilde{\mathcal{J}}[A;B]$, for thermodynamical models of QGIFs, and $\ _{q}^{\shortmid }\widetilde{\mathcal{W}}_{A}$ and $\ _{q}^{\shortmid }\widetilde{\mathcal{J}} _{\ \widetilde{\mathcal{W}}}[A;B],$ for quantum geometric evolution flows.

\paragraph{Monotonicity and monogamy of entanglement of relative entropy for QGIFs:}

The relative entropies for QGIFs are positive just as for the classical GIFs. Using $\ _{q}^{\shortmid }\widetilde{\mathcal{S}}(\ ^{\shortmid }\widetilde{\sigma }_{A|B}),$ we can prove \ that such a quantum entropy is also monotonic (for proofs, we can use the same methods as in \cite{lieb,witten18}, and posses also a strong subadditivity property as in \cite{witten2}). The intuition behind the classical theory of probability is not applicable in a direct way for geometric flows and/or quantum systems. In this sense, the monotonicity of quantum relative entropies is a miracle.

Let us consider a very basic property of QGIFs described by the von Neumann entropy $\ _{q}^{\shortmid }\widetilde{\mathcal{S}}(\ ^{\shortmid }\widetilde{\sigma }_{A}).$ For a bipartite system $AB$ with two density matrices $\ ^{\shortmid }\widetilde{\rho }_{AB}$ and $\ ^{\shortmid } \widetilde{\sigma }_{AB},$ we can define the corresponding reduced density matrices on $A,$ $\ ^{\shortmid }\widetilde{\rho }_{A}=Tr_{B}$ $(\ ^{\shortmid }\widetilde{\rho }_{AB})$ and $\ ^{\shortmid }\widetilde{\sigma }%
_{A}=Tr_{B}(\ ^{\shortmid }\widetilde{\sigma }_{AB}).$ The partial trace can only reduce the relative quantum entropy, \
\begin{equation}
\ _{q}^{\shortmid }\widetilde{\mathcal{S}}(\ ^{\shortmid }\widetilde{\rho }%
_{AB}\shortparallel \ ^{\shortmid }\widetilde{\sigma }_{AB})\geq \
_{q}^{\shortmid }\widetilde{\mathcal{S}}(\ ^{\shortmid }\widetilde{\rho }%
_{A}\shortparallel \ ^{\shortmid }\widetilde{\sigma }_{A}).
\label{relatentrneq}
\end{equation}%
see also (\ref{secondthlaw}) and (\ref{condhamentr}).

For a tripartite system $ABC$ with QGIF density matrix $\ ^{\shortmid }%
\widetilde{\rho }_{ABC}$ and above montonicity property, we can proved a
strong subadditivity property for geometric flows of quantum mechanical
Hamilton systems. There are used reduced density matrices and corresponding
second density matrices
\begin{eqnarray}
\ ^{\shortmid }\widetilde{\rho }_{A} &=&T_{BC}\ ^{\shortmid }\widetilde{\rho
}_{ABC},\ ^{\shortmid }\widetilde{\rho }_{BC}=T_{A}\ ^{\shortmid }\widetilde{%
\rho }_{ABC},\ ^{\shortmid }\widetilde{\rho }_{AB}=T_{C}\ ^{\shortmid }%
\widetilde{\rho }_{ABC}\mbox{ and }  \notag \\
\ ^{\shortmid }\widetilde{\sigma }_{ABC} &=&\ ^{\shortmid }\widetilde{\rho }%
_{A}\otimes \ ^{\shortmid }\widetilde{\rho }_{BC},\ ^{\shortmid }\widetilde{%
\sigma }_{AB}=T_{C}\ ^{\shortmid }\widetilde{\sigma }_{ABC}=\ ^{\shortmid }%
\widetilde{\rho }_{A}\otimes \ ^{\shortmid }\widetilde{\rho }_{B}.
\label{aux04}
\end{eqnarray}
Using above monotonicity property, we can write \
\begin{equation*}
\ _{q}^{\shortmid }\widetilde{\mathcal{S}}(\ ^{\shortmid }\widetilde{\rho }%
_{ABC}\shortparallel \ ^{\shortmid }\widetilde{\sigma }_{ABC})\geq \
_{q}^{\shortmid }\widetilde{\mathcal{S}}(\ ^{\shortmid }\widetilde{\rho }%
_{AB}\shortparallel \ ^{\shortmid }\widetilde{\sigma }_{AB})
\end{equation*}%
and/or as the \textit{monotonicity of mutual information}
\begin{equation}
\ _{q}^{\shortmid }\mathcal{\breve{J}}(A;BC)\geq \ _{q}^{\shortmid }\mathcal{%
\breve{J}}(A;B),  \label{mmi}
\end{equation}%
which is equivalent to the condition of \textit{strong subadditivity}%
\begin{equation}
\ _{q}^{\shortmid }\widetilde{\mathcal{S}}_{AB}+\ _{q}^{\shortmid }%
\widetilde{\mathcal{S}}_{BC}\geq \ _{q}^{\shortmid }\widetilde{\mathcal{S}}%
_{B}+\ _{q}^{\shortmid }\widetilde{\mathcal{S}}_{ABC}.  \label{ssad}
\end{equation}%
These formulas follow from (\ref{aux04}); notations of type $\
_{q}^{\shortmid }\widetilde{\mathcal{S}}(\ ^{\shortmid }\widetilde{\sigma }%
_{A})=\ _{q}^{\shortmid }\mathcal{\breve{S}}_{A},\ _{q}^{\shortmid }%
\widetilde{\mathcal{S}}(\ ^{\shortmid }\widetilde{\sigma }_{AB})=\
_{q}^{\shortmid }\mathcal{\breve{S}}_{AB},\ _{q}^{\shortmid }\widetilde{%
\mathcal{S}}(\ ^{\shortmid }\widetilde{\sigma }_{ABC})=\ _{q}^{\shortmid }%
\mathcal{\breve{S}}_{ABC};$ and definitions
\begin{eqnarray*}
\ _{q}^{\shortmid }\widetilde{\mathcal{S}}(\ ^{\shortmid }\widetilde{\rho }%
_{ABC} \shortparallel \ ^{\shortmid }\widetilde{\sigma }_{ABC})&=&\
_{q}^{\shortmid }\widetilde{\mathcal{S}}(\ ^{\shortmid }\widetilde{\rho }%
_{ABC}\shortparallel \ ^{\shortmid }\widetilde{\rho }_{A}\otimes \
^{\shortmid }\widetilde{\rho }_{BC})=\ _{q}^{\shortmid }\mathcal{\breve{J}}%
(A;BC):=\ _{q}^{\shortmid }\mathcal{\breve{S}}_{A}+\ _{q}^{\shortmid }%
\mathcal{\breve{S}}_{BC}-\ _{q}^{\shortmid }\mathcal{\breve{S}}_{ABC}; \\
\ _{q}^{\shortmid }\widetilde{\mathcal{S}}(\ ^{\shortmid }\widetilde{\rho }%
_{AB} \shortparallel \ ^{\shortmid }\widetilde{\sigma }_{AB}) &=&\
_{q}^{\shortmid }\widetilde{\mathcal{S}}(\ ^{\shortmid }\widetilde{\rho }%
_{AB}\shortparallel \ ^{\shortmid }\widetilde{\rho }_{A}\otimes \
^{\shortmid }\widetilde{\rho }_{B})=\ _{q}^{\shortmid }\mathcal{\breve{J}}%
(A;B):=\ _{q}^{\shortmid }\mathcal{\breve{S}}_{A}+\ _{q}^{\shortmid }%
\mathcal{\breve{S}}_{B}-\ _{q}^{\shortmid }\mathcal{\breve{S}}_{AB}.
\end{eqnarray*}

The von Neumann entropy for QGIFs allows us to deduce an important property
related to the monogamy of entanglement when a given qubit in a QGIF system $%
\widetilde{C}$ can be entangled with $\widetilde{D}$ (reducing $\
_{q}^{\shortmid }\mathcal{\breve{S}}_{CD})$ or with $\widetilde{B}$
(reducing $\ _{q}^{\shortmid }\mathcal{\breve{S}}_{BC}),$ but not with both
systems for set of 4 QGIF systems $\widetilde{A}\widetilde{B}\widetilde{C}%
\widetilde{D}$ (for mechanical systems, we can use tilde on symbols, which
can be omitted for general GIFs). This follows from the possibility of
purification of this type of entropy, which allows to find various
equivalent systems. If we consider $ABCD$ in a pure state, then $\
_{q}^{\shortmid }\mathcal{\breve{S}}_{AB}=\ _{q}^{\shortmid }\mathcal{\breve{%
S}}_{CD},\ _{q}^{\shortmid }\widetilde{\mathcal{S}}_{ABC}=\ _{q}^{\shortmid }%
\mathcal{\breve{S}}_{D}.$ The inequality (\ref{ssad}) becomes $\
_{q}^{\shortmid }\mathcal{\breve{S}}_{CD}+\ _{q}^{\shortmid }\mathcal{\breve{%
S}}_{BC}\geq \ _{q}^{\shortmid }\mathcal{\breve{S}}_{B}+\ _{q}^{\shortmid }%
\mathcal{\breve{S}}_{D}.$ We can consider, for instance, that $\
_{q}^{\shortmid }\mathcal{\breve{S}}(C|D)=\ _{q}^{\shortmid }\mathcal{\breve{%
S}}_{CD}-\ _{q}^{\shortmid }\mathcal{\breve{S}}_{D}<0,$ or $\
_{q}^{\shortmid }\mathcal{\breve{S}}(C|B)=\ _{q}^{\shortmid }\mathcal{\breve{%
S}}_{BC}-\ _{q}^{\shortmid }\mathcal{\breve{S}}_{B}<0,$ when the \textit{%
monogamy of entanglement} follows from the non negative condition
\begin{equation}
\ _{q}^{\shortmid }\mathcal{\breve{S}}(C|D)+\ _{q}^{\shortmid }\mathcal{%
\breve{S}}(C|B)\geq 0.  \label{monog}
\end{equation}

Above important conditions (\ref{mmi}), (\ref{ssad}) and (\ref{monog}) for
the von Neumann entropy for QGIFs can be proven in a standard form for
quantum information theory \cite{lieb,witten18,witten2}. It is not clear if
similar results can be proven for the thermodynamic entropy $\
_{q}^{\shortmid }\widetilde{\mathcal{S}}_{A}$ or W-entropy $\
_{q}^{\shortmid }\widetilde{\mathcal{W}}_{A}.$ In principle, such values
characterize certain complementary properties of QGIFs and relativistic
quantum mechanical systems.

\subsubsection{Measurements for QGIFs and quantum channels}

In QM, measurements involve projection onto orthogonal subspaces of a Hilbert space $\mathcal{H}_{A}.$ The same formalism can be applied to QGIFs if we work with a density matrix
$\ ^{\shortmid }\widetilde{\sigma }$ of type (\ref{aux01}) or (\ref{aux02}).

\paragraph{Generalized measurements for QGIFs of mechanical systems:}

Let us introduce a system of \underline{$s$}$=1,...,\underline{k}$
orthogonal Hermitian projection operators $\pi _{\underline{s}}$ subjected
to the conditions $\sum_{\underline{s}=1}^{\underline{k}}\pi _{\underline{s}%
}=1;(\pi _{\underline{s}})^{2}=\pi _{\underline{s}};$ and $\pi _{\underline{s%
}}\pi _{\underline{s}^{\prime }}=0$ for $\underline{s}\neq \underline{s}%
^{\prime }.$ Applying such a $\pi _{\underline{s}}$ to a pure quantum system
$|\psi >\in \mathcal{H},$ we obtain an outcome $\underline{s}$ with
probability $p_{\underline{s}}=<\psi |\pi _{\underline{s}}|\psi >,$ when the
properties of $\pi _{\underline{s}}$ result in $\sum_{\underline{s}=1}^{%
\underline{k}}p_{\underline{s}}=1.$ If a system $\widetilde{A}$ encodes a
QGIF of a mechanical system characterized by a density matrix $\ ^{\shortmid
}\widetilde{\sigma },$ the outcome $\underline{s}$ is $\ ^{\shortmid }%
\widetilde{p}_{\underline{s}}=Tr_{\mathcal{H}}\pi _{\underline{s}}\
^{\shortmid }\widetilde{\sigma }.$ We endow such a probability $\
^{\shortmid }\widetilde{p}_{\underline{s}}$ with a typical label for a
canonical Hamilton quantum system and respective geometric flows. A
measurement with an outcome $\underline{s}$ changes the QGIFs and results in
a new density matrix%
\begin{equation}
\ ^{\shortmid }\widetilde{\sigma }_{\underline{s}}=\pi _{\underline{s}}\
^{\shortmid }\widetilde{\sigma }\pi _{\underline{s}}/\ ^{\shortmid }%
\widetilde{p}_{\underline{s}}  \label{aux05}
\end{equation}%
encoding quantum information both from the geometric flows and the
mechanical Hamilton structure.

In a more general context, measurements can be performed using an auxiliary
system $C.$ Such a system is not obligatory a mechanical one, of type $%
\widetilde{C}$ (it can be an electric device etc.). A procedure with
auxiliary $C$ is called a "positive operator-valued measurement" or POVM)
with Hilbert space $\mathcal{C}.$ Conventionally, such a $\widetilde{C}$ is $%
k$-dimensional with a basis consisting from \underline{$k$} vectors/states $|%
\underline{s}>\in \mathcal{C},$ for \underline{$s$}$=1,2,...,$\underline{$k$}%
. We can initialize such a $\mathcal{C}$-system in the state $|1>,$ then
consider a combined system $\mathcal{C}\otimes \mathcal{H}$ and a
corresponding unitary transform $U$ which, for instance, adjusts a time- and
flow parameter - dependent Hamiltonian $H$ (if we a going to study quantum
geometric flows of mechanical systems). The operator $U $ can be chosen that
for any $\psi \in \mathcal{H}, $ the result of such a transform is
parameterized using arbitrary linear operators $E_{\underline{s}},$
\begin{equation}
U(|1>\otimes \psi )=\sum_{\underline{s}=1}^{\underline{k}}|\underline{s}%
>\otimes E_{\underline{s}}\psi \mbox{ when }\text{ }\sum_{\underline{s}=1}^{%
\underline{k}}E_{\underline{s}}^{\dag }E_{\underline{s}}=1  \label{aux06}
\end{equation}%
follows from the condition of unitarity (the symbol $\dag $ is used for the
Hermitian conjugation). We can label such values with "tilde" if they are
considered for geometric mechanical flows, for instance, using $\widetilde{U}
$ and $\widetilde{E}_{\underline{s}}.$ In princile, one can be used
arbitrary operators, $U$ and $E_{\underline{s}},$ even the quantum density
matrices, see below, will be taken for QGIFs. In general, projective
measurements of the system $\mathcal{C}\otimes \mathcal{H}$ can be performed
using the commuting projection operators $\pi _{\underline{s}}=|\underline{s}%
><\underline{s}|\otimes 1$ when the probability of outcome $\underline{s}$
is $p_{\underline{s}}=|E_{\underline{s}}|\psi >|^{2}=<\psi E_{\underline{s}%
}^{\dag }E_{\underline{s}}|\psi >.$

The described above POVM procedure can be applied for measurements of a QGIF
system defined by a density matrix $\ ^{\shortmid }\widetilde{\sigma },$
when the probability of outcome $\underline{s}$ is $\ ^{\shortmid }%
\widetilde{p}_{\underline{s}}=Tr_{\mathcal{H}}E_{\underline{s}}^{\dag }E_{%
\underline{s}}\ ^{\shortmid }\widetilde{\sigma }.$ We can treat the numbers $%
\ ^{\shortmid }\widetilde{p}_{\underline{s}}$ as probabilities for any $\
^{\shortmid }\widetilde{\sigma }$ because $E_{\underline{s}}^{\dag }E_{%
\underline{s}}\geq 0$ for any $\underline{s}$ and (together with (\ref{aux06}%
)) this results in $\sum_{\underline{s}=1}^{\underline{k}}\ ^{\shortmid }%
\widetilde{p}_{\underline{s}}=1.$ It should be noted that $E_{\underline{s}%
}^{\dag }E_{\underline{s}}$ are nonnegative Hermitian operators that add to
1 but not orthogonal projection ones. After a measurement with an outcome $%
\underline{s},$ the combined system $\mathcal{C}\otimes \mathcal{H}$ can be
described by the density matrix for a "pure" quantum system. It can be
parameterized in the form $(p_{\underline{s}})^{-1}|\underline{s}><%
\underline{s}|\otimes E_{\underline{s}}|\psi ><\psi E_{\underline{s}}^{\dag
},$ see (\ref{aux05}), and, taking the partial trace over $\mathcal{C},$ we
obtain a conventional density matrix $(p_{\underline{s}})^{-1}E_{\underline{s%
}}|\psi ><\psi E_{\underline{s}}^{\dag }$ for the orginal system $\mathcal{H}%
.$ QGIFs of such quantum mechanical systems can be described by mixed stated
with density \ matrix $\ ^{\shortmid }\widetilde{\sigma },$ when $(%
\widetilde{p}_{\underline{s}})^{-1}E_{\underline{s}}\ ^{\shortmid }%
\widetilde{\sigma }E_{\underline{s}}^{\dag }$ results in an outcome $%
\underline{s}.$

Finally, we note that above POVM constructions can be generalized for any
Hilbert space of type $\mathcal{C}\otimes (\mathcal{H\oplus H}^{\prime })$
with linear transforms $E_{\underline{s}}:\mathcal{H\rightarrow H}^{\prime
}, $ which is useful for elaborating on generalized quantum models and
information theory.

\paragraph{Quantum channels for QGIFs:}

For modeling quantum information flow theories, a corresponding density
matrix evolves both in a QM form and as a geometric flow evolution process.
The usual Hamiltonian evolution of a state $|\psi >\rightarrow U|\psi >$ can
be described by a unitary operator $U$ a Hamiltonian $\widehat{H}$
corresponding to a canonical relativistic Hamiltonian $\widetilde{H}$ (and
respective Hessian $\ ^{\shortmid }\widetilde{g}^{ab}(x,p)$ (\ref{hesshs})
and canonical d-metric $\ ^{\shortmid }\widetilde{\mathbf{g}}$ (\ref{cdmds}%
)) or by a thermodynamic GIF system $\widetilde{A}=\left[ \ ^{\shortmid }%
\widetilde{\mathcal{E}},\ ^{\shortmid }\widetilde{\mathcal{S}},\ ^{\shortmid
}\widetilde{\eta }\right] $ (\ref{8rdthvhs}). In all cases, we can introduce
the von Neumann entropy $\ _{q}^{\shortmid }\widetilde{\mathcal{S}}(\
^{\shortmid }\widetilde{\sigma }_{A})$ (\ref{neumgfentr}), and conditional
entropy $\ _{q}^{\shortmid }\widetilde{\mathcal{S}}\ [\widetilde{A}|%
\widetilde{B}]$ (\ref{condqentr}), which are invariant under unitary
transforms $\ ^{\shortmid }\widetilde{\sigma }_{A}\in U\ ^{\shortmid }%
\widetilde{\sigma }_{A}U^{-1}.$ Such QGIFs are also characterized by
W-entropy $\ _{q}^{\shortmid }\widetilde{\mathcal{W}}_{A}$ (\ref{wfperelmctl}%
) and or $\ _{q}^{\shortmid }\widetilde{\mathcal{S}}_{A}$ (\ref{8rdthvhs}).

Let us analyze how the notion of quantum channels can be elaborated for
QGIFs of mechanical Hamilton systems. We consider again an extended system $%
\mathcal{C}\otimes \mathcal{H}$ enabled with a density matrix $\ ^{\shortmid
}\breve{\sigma}=|1><1|\ ^{\shortmid }\widetilde{\sigma },$ where $\
^{\shortmid }\widetilde{\sigma }$ is a density matrix on $\mathcal{H}.$
Unitary maps $\ ^{\shortmid }\breve{\sigma}\rightarrow \ ^{\shortmid }\breve{%
\sigma}^{\prime },$ and with a trace \ induced matrix $\ ^{\shortmid }%
\widetilde{\sigma }^{\prime }$ on $\mathcal{H}$, can be parameterized in the
form (\ref{aux06}),%
\begin{equation*}
\ ^{\shortmid }\breve{\sigma}^{\prime }=U\ ^{\shortmid }\breve{\sigma}%
U^{-1}=\sum_{\underline{s},\underline{s}^{\prime }=1}^{\underline{k}}|%
\underline{s}><\underline{s}^{\prime }|\otimes E_{\underline{s}}\
^{\shortmid }\widetilde{\sigma }E_{\underline{s}}^{\dag }\mbox{ and }\
^{\shortmid }\sigma ^{\prime }=Tr_{\mathcal{C}}\ ^{\shortmid }\breve{\sigma}%
^{\prime }=\sum_{\underline{s}=1}^{\underline{k}}E_{\underline{s}}\
^{\shortmid }\widetilde{\sigma }E_{\underline{s}}^{\dag }.
\end{equation*}%
In result, we can define certain "quantum channels" for evolution of QGIF
density matrices for mechanical systems as operations $\ ^{\shortmid }%
\widetilde{\sigma }\rightarrow \ \sum_{\underline{s}=1}^{\underline{k}}E_{%
\underline{s}}\ ^{\shortmid }\widetilde{\sigma }E_{\underline{s}}^{\dag },$
where the so-called Kraus operators $E_{\underline{s}}$ are subjected to the
condition $\ \sum_{\underline{s}=1}^{\underline{k}}E_{\underline{s}}E_{%
\underline{s}}^{\dag }=1.$ If we consider only one Kraus operator, we obtain
as a special case the unitary evolution of a QGIF system.

We can consider quantum channels for the relative entropy and respective
inequality conditions (\ref{relatentrneq}) which are written in the form
\begin{equation*}
\ _{q}^{\shortmid }\widetilde{\mathcal{S}}(\ ^{\shortmid }\widetilde{\rho }%
\shortparallel \ ^{\shortmid }\widetilde{\sigma })\geq \ _{q}^{\shortmid }%
\widetilde{\mathcal{S}}(\ ^{\shortmid }\widetilde{\rho }\shortparallel \
^{\shortmid }\widetilde{\sigma })
\end{equation*}%
for $\ ^{\shortmid }\widetilde{\rho }\rightarrow \ \sum_{\underline{s}=1}^{%
\underline{k}}E_{\underline{s}}\ ^{\shortmid }\widetilde{\rho }E_{\underline{%
s}}^{\dag }$ and $\ ^{\shortmid }\widetilde{\sigma }\rightarrow \ \sum_{%
\underline{s}=1}^{\underline{k}}E_{\underline{s}}\ ^{\shortmid }\widetilde{%
\sigma }E_{\underline{s}}^{\dag },$ when the fist step of initialization
consists in replacing $\ ^{\shortmid }\widetilde{\rho }$ and$\ ^{\shortmid }%
\widetilde{\sigma },$ respectively, by $|1><1\otimes \ ^{\shortmid }%
\widetilde{\rho }$ and $|1><1\otimes \ ^{\shortmid }\widetilde{\sigma }.$
This is a very general statement on monotonicity of relative entropy and the
von Neumann entropy for QGIFs of mechanical systems. The properties of Kraus
operators for quantum channels are similar to those outlined in paragraphs
(1)-(6) in section 3.7 of \cite{witten18}, see also references therein.
There are two differences: the first one is that we consider geometric flow
evolution of density matrices and that such rich quantum and geometric flow
evolutions are characterized by additional inequalities for the quantum
versions of thermodynamic entropy and W-entropy.

\paragraph{Thermodynamics of QGIFs and quantum channels:}

Let us consider a thermal quantum density matrix as in QM, $\
_{q}^{\shortmid }\widetilde{\sigma }=\ _{q}^{\shortmid }\widetilde{\mathcal{Z%
}}^{-1}e^{-\beta \widetilde{H}},$ with $\beta =1/T,$ $\tau =T.$ We define
for the conditional quantum entropy for geometric flows of Hamilton
mechanical systems $
\ _{q}^{\shortmid }\widetilde{\mathcal{S}}(\ _{q}^{\shortmid }\widetilde{%
\rho }\shortparallel \ _{q}^{\shortmid }\widetilde{\sigma })=\beta \lbrack \
_{q}^{\shortmid }\widetilde{\mathcal{F}}(\ _{q}^{\shortmid }\widetilde{\rho }%
)-\ _{q}^{\shortmid }\widetilde{\mathcal{F}}(\ _{q}^{\shortmid }\widetilde{%
\sigma })]$, 
where the free energy corresponding to a second density matrix $\ \
_{q}^{\shortmid }\widetilde{\rho }$ is $\ _{q}^{\shortmid }\widetilde{%
\mathcal{F}}(\ _{q}^{\shortmid }\widetilde{\rho }):=\ _{q}^{\shortmid }%
\widetilde{\mathcal{E}}(\ _{q}^{\shortmid }\widetilde{\rho })-T\
_{q}^{\shortmid }\widetilde{\mathcal{S}}(\ _{q}^{\shortmid }\widetilde{\rho }%
).$ The energy operator is defined and computed as $\ _{q}^{\shortmid }%
\widetilde{\mathcal{E}}(\ _{q}^{\shortmid }\widetilde{\rho })=Tr[(\
_{q}^{\shortmid }\widetilde{\rho })\widetilde{H}]$ and the thermodynamic
entropy is
\begin{equation*}
\ _{q}^{\shortmid }\widetilde{\mathcal{S}}(\ _{q}^{\shortmid }\widetilde{%
\rho }):=\beta \ _{q}^{\shortmid }\widetilde{\mathcal{E}}(\ _{q}^{\shortmid }%
\widetilde{\rho })+\log \ _{q}^{\shortmid }\widetilde{\mathcal{Z}}(\
_{q}^{\shortmid }\widetilde{\rho }).
\end{equation*}%
If $\log \ _{q}^{\shortmid }\widetilde{\mathcal{Z}}$ is independent on $\
_{q}^{\shortmid }\widetilde{\rho },$ we obtain $\ _{a}^{\shortmid }%
\widetilde{\mathcal{S}}(\ _{q}^{\shortmid }\widetilde{\sigma }\shortparallel
\ _{q}^{\shortmid }\widetilde{\sigma })=0.$ For any quantum channel
preserving the thermal equilibrium at temperature $T,$ there is a map $\
_{q}^{\shortmid }\widetilde{\sigma }$ to itself and transforms $\
_{q}^{\shortmid }\widetilde{\rho }$ to a general density matrix $\
_{q}^{\shortmid }\widetilde{\rho }^{\prime }.$ In such a quantum channel the
entropy decreases following formulas
\begin{equation*}
\ _{q}^{\shortmid }\widetilde{\mathcal{S}}(\ _{q}^{\shortmid }\widetilde{%
\rho }\shortparallel \ _{q}^{\shortmid }\widetilde{\sigma })\geq \
_{q}^{\shortmid }\widetilde{\mathcal{S}}(\ _{q}^{\shortmid }\widetilde{\rho }%
^{\prime }\shortparallel \ _{q}^{\shortmid }\widetilde{\sigma })\mbox{ and }%
\ _{q}^{\shortmid }\widetilde{\mathcal{F}}(\ _{q}^{\shortmid }\widetilde{%
\rho })\geq \ _{q}^{\shortmid }\widetilde{\mathcal{F}}(\ _{q}^{\shortmid }%
\widetilde{\rho }^{\prime }).
\end{equation*}

For quasi-classical approximations, we consider that such formulas transform
into similar ones, see (\ref{condhamentr}), for the state densities of type $%
\ ^{\shortmid }\widetilde{\sigma }$ (\ref{statedens}).

\section{Outlook and conclusions}

\label{s5} In this paper, we put emphasis on the roles of entropic values  derived from Perelman-Lyapunov type functionals \cite{perelman1,lyapunov1892} in elaborating relativistic models of geometric flow evolution of Lagrange-Hamilton mechanical systems and possible applications in classical and quantum information theory. Our aim was to seek answer to wether the incorporation of fundamental geometric objects in relativistic mechanics into canonical noholonomic structures on (co) tangent Lorentz bundles allow a new (J. Kern type) geometrization in terms of certain generalized (pseudo) Riemannian and Finsler-Lagrange-Hamilton spaces \cite{kern74,matsumoto86,vacaru18,bubuianu18}. Due to Grigory Perelman, such geometric constructions can be characterized by W-entropy functionals and respective statistical/ geometric thermodynamic functionals like average flow energy, flow entropy and flow fluctuation, see further developments and applications in physics \cite{vacaru2000,vjmp08,vrmp09,vacaru11,vacaru13,ruchin13,gheorghiu16,alexiou}.

Here it should be emphasized that such concepts of "non-area, non-holographic, non-conformal ... " entropy are more general that those based on the Bekenstein-Hawking thermodynamics
\cite{bekenstein72,bekenstein73,bardeen73,hawking75}. In our approach, the fundamental geometric and physical objects are defined by analogous metrics, nonlinear and linear connections, and their curvatures, canonically determined by Hessians of respective Lagrange and/or Hamilton generating functions. Corresponding entropic and thermodynamic type values can be computed for various classes of exact and parametric solutions (not only black hole type ones) in geometric flow evolution and (modified) gravity
theories.

The work presented here indicates that G. Perelman's ideas and geometric methods with W-entropy and associated thermodynamic models for Ricci flows presented not only an important tool for proving the Poincar\'{e}-Thurston hypothesis. The constructions can be generalized for various types of relativistic and/or non-Riemannian geometries which allow to elaborate on further developments for noncommutative, supersymmetric, stochastic and quantum geometries \cite{vacaru09,vacaru2012,vacaru2013,
rajpoot17,bubuianu19}. Although in this paper we investigated only flows of geometric mechanical Lagrange-Hamilton models elaborated on (co) tangent Lorentz bundles, and did the hole analysis based on classical and quantum mechanical Hamilton structures, our study sheds light on the importance of
such constructions in elaborating new directions in quantum information theory \cite{preskill,witten18,nielsen,cover,wilde} . We note that the conjecture that gravity can be thought of as an entropic force \cite{verlinde10,verlinde16} can be proven for certain classes of nonholonomic deformations of G. Perelman's functionals \cite{bubuianu19,vacaru19,vacaru19a}. Using the results of this and partner works
\cite{bubuianu19,vacaru19,vacaru19a}, we conclude that such proofs can be performed for the emergent gravity from classical and quantum mechanical Lagrange-Hamilton theories.

The results of section \ref{s4} support also the conclusion that using advanced geometric methods we can elaborate on basic ingredients of the geometric flow information, QGIF, theory. We close with the remark that in our future works there will be considered some more special topics of QGIFs such as teleportation and conditional geometric flow entropy; relative entropy and hypothesis geometric flow testing; how to encode classical geometric flow information in quantum states; geometric classical and
quantum flow entanglement and emergent gravity theories.

\vskip3pt

\textbf{Acknowledgments:} This research develops author's former programs partially supported by IDEI, PN-II-ID-PCE-2011-3-0256, CERN and DAAD and contains certain results for new grant proposals. The UAIC co-affiliation refers to the Project IDEI hosted by that University during 2012-2015, when the bulk of main ideas and results of this and partner works were elaborated. The third co-affiliation reflects a present visiting position at Yu. Fedkovych Chernivtsi National University.  The author is grateful to D. Singleton,  P. Stavrinos, M. V. Tkach and Ju. O. Seti for collaboration and supporting his research on geometric and quantum methods in physics and information theory.

\end{document}